\documentclass[letterpaper, 12pt]{article}
\pdfoutput = 1

\usepackage{shortcuts}

\usepackage[margin = 2.5cm]{geometry}
\titleformat{\subsubsection}
  {\normalfont\normalsize \it}{\thesubsubsection}{1em}{}

\numberwithin{equation}{section}

\usepackage[nottoc]{tocbibind}

\begin{document}

\thispagestyle{empty}

\begin{flushright}
\texttt{BRX-TH-6681}
\end{flushright}

\begin{center}

\vspace*{0.02\textheight}

{\Large \textbf{The Ultraviolet Structure of Quantum Field Theories} \\ \medskip
  Part 2: What is Quantum Field Theory?}

\vspace{1cm}

{\large \DJ or\dj e Radi\v cevi\'c}
\vspace{1em}

{\it Martin Fisher School of Physics\\ Brandeis University, Waltham, MA 02453, USA}\\ \medskip
\texttt{djordje@brandeis.edu}\\

\vspace{0.04\textheight}

\begin{abstract}
{This paper proposes a general framework for nonperturbatively defining continuum quantum field theories. Unlike most such frameworks, the one offered here is finitary: continuum theories are defined by reducing large but finite quantum systems to subsystems with conserved entanglement patterns at short distances. This makes it possible to start from a lattice theory and use rather elementary mathematics to isolate the entire algebraic structure of the corresponding low-energy continuum theory.

The first half of this paper illustrates this approach through a persnickety study of $(1 + 1)$D continuum theories that emerge from the $\Z_K$ clock model at large $K$. This leads to a direct lattice derivation of many known continuum results, such as the operator product expansion of vertex operators in the free scalar CFT. Many new results are obtained too. For example, self-consistency of the lattice-continuum correspondence leads to a rich, novel proposal for the symmetry breaking structure of the clock model at weak coupling, deep in the BKT regime. This also makes precise what one means by ``continuous'' when saying that continuous symmetries cannot be broken in $(1+1)$D.

The second half of this paper is devoted to path integrals for continuum theories of bosons and fermions defined in this finitary formalism. The path integrals constructed here have both nonperturbative lattice definitions and manifest continuum properties, such as symmetries under infinitesimal rotations or dilatations. Remarkably, this setup also makes it possible to generalize Noether's theorem to discrete symmetries.}
\end{abstract}

\vfill
\textit{Dedicated to the fighters against COVID-19.}
\end{center}

\newpage


\tableofcontents

\newpage

\section{Introduction}

Hinchcliffe's rule states that if the title of a paper is in the form of a yes/no question, the answer is always revealed to be ``no.'' One of its informal generalizations is that the answer to titular questions of the form ``what is quantum field theory?''  is ``we don't know'' \cite{ArkaniHamed:2014, Seiberg:2014}. This paper is a counterexample to this generalized Hinchcliffe's rule.

This paper is also the second, and most fundamental, part of a series dedicated to precisely understanding the lattice-continuum correspondence in quantum field theory (QFT) \cite{Radicevic:1D, Radicevic:3D, Radicevic:4D}. The twin r$\hat{\trm o}$les of this work are closely connected. The overarching goal of this series is to quantitatively understand how a theory with a large but finite Hilbert space exhibits emergent continuum behavior in an appropriate subspace. Having this understanding is equivalent to rigorously defining this emergent QFT and hence answering the question posed by this paper.

That this kind of emergence is \emph{possible} is not surprising. For example, take a chunk of metal with $N \sim 10^{23}$ electrons. It is described by a theory with a finite, $2^N$-dimensional Hilbert space. Nevertheless, it can encode a plethora of continuum QFTs (cQFTs)\cite{Sachdev:2011}.

That this kind of emergence is \emph{fundamentally useful} is perhaps less evident. Still, the ease of defining a finite quantum theory stands in stark contrast to the sophistication needed to work with continuum theories with any semblance of rigor \cite{Haag:1996}. It thus behooves us to ask whether it might be possible to give precedence to finite theories, and to view every cQFT as a controlled, well defined reduction of a suitable lattice theory.

Variants of this question have been posed many times; see \cite{Seiler:1982pw, Balaban:1983dd, Federbush:1984ef, Balaban:1985yy, Bimonte:1995jk, Jones:2014, Milsted:2016jms, Chatterjee:2018, Osborne:2019bsq, Brothier:2019asa, IfQ} for a broad but very incomplete sampling of the literature. However, progress on practical aspects of the lattice-continuum correspondence has  been surprisingly slow. Take the following three examples. First, the notion of chiral continuum theories emerging from a lattice still causes unease due to the famous no-go theorems of Nielsen and Ninomiya \cite{Nielsen:1980rz, Nielsen:1981xu}. Second, we still do not possess an acceptable lattice realization of Chern-Simons theory, to say nothing of more complicated beasts like the 6D (2,0) theory or Einstein gravity. Third, many ``advanced'' structures found in a continuum QFT, e.g.\ current algebras or operator product expansions, have only recently been quantitatively understood from a lattice viewpoint in the simplest possible examples \cite{Mong:2014ova, Milsted:2017csn, OBrien:2017wmx, Zou:2019dnc, Zou:2019iwr, Radicevic:2019jfe, Radicevic:2019mle}. We may know much, but our understanding is still fragmentary at best.

This paper aims to make progress by proposing a very general procedure to isolate the cQFT that may reside in a given lattice theory. This procedure generalizes the Wilsonian renormalization group (RG) by allowing certain short-distance degrees of freedom to be partially integrated out, so that they become \emph{classical}. This causes all states in the reduced lattice theory to have a fixed amount of entanglement at short distances. This single property will here be shown to imply many (and possibly all) familiar hallmarks of cQFT.

More concretely, the idea is as follows. The fundamental property that allows a finite quantum theory to be recast as a cQFT is the existence of a \emph{complete} set of a large number of commuting operators $n_k$ that all commute with the Hamiltonian. Without loss of generality, these operators can be assumed to be Hermitian and to have integer eigenvalues, and so they will be called \emph{particle number operators}. Their labels $k$ will be called \emph{momenta}, and operators that map $n_k$-eigenstates to each other will be called \emph{ladder operators}. This terminology makes explicit that the $n_k$'s generalize the textbook notion of occupation numbers in a Fock space. The present definition differs from customary constructions in at least three important ways. First, there is no concept of particle statistics, as ladder operators obey no specific commutation relations. Second, more broadly, there is no concept of identical particles, and in particular the maximal allowed particle numbers may differ at different $k$. Third, the momenta are not necessarily defined via any kind of Fourier transform in the starting theory.

The ladder operators will be said to generate a \emph{precontinuum basis}. The existence of a precontinuum basis is a necessary but not sufficient condition for the existence of a cQFT. A certain kind of structure on the space $\Pbb$ of momenta $k$ is also necessary. Roughly, this means that there should exist a small subset $\Pbb\_S \subset \Pbb$ such that changing the particle numbers at any $k \notin \Pbb\_S$ costs a large amount of energy. The set of operators obtained by removing all ladder operators at momenta outside of $\Pbb\_S$ from the precontinuum basis will be called a \emph{continuum basis}. Its elements span the operator algebra of the cQFT. The operators $n_k$ for $k \notin \Pbb\_S$ are, by definition, in the center of this continuum algebra. They are thus effectively classical. This implements the intuitive idea that a cQFT has an extensive number of degrees of freedom whose entanglement pattern is completely fixed. The notion that this entanglement is found at ``short distances'' can then be given currency by defining a \emph{position space} as a Fourier transform of $\Pbb$, arranged so that $\Pbb\_S$ is the set of ``low momenta.''

The large energy gap that is associated to excitations at $k \notin \Pbb\_S$ is an independent parameter that necessarily enters the definition of any cQFT. One can think of it as an energy scale $\E\_S$ that is much lower than the largest energy gaps in the spectrum, while still being much larger than the smallest energy gaps for momenta in $\Pbb\_S$. (In a theory with lattice spacing $a$ and system size $L$, this new parameter would obey $1/L \ll \E\_S \ll 1/a$.) In position space, defined as in the previous paragraph, $\E\_S$ can be understood as the parameter that governs derivative expansions. It is for this reason that this parameter was called a ``string scale'' in two earlier papers \cite{Radicevic:2019jfe, Radicevic:2019mle}. More generally, one can think of it as parameterizing the smearing of a lattice field that is needed to obtain a continuum field.

Computing cQFT quantities via some kind of lattice smearing is not novel, see e.g.\ \cite{Teper:1998te, Bernard:1999kc, Morningstar:2003gk, Davoudi:2012ya, Monahan:2015lha}. The emphasis of this paper is that a precise smearing is, in fact, \emph{the} ingredient needed to define a continuum algebra. The proposed procedure is not unique, but it is the simplest known way to rigorously define a cQFT based on a large but finite quantum theory.

\textbf{The goal of this paper} is to pedagogically work out the details of the above prescription in two familiar lattice theories: the $\Z_K$ clock model and the Dirac fermion in $(1 + 1)$D. (The continuum limit of the Dirac fermion was already formulated in \cite{Radicevic:2019jfe} using the canonical formalism; here this program is supplemented by a path integral construction.) The overall lesson will be that this way of thinking leads both to new insights and to clear and rigorous lattice-based (finitary) derivations of many known continuum results.

\subsection*{Summary of the paper}

\textbf{Section \ref{sec definitions}} will flesh out the formal ideas needed to define a generic cQFT. Precontinuum and continuum bases will be defined in detail. Importantly, this construction is not limited to free theories: many interacting cQFTs can be defined at the same level of rigor as free scalars and fermions. It will also be shown that many important notions --- such as chiral theories, the renormalization group, and hydrodynamics --- can be naturally expressed using the algebraic language of (pre)continuum bases.

In \textbf{Section \ref{sec scalars}}, the continuum behavior of a well known lattice theory, the $(1+1)$D nonchiral clock model, will be described in detail that was not available before. This theory, defined on a spatial lattice with $N$ sites arranged in a circle, has a $\Z_K$ target space (a ``clock'' with $K$ positions) at each site, and a single tunable coupling $g$ in the Hamiltonian \eqref{def H micro}. At $K \rar \infty$ and $g = O(K^0)$, the clock model becomes the quantum rotor (or O(2)) model. It is famed for the existence of the BKT line of fixed points \cite{Berezinsky:1970fr, Kosterlitz:1973xp} and for the absence of spontaneous symmetry breaking, which will here be referred to as the CHMW theorem \cite{Mermin:1966fe, Hohenberg:1967zz, Coleman:1973ci}.

\begin{figure}
\begin{center}
\begin{tikzpicture}[scale = 1]
  \contourlength{1pt}

  \fill[orange!20!pink!30] (0, 0) rectangle (3, 1);
  \draw (1.5, 0.5) node {$\Z_K$ broken};

  \fill[orange!20!pink!40] (3, 0) rectangle (4.5, 1);
  \fill[orange!30!pink!50] (4.5, 0) rectangle (5.5, 1);
  \fill[orange!40!pink!60] (5.5, 0) rectangle (6.5, 1);
  \fill[orange!60!pink!60] (6.5, 0) rectangle (8.5, 1);
  \fill[orange!70!pink!70] (8.5, 0) rectangle (9.5, 1);
  \fill[orange!80!pink!80] (9.5, 0) rectangle (10.5, 1);
  \fill[orange!90!pink!90] (10.5, 0) rectangle (12, 1);

  \draw (7.5, 0.5) node {compact scalar CFT};

  \fill[orange] (12, 0) rectangle (14.9, 1);
  \draw (13.5, 0.5) node {$\Z_K$ unbroken};


  \draw[->, thick] (0, 0) -- (15, 0);
  \draw (15, 0) node[below] {$g$};
  \draw[thick] (0, -0.1) -- (0, 0.1);
  \draw (0, 0) node[below] {0};

  \draw[thick] (2.5, -0.1) -- (2.5, 0.1);
  \draw (2.5, 0) node[below] {$g\_{KT}^\vee \sim 1/K$};

  \draw[thick] (12.5, -0.1) -- (12.5, 0.1);
  \draw (12.5, 0) node[below] {$g\_{KT} \sim 1$};

  \draw[thick] (7.5, -0.1) -- (7.5, 0.1);
  \draw (7.5, 0) node[below] {$g_\star = \sqrt{2\pi/K}$};

  \draw[thick] (3, 0) -- (3, 0.05);
  \draw[thick] (4.5, 0) -- (4.5, 0.05);
  \draw[thick] (5.5, 0) -- (5.5, 0.05);
  \draw[thick] (6.5, 0) -- (6.5, 0.05);

  \draw[thick] (8.5, 0) -- (8.5, 0.05);
  \draw[thick] (9.5, 0) -- (9.5, 0.05);
  \draw[thick] (10.5, 0) -- (10.5, 0.05);
  \draw[thick] (12, 0) -- (12, 0.05);


\end{tikzpicture}
\end{center}
\caption{\small A sketch of the proposed phase diagram for the clock model at large $K$. The $g$-axis is not drawn on a linear scale. Each shade corresponds to a different $\Z_{K'} \subset \Z_K$ being broken, i.e.\ to a different number $K'$ of approximately degenerate ground states. The Kosterlitz-Thouless critical point $g\_{KT}$ and its dual $g^\vee\_{KT}$ do not necessarily represent symmetry breaking points.}
\label{fig phases}
\end{figure}
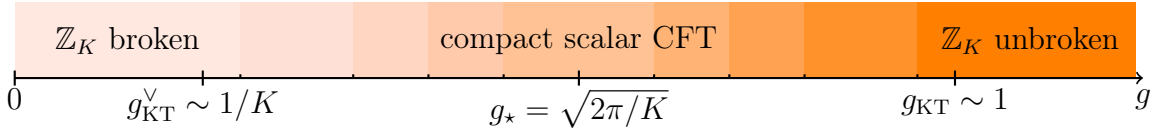

This paper will precisely define the cQFT that describes small fluctuations within the clock model at $K \gg 1$ and $N \gg 1$. This will synergize with the fact that the proof of the CHMW theorem becomes invalid as the coupling $g$ is decreased past a certain value that is small but still $O(K^0)$. At such small couplings, the $\lim_{K \rar \infty} \Z_K = $ U(1) shift symmetry \emph{does} spontaneously break. In fact, self-consistency of the cQFT analysis suggests that the breaking pattern is quite interesting.  Instead of the symmetry completely breaking at one critical value of $g$, as happens e.g.\ with the $\Z_2$ symmetry in the $(1+1)$D Ising model, here there is a \emph{cascade} of symmetry breakings as $g$ is dialed from $O(1/K)$ up to $O(K^0)$, with the spontaneously broken groups forming a sequence $\Z_K \supset \Z_{K'} \supset \ldots \supset \Z_1$ (Fig.\ \ref{fig phases}). Thus the BKT line of critical points represents a ``congealing'' of many symmetry breaking transitions --- or, more precisely, crossovers --- over an interval of $O(1)$ length in parameter space. This analysis also indicates that the BKT regime only exists at $g \geq g\_{KT}^\vee \sim 1/K$, which is the Kramers-Wannier dual of the Kosterlitz-Thouless transition point $g\_{KT} \sim 1$. (Na\"ively taking $K \rar \infty$ hides the symmetry breaking pattern and merely indicates that the BKT line extends all the way to $g = 0$.) At couplings $g < g\_{KT}^\vee$, the $\Z_K$ symmetry is fully broken, the system is in a ``ferromagnetic'' phase, and a cQFT description no longer applies.

Deep inside the BKT line in parameter space, near the self-dual point $g_\star = \sqrt{2\pi/K}$, the clock model can be precisely described by a cQFT of small bosonic fluctuations (supplemented by two sets of classical degrees of freedom, the momentum and winding modes). This is the regime in which the familiar compact boson CFT emerges, with the boson radius defined by rescaling the clock coupling, $R \equiv g\sqrt{K/\pi}$. The approach of this paper not only allows this cQFT to be defined using the starting clock variables, it also leads to a fully lattice-based derivation of all associated operator product expansions, including those of vertex operators.

All the results mentioned above are presented in the canonical formalism. Analogous definitions can be formulated using path integrals. \textbf{Section \ref{sec bos path int}} will show how this is done in exhaustive detail, starting from the clock model at appropriate couplings and rather rigorously deriving (Euclidean) continuum path integrals for scalar fields.

This derivation will result in the familiar free scalar action \eqref{def S nc tilde}. Its schematic form is
\bel{
  \frac1 \hbar \int \d x \, \d\tau \, (\del_\mu \varphi)^2.
}
The novelty here is that this action will be completely and explicitly regularized. The full regularization goes far beyond simply replacing the integral over spacetime with a sum over a grid of points. Besides the usual lattice spacings in time and space (which will be kept different throughout, indicating their fundamentally different statuses), the other cutoffs that must be specified are the maximal value that $|\varphi|$ can attain, the ``target space lattice spacing'' $\d\varphi$, and the smearing lengths that govern the smoothness of the scalar field along spatial and temporal directions. Apart from the cutoffs pertaining to the time axis, all regulators have direct counterparts in the canonical formalism.

The fact that a scalar continuum path integral involves half a dozen different cutoffs is important. The action features various parameters --- couplings, the temperature, the spatial size of the system. If they exceed any of these cutoffs, the continuum path integral will no longer compute the correct microscopic results. Such parametric limits will be estimated and explicitly stated. These bounds are important when a rigorous continuum path integral is applied to a situation in which a spacetime dimension is compactified or a coupling vanishes.

\newpage

Another unorthodox observation is that there is no method in the canonical formalism to make fields in a path integral be smooth functions of time. It is the restriction to such smooth fields that precipitates the introduction of counterterms in the action, quite independently of any renormalization group manipulations that may be done. This idea was worked out in detail in the case of fermion quantum mechanics \cite{Radicevic:1D}. Its importance to field theoretic path integrals will be briefly described here.

An analogous construction will be given in \textbf{Section \ref{sec ferm path int}} for continuum path integrals of fermionic lattice models, stressing the similarities and (important) differences between scalar and Berezin path integrals. Purely fermionic cQFTs are easier to define than purely scalar ones, and the corresponding path integrals will be defined completely rigorously. As such, the approach of this paper forms an alternative to some recent rigorous definitions of fermionic continuum path integrals that use discrete holomorphy (see \cite{Cardy:2009, DuminilCopin:2012} for reviews).

\textbf{Section \ref{sec symmetries}} will discuss symmetries of the rigorously defined continuum path integrals. The focus will be on fermionic theories, though the same methods will apply to bosonic ones. This Section will highlight how the various symmetries and their currents are defined in the manifest presence of the lattice. The key insight here, similar in spirit to \cite{Davoudi:2012ya}, is simple: requiring that path integral variables be smooth lattice fields allows one to define continuous symmetry transformations while keeping the lattice manifest. Another lesson will be that it is important to distinguish between symmetries of the Hamiltonian and symmetries of the action. The latter are a superset of the former. Symmetries of the action but not of the Hamiltonian --- such as spacetime rotations/boosts, or dilatations --- are approximate symmetries only, with no natural counterparts in the canonically defined lattice theory.

This Section will also derive Noether currents for all the listed symmetries. This derivation differs from textbook ones because it treats the lattice (and the smearing) explicitly. As a consequence, it applies to all symmetries, \emph{including discrete ones}. For example, currents of chiral fermion number symmetries will be derived assuming arbitrary phase rotations of corresponding Grassmann fields. Even more interestingly, the Noether current of the $\Z_2$ charge conjugation symmetry of the scalar cQFT will be shown to take the form $J_\mu = \varphi \del_\mu \varphi$.

\textbf{Warning:} A cursory reading of this paper may lead the reader astray in at least two ways. First, the didactic focus on free cQFTs may make it seem that this paper merely rederives known results using unusual notation. To avoid this snare, keep in mind that, contrary to custom, every single quantity in this paper is \emph{rigorously} defined, and all approximations (except for those invoking universality at the end of the paper) are under analytic control. Second, deeper thinkers may note that the smoothing philosophy of this paper is not novel because it is implicit in existing axiomatic approaches. While true, this misses the point that the present definition of cQFT is \emph{elementary}, and does not rely on anything more complicated than linear algebra and quantum mechanics in a finite-dimensional Hilbert space.

\newpage

\section{General concepts} \label{sec definitions}

\subsection{A finitary definition of continuum QFT} \label{subsec definition}

Consider a lattice theory whose Hilbert space $\H$ is a (possibly graded) product of $K$-dimensional Hilbert spaces $\H_v$ associated to vertices $v$ of some finite graph $\Mbb$. This is not the most general possible setup for what follows: one can also consider Hilbert spaces associated to links, faces, and other parts of the graph, and the various local Hilbert spaces need not all have the same dimension. However, the simple model of $K$ degrees of freedom per vertex will suffice to illustrate the needed ideas.

Let $\{\O_v^i\}$ be a basis of the operator algebra $\A$ acting on this theory, with $i$ labeling ``local'' operators associated to site $v \in \Mbb$.  In the simplest case, $i$ takes $K^2$ different values, and the corresponding algebra at $v$ is isomorphic to $\C^{K\times K}$. The set of local operators at each site may also be restricted via gauging, orbifolding, taming, etc. Instead of specifying a basis, it is enough to simply specify the \emph{generators} of an algebra --- a particular set of operators whose all possible products form the desired basis.

A \emph{precontinuum basis} of $\A$ is a basis generated by a set of \emph{ladder operators} $c_k$ and $c_k\+$. The label $k$ will be called the \emph{spatial momentum}, or just \emph{momentum}, and the set $\bb P$ of its possible values will be called the \emph{momentum space}. Ladder operators obey the following conditions:
\begin{enumerate}
  \item The operators $n_k \equiv c_k\+ c_k$ at different $k$ must all commute with each other.
  \item Each $n_k$ is a particle number operator, so that, for any $k \in \bb P$, diagonalizing $n_k$ yields
      \bel{
        n_k = \trm{diag}(0, 1, \ldots, J_k - 1) \otimes \bigotimes_{\substack{k' \in \bb P \\ k' \neq k}} \1_{k'} \quad \trm{for}\quad J_k \geq 2.
      }

  \item The ladder operators must map the eigenstates of $n_k$ to each other: they raise or lower the particle number (the eigenvalue of $n_k$) by one, and annihilate the eigenstate when further raising or lowering is impossible. They thus necessarily obey $c_k^{J_k} = (c_k\+)^{J_k} = 0$, but do \emph{not} obey the canonical commutation relations for bosons, $[c_k, c\+_k] = \1$, even when $J_k \gg 1$ --- indeed, no finite matrices can obey this commutation relation. When $J_k = 2$ the above requirements do imply the fermionic commutation relation, $\{c_k, c_k\+\} = \1$. The $c_k$'s at different momenta $k$ need not obey any particular (anti)commutation relations. In this sense the precontinuum basis generalizes the usual notion of a Fock basis.
  \item Lastly, the $n_k$'s are required to be symmetries of a lattice Hamiltonian $H \in \A$. This means that different lattice theories can have different precontinuum bases (and hence different cQFT descriptions) even if their underlying Hilbert spaces are isomorphic.
\end{enumerate}

If a precontinuum basis with a large number of different momenta $k$ exists, the lattice theory can have a continuum limit. A \emph{continuum basis} \label{ref cont basis} is obtained by removing ladder operators at certain momenta from the precontinuum basis, while keeping the associated particle number operators. In order to do this in a controlled way, it is necessary to impose some kind of ordering on the momentum space $\Pbb$, and to then use it to differentiate between discarded and preserved momenta. A natural choice here can be induced by the Hamiltonian.


Before specifying the requirements for a continuum basis, it will be useful to elaborate on the Hamiltonians of interest. Since the $n_k$'s are symmetries and since their constituent ladder operators generate the entire algebra $\A$, any Hamiltonian with a precontinuum basis takes the form
\bel{\label{def H general}
  H = h^{(0)} + \sum_{k \in \Pbb} \left(h_k^{(1)} n_k + h_k^{(2)} n_k^2 + \ldots \right) + \sum_{k,\, l \in \Pbb} \left(h_{k,\, l}^{(1, 1)} n_k n_l + h_{k,\, l}^{(2, 1)} n_k^2 n_l +  \ldots \right) + \ldots
}
The omitted terms contain all possible powers of each $n_k$ going as far as $n_k^{J_k - 1}$; any higher powers can be expressed in terms of the lower ones. A theory will be called \emph{free} if all coefficients in \eqref{def H general} are zero except for the $h_k^{(1)}$'s and the insignificant additive term $h^{(0)}$.

Henceforth the $h_k^{(1)}$'s will be called \emph{dispersions}. In a free theory, the distribution of dispersions may be used to endow the space $\Pbb$ with a lattice structure. The situation is particularly simple if the dispersion function can be written in a parametric form like
\bel{\label{sample distributions}
  h_k^{(1)} = \omega k \quad \trm{or} \quad h_k^{(1)} = |\sin \omega k|, \quad -N \leq k < N,
}
for some sufficiently large integer $N$. In such cases it is reasonable to think of the parameters $k$ as elements of the momentum space $\Pbb$ equipped with a natural ordering. More generally, the dispersion function may need to be parameterized by multiple integers $k_i$, in which case $\Pbb$ is to be viewed as a higher-dimensional lattice with a possibly nontrivial topology.

Matters are more complicated if there are no few-parameter fits to the distribution of the $h_k^{(1)}$'s. One way to proceed may be to view any such distribution as a $k$-dependent perturbation of simple distributions shown in \eqref{sample distributions}. More conservatively, one may simply posit that any theory with such disorderly energy levels does not have a continuum limit.

The same approach to ordering momenta can be used if all the $h$-coefficients in \eqref{def H general} vanish except for $h_k^{(n)}$ for some $n > 1$. However, there are no known local QFTs which have a Hamiltonian of this form.

Finally, if multiple $h$-coefficients have the same order of magnitude, there is in general no natural way to impose a lattice structure on $\Pbb$. One (rather arbitrary) way forward is to demand that the lattice structure persist as all coefficients except for $h_k^{(1)}$ are smoothly tuned to zero along a chosen trajectory in parameter space.

For now, start by assuming that the Hamiltonian describes a free theory. Its ground state is easily determined: it has $n_k = 0$ for all $k$ such that $h_k^{(1)} > 0$, and $n_k = J_k - 1$ for all $k$ such that $h_k^{(1)} < 0$. To keep things simple, assume that there is no $k$ with $h_k^{(1)} = 0$, so the ground state is unique. In addition, choose $h^{(0)}$ such that the ground state has zero energy.

Next, assume that dispersions $h_k^{(1)}$ follow a simple distribution. This means that the momentum space $\Pbb$ is a lattice in which $\big |h_k^{(1)} - h_l^{(1)} \big|$ is much smaller than the typical dispersion difference whenever the minimal number of links $||k - l||$ connecting sites $k$ and $l$ is of $O(1)$. In this setup, it is possible to pick a ``spherical'' subregion $\Pbb_E \subset \Pbb$ defined by
\bel{
  \Pbb_E \equiv \left\{\trm{all } k \in \Pbb \trm{ such that } \big |h_k^{(1)} \big| \leq E \right\}.
}
States that differ by a small number of excitations at momenta in $\Pbb_E$ are then necessarily close to each other in energy. Said more precisely, for any energy eigenstate $\qvec \Psi$ and for any momentum $k \in \Pbb_E$, the states $c_k\+ \qvec \Psi$ and $c_k \qvec \Psi$ --- at least one of which must be nonzero --- necessarily have energies that differ from that of $\qvec \Psi$ by not more than $E$.

Furthermore, if there exist many energy eigenstates with energy below $E$, it must be possible to create many excitations of the ground state at momenta deep in the interior of $\Pbb_E$ (i.e.\ at momenta $k \in \Pbb_{E'}$ for $E' \ll E$) while still keeping the energy of the resulting state much below $E$. Thus the states created from the ground state by ladder operators at $k \in \Pbb_{E'}$ have a special status: they span a large energy eigenspace that, to leading order in $J_k E'/E$, must remain invariant under small perturbations of the free Hamiltonian. This, in turn, guarantees that restricting the algebra $\A$ to the subalgebra generated by ladder operators at momenta in $\Pbb_{E'}$ will produce a unitary effective theory --- even if the original theory is deformed away from a free theory.

This conclusion can be summarized as follows: \emph{if a theory admits a precontinuum basis, and if the dispersions $h^{(1)}_k$ induce a finite-dimensional lattice structure on $\Pbb$ while other $h$-coefficients are parametrically small, then this theory can be robustly reduced to a unitary effective theory involving degrees of freedom on a small subset of momentum space}. In a generic theory, there are two main ways the needed conditions may fail to be fulfilled. First, if the theory has no precontinuum basis, there is no notion of momentum at all. Second, if the precontinuum basis exists but only has $O(1)$ different momenta, or if the Hamiltonian induces no reasonable structure on $\Pbb$, the effective theory obtained by restricting to a subset of spatial momenta is a brittle object whose unitarity may be ruined by small perturbations of the Hamiltonian.

The ability to restrict to an effective theory on a momentum subspace is what allows an orderly approach to the continuum. This will be described next. The motivation for this definition will be explicated in the following Subsections.

\newpage

The definition of a continuum basis, anticipated on page \pageref{ref cont basis}, is precisely stated as follows. Assume that a theory comes with a lattice structure on the momentum space, as described above. Pick an energy scale $\E\_S$ that is much smaller than the typical eigenenergies found in the spectrum of $H$. Then a \emph{continuum basis} is generated by the set of operators
\bel{
  \{c_k, c_k\+\}_{k \in \Pbb_{\E\_S}} \cup \{n_k\}_{k \notin \Pbb_{\E\_S}}.
}

It is also possible to define a continuum basis by specifying an explicit cutoff in momentum space. For instance, if the the momentum space is viewable as a one-dimensional lattice labelled by an integer $k$ in the range $- N \leq k < N$, as in \eqref{sample distributions}, an acceptable continuum basis can be generated by
\bel{
  \{c_k, c_k\+\}_{k = -k\_S}^{k\_S - 1} \cup \{n_k\}_{k = - N}^{-k\_S - 1} \cup \{n_k\}^{N-1}_{k = k\_S}, \quad k\_S \ll N.
}
(It will also be very convenient to take $k\_S \gg 1$ in a setup like this, but it is not strictly necessary.) The advantage of this definition is that it can be applied to any theory with a one-dimensional momentum space, without needing to specify the dispersion function in further detail. To simplify notation, the set of momenta whose ladder operators are retained in the continuum basis will be denoted $\Pbb\_S$, whether this is defined using an energy cutoff $\E\_S$ or a momentum cutoff like $k\_S$.

The algebra $\A\_S$ spanned by the continuum basis has a large center generated by all the $n_k$'s whose ladder operators have been removed. Any operator $\O \in \A\_S$ can thus be written as a block-diagonal matrix,
\bel{
  \O = \bigoplus_{\{n_k\}_{k \notin \Pbb\_S}} \O_{\{n_k\}},
}
where each term in the direct sum acts on a different \emph{superselection sector}.

Each superselection sector can be viewed as a Hilbert space of a separate cQFT. \emph{This will be taken as the definition of a cQFT Hilbert space in this paper.} The cQFT algebra of operators is then simply the algebra of matrices that fit into the block labeled by a particular set of values $\{n_k\}_{k \notin \Pbb\_S}$.

As the number operators are symmetries of the microscopic Hamiltonian, the labels $\{n_k\}_{k \notin \Pbb\_S}$ are guaranteed to remain constant. From the cQFT point of view, these give rise to various structure constants, OPE coefficients, etc. These numbers in turn quantify the entanglement of cQFT degrees of freedom at high momenta. Each superselection sector corresponds to a cQFT with different high-momentum entanglement patterns; a cQFT in which $n_k = 0$ for all $k \notin \Pbb\_S$ will be very different from a cQFT in which half the $n_k$'s are nonzero. The overall ground state will necessarily belong to one of these superselection sectors, and unless otherwise stated it will be assumed that this is the sector under discussion.

\subsection{Original space, momentum space, and position space} \label{subsec spaces}

Subsection \ref{subsec definition} studiously avoided elaborating on the connection between the original lattice $\Mbb$ and the momentum lattice $\Pbb$. A priori, there need not be a simple relation between these two spaces. However, in essentially all examples over which we have microscopic computational control, the quantum fields on these spaces are related by Fourier transforms.

By far the simplest example of this connection is provided by a theory of a free Dirac fermion in $(1+1)$D. Here the full Hilbert space is a $\Z_2$-graded product of two-dimensional Hilbert spaces associated to $2N$ sites lying along a circle. The algebra is generated by (one-component) fermion fields $\psi_v$ and $\psi_v\+$ at each site, and a conveniently normalized Hamiltonian is
\bel{\label{def H Dirac}
  H = \i \sum_{v = 1}^{2N} \left(\psi_v\+ \psi_{v + 1} - \psi_{v + 1}\+ \psi_v \right).
}
The Fourier transform
\bel{
  \psi_v \equiv \frac1{\sqrt {2N}} \sum_{k = -N}^{N - 1} \e^{\frac{2\pi \i}{2N} k v} \psi_k
}
defines the momentum space ladder operators $\psi_k$ and $\psi_k\+$ that play the role of $c_k$ and $c_k\+$ from the previous Subsection. The Hamiltonian then becomes the manifestly free theory
\bel{
  H = 2\sum_{k = -N }^{N - 1} n_k \, \sin\frac{\pi k}N,
}
with the momentum space $\Pbb$ being a ring of $2N$ sites --- a dual of the original space $\Mbb$. The operators $\psi_k$ and $\psi_k\+$ generate a precontinuum basis that has an obvious reduction to a continuum basis via a restriction to momenta near $k = 0$ and $k = N$, the points at which the dispersion function $2\sin\frac{\pi k}N$ is zero. This way of constructing a continuum fermion theory was explored in detail in \cite{Radicevic:2019jfe}.

A more intricate example is provided by the critical Ising model in $(1+1)$D, which features an ordinary product of two-dimensional Hilbert spaces on $N$ sites along a circle. This system can be exactly dualized to a fermionic theory via the Jordan-Wigner transformation. The fermionic fields can then be Fourier-transformed, as above, to obtain another free theory. The nonlocality of the Jordan-Wigner transformation means that  generic local operators in the original space $\Mbb$ are not simply related to the elements of the precontinuum basis, and that the spaces $\Mbb$ and $\Pbb$ are not dual to each other in the same sense as above. This has been described in \cite{Radicevic:2019mle}, where the effect of the nontrivial transformation between the two spaces was shown to lead to nontrivial OPE coefficients and scaling dimensions associated to the Ising conformal field theory.

A well known piece of lore is that, on a large lattice, a lattice theory near a critical point (i.e.\ near a second-order phase transition) is well described by a continuum theory. The free Dirac fermion and the critical Ising model are just two examples of this fact. There is no reason to expect that the precontinuum generators of a generic near-critical lattice theory can be obtained by a simple Fourier transform of the lattice fields. Unfortunately, in the vast majority of cases, there is no other known method to dualize the lattice variables to the precontinuum basis. The lone --- and extremely nontrivial --- explicit example of such a map has recently been worked out in a variant of the three-state Potts model, whose continuum description is a parafermionic field theory \cite{Fendley:2013snq}.

Remarkable examples of the potential complexity of the map between the original local operators and the precontinuum basis are supplied by various holographic dualities \cite{Maldacena:1997re}. Their existence suggests that certain strongly coupled lattice theories on a $d$-dimensional lattice are viewable as cQFTs on a curved $(d + 1)$-dimensional manifold. In other words, the duality between $\Mbb$ and $\Pbb$ need not preserve the dimensionality of these spaces. Holography aside, string theory itself is a vivid illustration of this concept: a near-critical lattice theory in $(1+1)$D is there understood to be describable as a cQFT on a fluctuating spacetime in a dimension determined by the central charge of the lattice theory.

There exists a toy version of this last example that lucidly illustrates the idea in question. Consider the free fermion theory \eqref{def H Dirac}. This theory has a conserved particle number $N\^F \equiv \sum_v \psi_v\+ \psi_v$. The Hamiltonian $H_\lambda \equiv H + \lambda (N\^F - \1)^2$, for $\lambda \rar \infty$, at low energies describes a free fermion theory projected to the $N\^F = \1$ sector. Viewed as a lattice QFT in $(1+1)$D, this sector on its own represents a strongly coupled theory. A precontinuum basis nevertheless exists and is very simple. The momentum space $\Pbb$ consists of a single point; the particle number $n \equiv \sum_{k = -N}^{N - 1} (k + N)\, \psi_k\+ \psi_k$ measures the momentum of the solitary fermion along the spatial circle; and ladder operators increase or decrease this momentum (for example the lowering operator is $c \equiv \sum_{k = -N}^{N - 2} \psi_k\+ \psi_{k + 1}/\sqrt{k + N + 1}$). This precontinuum basis does not have a large number of momenta and so cannot be reduced to a continuum basis. In other words, there is no $(1+1)$D cQFT that describes the low-energy behavior of the $N\^F = \1$ fermion on the lattice. However, there \emph{does} exist a continuum quantum mechanics theory --- a cQFT in $(0 + 1)$D --- which can be used to describe the relevant low energy behavior \cite{Radicevic:1D}. This is thus a simple example of a situation where the low-energy sector of a strongly coupled lattice theory is described by a continuum theory in a different number of dimensions.

The momentum space $\Pbb$ can often be Fourier-transformed to a space $\Mbb^\star$. More generally, it may be possible to define
\bel{
  c_x \equiv \sum_{k \in \Pbb} G_{k, \, x} \, c_k , \quad \trm{or} \quad c_k \equiv \sum_{x \in \Mbb^\star} \~G_{k, \, x} \, c_x,
}
where $G_{k, \, x}$ is a function chosen so that the Hamiltonian of a free theory is local in terms of the $c_x$ fields. This means that $\~G_{k, \, x}$ must satisfy a relation of the form
\bel{
  \sum_{k \in \Pbb} \~G_{k, \, x}^* \~G_{k, \, y} h^{(1)}_k = \alpha_0 \delta_{x, \, y} + \sum_{||y' - y|| = 1} \alpha_{y' - y} \delta_{x,\, y'} + \ldots
}
where the number of terms is much smaller than the number of sites in $\Pbb$. The lattice $\Mbb^\star$ will be called the \emph{position space}. When $\Pbb$ and $\Mbb$ are related by a Fourier transform of the original fields, $\Mbb$ and $\Mbb^\star$ naturally coincide after setting $G_{k,\, x} \sim \e^{\i k x}$. In more complicated examples, e.g.\ when holography is involved, $\Mbb$ and $\Mbb^\star$ need not even have the same dimension.

Expressing all fields in position space clarifies the physical meaning of the continuum basis. The projection $\A \mapsto \A\_S$ that is induced by restricting from the precontinuum operators to the continuum ones induces a projection to \emph{continuum fields},
\bel{
  c_x \mapsto c(x) \equiv \sum_{k \in \Pbb\_S} G_{k, \, x} c_k =  \sum_{y \in \Mbb^\star} \sum_{k \in \Pbb\_S} G_{k, \, x} \~G_{k,\, y}\, c_y.
}
The expression on the right can be interpreted as a smearing of lattice fields $c_y$ over $\sim N/k\_S$ sites in each direction, where $N$ is the linear size of the lattice and $k\_S$ is the cutoff used to define the space $\Pbb\_S$. In other words, the continuum field $c(x)$ must satisfy a \emph{constraint} of the form
\bel{\label{smooth constr general}
  c(x') = c(x) + (x' - x) \hat\del c(x) + \ldots
}
where $\hat\del c(x)$ is an operator whose entries are $O(k\_S/N)$. Whenever $||x' - x|| \ll N/k\_S$, the second term in this formula is small, and the others can be ignored. In short, $k\_S$ controls the derivative expansion of continuum fields $c(x)$. No such expansion applies to the starting real-space fields $c_x$. This is why projecting to the continuum basis was called \emph{smoothing} when developed in the context of fermionic theories \cite{Radicevic:2019jfe, Radicevic:2019mle}. The same name will be used here.

Defining continuum fields via smoothing immediately implies the existence of operator product expansions (OPEs) in cQFT. The crucial fact is that products of smooth fields are not equal to smoothings of products of fields. The OPE encodes the difference between these products. For elementary fields $c_x$, the OPE defined this way is
\bel{\label{OPE general}
  c_x\+ \times c_y \equiv c\+c(x, y) - c\+(x) c(y) = \sum_{k \notin \Pbb\_S} n_k \, G_{k, \, x}^* G_{k, \, y}.
}
When $x$ and $y$ are within a smearing length $N/k\_S$ of each other, the OPE can usually be expressed as a Laurent series in $||x - y||$, with sums over high momenta $k \notin \Pbb\_S$ giving rise to ``singular'' terms of the form $1/||x - y||^\Delta$ for $x \neq y$ and $\Delta > 0$. These ``singularities'' are of course all regularized by the fact that $||x - y|| \geq 1$ for $x \neq y$.

\subsection{Chiral theories} \label{subsec chiral}

The goal of this Subsection is to elucidate the tension felt when trying to talk about chiral theories in a lattice framework. The notion of chirality is most commonly presented in the context of fermionic cQFTs, but in fact it can be defined quite generally on spatial lattices of odd dimension, without any reference to Dirac matrices or the like. For simplicity, the focus here will be on systems with one spatial dimension.

Consider a free theory whose dispersion function $h_k^{(1)}$ endows the momentum space with a one-dimensional lattice structure, with momenta $-N \leq k < N$. As mentioned in the previous Subsection, the lattice $\Pbb$ is in essence chosen so that the dispersion function satisfies a type of ``uniform continuity.'' This can be expressed as the condition that the absolute value of the ``gradient''
\bel{
  \frac{h_k^{(1)} - h_l^{(1)}}{k - l}
}
for all $|k - l| = O(1)$ remains much smaller than some global scale featured in the distribution of dispersions, for instance $\max_{k,\, l \in \Pbb} \big |h_k^{(1)} - h_l^{(1)} \big|$.

It is particularly natural to focus on points where the nearest-neighbor derivative
\bel{
  \del h_k^{(1)} \equiv h_{k + 1}^{(1)} - h_k^{(1)}
}
changes sign. When the lattice $\Pbb$ has the topology of a circle, so that $h_N^{(1)} \equiv h_{-N}^{(1)}$, there must exist an even number of points $k_\star$ such that $\sgn \del h_{k_\star}^{(1)} = - \sgn \del h_{k_\star + 1}^{(1)}$.  The momentum space can thus be split into three subspaces based on the value of $\sgn \del h_k^{(1)}$,
\bel{
  \Pbb = \Pbb_+ \cup \Pbb_- \cup \Pbb_0.
}
The set $\Pbb_0$ of points with vanishing derivative will not be further considered here.

This decomposition is especially simple when the derivative does not change sign too often, so that $\Pbb_+$ and $\Pbb_-$ are unions of long segments with sequential momenta. In the free theory \eqref{def H Dirac}, for instance, $\Pbb_+$ is a single segment of momenta between $-N/2$ and $N/2$. Other reasonable theories will likewise have $O(1)$ segments in $\Pbb_+$.

The segments of greatest interest for continuum physics are those that contain \emph{nodes}, i.e.\ points where the dispersion function itself becomes zero or changes sign. By construction, $\Pbb_+$ and $\Pbb_-$ contain an equal number of nodes. If the smoothing procedure keeps the ladder operators only in neighborhoods of nodes where the dispersion function is approximately linear, the resulting cQFT can be represented as having a separate continuum field for each node. Fields around nodes that belong to the same subset $\Pbb_\alpha$ are said to have \emph{chirality $\alpha$}, for $\alpha \in \{+, -\}$. Every cQFT must have the same number of fields of either chirality.

The operator algebra of a \emph{chiral cQFT} can be defined as the collection of operators associated to momenta in one subset $\Pbb_\alpha$ (with ladder operators kept only in the vicinity of nodes, and particle number operators kept at all other momenta). The corresponding Hilbert space is analogously defined as a (potentially graded) direct product of individual spaces at all momenta in $\Pbb_\alpha$.

The operators and states of a free chiral cQFT can thus be perfectly well defined by starting from a lattice. In particular, no assumption needs to be made about the statistics followed by elementary fields/ladder operators, and so it is equally easy to define chiral theories of bosons, fermions, or fields of more exotic statistics.

The problem with taking a chiral cQFT seriously is that it is not robust under generic small perturbations of the original lattice theory. A local interaction in the original space will typically cause states of different chiralities but similar energies to mix in perturbation theory. The arguments above --- in line with the Nielsen-Ninomiya theorem \cite{Nielsen:1980rz, Nielsen:1981xu} --- indicate that, \emph{by the definition of a cQFT presented here}, there is no way to define a lattice theory on a circular momentum space such that the theory  has only one chirality and a cQFT subtheory robust under arbitrary perturbations.

One way to proceed is to only work with interactions that do not couple different chiralities. It is easy to write down such Hamiltonians, at least in momentum space. In the original or position spaces, the restriction may seem unnatural. In particular, gauging the particle number symmetry is one extremely familiar operation that would have to be disallowed because it couples different chiralities. This is reflected by the existence of the ABJ anomaly \cite{Adler:1969, Bell:1969}, which was analyzed in a manifestly finite context in \cite{Radicevic:2018zsg}.

Another alternative is to simply fix a specific microscopic theory without trying to deform it in any way. In this context it is perfectly legitimate to ask about just one chiral sector of the theory. For instance, starting from the free Dirac theory \eqref{def H Dirac}, one can write down a Hamiltonian that describes the unitary evolution of each chiral sector separately. The position space $\Mbb^\star$ on which each chiral subtheory lives contains $N$ points arranged on a ring, even though the original space had $2N$ points. From the point of view of the chiral theory, its fermions have to be ``staggered'' in order to endow them with a microscopic definition \cite{Kogut:1974ag, Susskind:1976jm}.

Finally, it may make sense to talk about a chiral cQFT when the momentum space is a lattice with boundaries. If $\Pbb$ is merely a line with $2N$ sites, dispersion relations of the form
\bel{
   h_k^{(1)} = \sin \frac{\pi k}{2N} \quad \trm{or} \quad h_k^{(1)} = \omega k, \quad -N \leq k < N
}
are compatible with the assumption of ``uniform continuity.'' Fermionic theories with such dispersions are known respectively as Wilson fermions \cite{Wilson:1975id} and SLAC fermions \cite{Drell:1976mj}. They clearly have only one chirality present.

\subsection{Effective descriptions of interacting theories} \label{subsec EFTs}

The definition given in Subsection \ref{subsec definition} ensures that small perturbations of a lattice theory can only induce small deviations from unitarity in the emergent cQFT. This is a crucial requirement for most ``real-world'' applications of QFT, where theories of interest are typically expressed as deformations of free theories. Indeed, the standard field-theoretic Hamiltonian encountered in the literature takes the momentum-space form\footnote{This is by no means the most general perturbation of a free theory one can write. Various Hamiltonians not of form \eqref{def H standard} are important in their own right; a famous and simple example is BCS theory, which involves ``Cooper pair'' terms $c_k c_{-k} + c\+_k c\+_{-k}$. Conversely, it is natural to constrain the tensorial interaction parameters $g^{(n)}$ such that the Hamiltonian is local in position space, and one typically further chooses the couplings to make the position space theory have various pleasant features such as Lorentz invariance or cluster decomposition \cite{Weinberg:1995mt}. The Hamiltonian \eqref{def H standard} is chosen here for concreteness, and the discussion can be easily altered to take these generalizations or constraints into account.}
\gathl{\label{def H standard}
  H = H\_{free} + H\_{int}, \\
  H\_{free} = \sum_{k \in \Pbb} h_k^{(1)} n_k,  \quad
  H\_{int} = \sum_{k_1, \, k_2 \in \Pbb} g^{(2)}_{k_1 k_2} c_{k_1}\+ c_{k_2} + \sum_{k_1, \ldots,\, k_4 \in \Pbb} g^{(4)}_{k_1 \ldots k_4} c_{k_1}\+ c_{k_2} c\+_{k_3} c_{k_4} + \ldots
}
The key problem here is to determine the character of this interacting theory: does it have a precontinuum basis, what are its $h$-coefficients, and what kind of cQFT (if any) emerges from it? These questions are sensible because, by construction, if $H\_{free}$ describes a cQFT and the interactions $g^{(n)}$ are sufficiently small, a unitary theory is guaranteed to emerge from $H$ at low energies.

Old-fashioned perturbation theory provides a direct way to study the interacting theory. It may be useful to give a brief review of this approach, adapted to the present nomenclature. The starting point is the assumption that the interacting theory has a set of conserved quantities $\~n_k$ that are deformations of the starting ones. A plausible Ansatz is
\bel{
  \~n_k \equiv n_k + \Delta n_k \equiv c_k\+ c_k + \sum_{p,\, q \in \Pbb} f^{pq}_k c_p\+ c_q + \sum_{p,\ldots,\, s \in \Pbb} f^{pqrs}_k c_p\+ c_q c_r\+ c_s + \ldots,
}
where the $f$'s are determined order by order from $[\~n_k, H] = [\~n_k, \~n_l] = 0$. To first order, this requirement is
\bel{
  [H\_{free}, \Delta n_k] + [H\_{int}, n_k] = 0.
}
This provides a set of equations for the $f$'s. For instance, if the only nonzero interaction is $g^{(4)}$ and the ladder operators are fermionic, the perturbation $\Delta n_k$ is schematically
\bel{\label{Delta n k}
  \Delta n_k \sim \sum_{p, \ldots,\, s\in \Pbb} g^{(4)}_{pqrs} \frac{\delta_{pk} - \delta_{qk} + \delta_{rk} - \delta_{sk}}{h_p^{(1)} - h_q^{(1)} + h_r^{(1)} - h_s^{(1)}} c_p\+ c_q c_r\+ c_s.
}

The $\~n_k$'s, as defined by \eqref{Delta n k}, do not have integer eigenvalues and so they are not number operators themselves. If the interacting theory has a precontinuum basis, its number operators must be combinations like
\bel{
  \sum_{k \in \Pbb} \alpha_k \, \~n_k,
}
though terms with products of multiple $\~n_k$'s may also be included. In QFT jargon, these new number operators are usually said to count \emph{quasiparticles} --- as opposed to the original operators $n_k$ that count ``ordinary'' particles. Importantly, quasiparticles are spread over multiple momenta in $\Pbb$, and as interactions are increased this spread will change further. The lattice structure of the momentum space can thus change drastically at strong coupling.

Perturbation theory is truly guaranteed to work only when the couplings are much smaller than the scale set by the size of the original Hilbert space. Larger couplings can lead to nontrivial phenomena at high enough orders in the perturbation series. This issue aside, however, it is often simply too impractical to use perturbation theory to construct quasiparticle operators. An alternative is to \emph{guess} a consistent effective cQFT. Roughly speaking, there are two ways to do so: by using the Wilsonian renormalization group (RG), or by developing a hydrodynamic description based on symmetries of the starting theory.

The Wilsonian approach is predicated on reducing the operator algebra by removing all generators at momenta far away from nodes in $\Pbb$. Unlike smoothing, RG removes even the number operators, and effectively reduces the momentum space to a smaller lattice $\Pbb' \subset \Pbb$. Performing this decimation gradually yields a flow on the space of low-momentum couplings like $g^{(4)}_{k_1\ldots k_4}$ that feature in the effective Hamiltonian of the reduced algebra. The preservation of unitarity of the low-momentum theory hinges on the assumption that these couplings are small enough. The RG flow is controlled by its fixed points, so what emerges after many decimation steps is a small perturbation of an appropriate fixed point theory. RG fixed points are characterized by scale invariance and can be defined directly in the continuum, therefore sidestepping the need to explicitly find fixed-point lattice couplings $g^{(4)}$, the corresponding precontinuum bases, and the cQFTs obtained by their smoothing. Of course, controlling the RG flow is often impossible, and one then has to guess the right fixed point.

The hydrodynamic approach involves a different algebraic decimation. After identifying global symmetries of the interacting theory, it is possible to define local current operators.\footnote{In a free theory with a one-dimensional momentum space, global symmetries are of the form $Q_s \sim \sum_{k \in \Pbb} n_k \, \e^{\i k s}$ or $P_s \sim \sum_{k \in \Pbb} k^s n_k$; interactions typically preserve their conservation for a few values of $s$. The local currents themselves are not expressible via the $n_k$'s.} A hydrodynamic effective theory arises by restricting to the algebra generated just by the spatially smoothed versions of these currents. Unlike a Wilsonian effective theory, there is no guarantee that hydrodynamics is unitary even at small couplings, and indeed the hydrodynamic effective theory is generically dissipative.


\section{Scalar field theory} \label{sec scalars}

\subsection{The clock model: kinematics}

To concretely illustrate the rather abstract ideas presented so far, this Section will work out the lattice-continuum correspondence for scalar field theory in $(1+1)$D. The goal is to start from a finite, well defined lattice model and to precisely specify under which conditions it can be restricted to a subsector that exhibits all salient features expected of a cQFT.


Consider a theory consisting of $N \gg 1$ copies of a $K$-state clock model that are arranged on a circle $\Mbb$. Each copy has associated clock and shift operators $Z_x$ and $X_x$ for $x = 1, \ldots, N$. For $K = 2$, this is the $(1+1)$D Ising model, which can be described (when near a phase transition) by a cQFT with fermion fields. Here the focus will be on the opposite extreme, $K \gg 1$, which will give rise to a bosonic cQFT. Eigenvalues of $Z_x$ will be denoted $\e^{\i\phi_x}$, with
\bel{
  \phi_x = \frac{2\pi}K n_x \equiv n_x \d\phi, \quad n_x = 0, 1, \ldots, K - 1.
}

The maximal operator algebra  in this system is isomorphic to $\C^{D \times D}$, where $D \equiv K^N$ is the dimension of the full Hilbert space. This algebra is too large for the purposes of identifying the scalar cQFT. Instead, the starting point in this story will be the algebra $\A$ obtained by taking a direct product of \emph{tamed} clock algebras at each site.

In quantum mechanics, taming is a projection of a clock algebra to a subalgebra that preserves the smoothness and compactness of all wavefunctions \cite{Radicevic:1D}. In the field-theoretic context, the algebra $\A$ contains all operators that preserve the smoothness and compactness of wavefunctionals along the target space directions. A \emph{smooth} wavefunctional $\greek y[\phi]$ satisfies
\bel{
  \Big| \, \greek y[\phi + \delta^{(x)}\d\phi] -  \greek y[\phi] \, \Big| = O\left(\frac{p\_S}K\right) \quad \trm{for all }x \in \Mbb,
}
where $1 \ll p\_S \ll K$, and $\delta^{(x)}$ is a Kronecker delta supported at $x$. A \emph{compact} wavefunctional satisfies
\bel{
  \greek y[\phi]  = 0 \quad \trm{whenever}\quad |\phi_x| > \varphi\_T \quad \trm{for all }x \in \Mbb.
}
A \emph{tame} wavefunctional is both compact and smooth, with $\frac1{p\_S} \ll \varphi\_T \ll 1$. It is convenient to define $\d\varphi \equiv \frac{2\pi}{2p\_S}$ and an integer $n\_T \equiv \frac{\varphi\_T}{\d\varphi}$; tameness then requires $1 \ll n\_T \ll p\_S \ll K$. 

The subspace of tame wavefunctionals has dimension $(2n\_T)^N$. It has natural basis vectors $\qvec{\e^{\i\varphi}}$ formed by smearing eigenstates $\qvec{\e^{\i\phi_x}}$ of $Z_x$ by suitable functions $f_{\phi, \, \varphi}$ \cite{Radicevic:1D},
\bel{\label{def varphi}
  \qvec{\e^{\i\varphi}} \equiv \bigotimes_{x \in \Mbb} \qvec{\e^{\i\varphi_x}}, \quad \qvec{\e^{\i\varphi_x}} \equiv \sum_{\phi_x = \d\phi}^{2\pi} f_{\phi_x, \, \varphi_x} \qvec{\e^{\i\phi_x}}, \quad \varphi_x \equiv n_x \d\varphi, \quad -n\_T \leq n_x < n\_T.
}

When studying the tamed theory, it is convenient to work with operators
\bel{\label{def phi pi}
   \widehat\varphi_x \equiv \frac1{2\i} \left(Z_x - Z_x^{-1}\right)\_T, \quad \widehat\pi_x \equiv \frac{1}{2\i \, \d\phi} \left(X_x - X_x^{-1}\right)\_T,
}
where the ``T'' subscript indicates taming at site $x$. To leading order in $p\_S/K$ and $n\_T/p\_S$, these operators act on tame wavefunctionals as canonical position and momentum operators, namely
\bel{\label{phi pi canon actions}
  \widehat\varphi_x \qvec{\e^{\i\varphi}} \approx \varphi_x \qvec{\e^{\i\varphi}}, \quad \widehat\pi_x \qvec{\e^{\i\varphi}} \approx -\i \hat \del_{\varphi_x} \qvec{\e^{\i\varphi}}.
}
The symbol $\hat\del_{\varphi_x}$ in the action of $\widehat\pi_x$ denotes a formal derivative w.r.t\ $\varphi_x$; alternatively, $-\i\hat\del_{\varphi_x}$ denotes multiplication by the ``target momentum.'' The identification of $\widehat \pi_x$ with a derivative is only correct in states with $|\varphi_x| \ll \varphi\_T$. These issues are explained in detail in \cite{Radicevic:1D}. The carets will be dropped from $\widehat\varphi_x$ and $\widehat\pi_x$ whenever context makes it clear these are operators.

Here is a crucial convention that will be used henceforth: \emph{all products of $\varphi$ and $\pi$ operators tacitly assume that a product is taken \emph{before} taming}. For example, the  tamed product of two momentum operators will be denoted
\bel{
  \pi_x^2 =  -\frac1{4(\d\phi)^2} \left[ \left(X_x - X_x^{-1} \right)^2\right]\_T = \frac1{4(\d\phi)^2} \left(2 - X_x^2 - X_x^{-2}\right)\_T.
}
Meanwhile, the product of tamed operators can be denoted by a ``normal ordering'' symbol,
\bel{
  \tord{\, \pi_x^2 \,} \equiv -\frac1{4(\d\phi)^2} \left[ \left(X_x - X_x^{-1} \right)\_T\right]^2.
}
The operator $\pi_x^2 - \tord{\, \pi_x^2\,}$ has nonzero matrix elements in any tame state with $|\varphi_x| \sim \varphi\_T$. This means that expressions like $\pi_x^4$ or $\varphi_x \pi_x$ are, in a general tame state, not merely given by matrix multiplication of operators $\varphi_x$ and/or $\pi_x$ from \eqref{def phi pi}. It is thus important to keep this convention in mind. In particular, its use underlies the bosonic commutation relation
\bel{\label{comm rel bos}
  [\varphi_x, \pi_y] \approx \i  \delta_{xy},
}
which holds when acting on tame wavefunctionals: the $O(1/p\_S)$ corrections to the operator actions \eqref{phi pi canon actions} are needed in order to obtain \eqref{comm rel bos}, and these corrections only sum to a nontrivial result if the operators are multiplied before taming.

The principal reason for working with the algebra $\A$ is that it approximately decomposes into a direct product of subalgebras labelled by spatial momenta.
This happens because the algebra is generated by operators whose commutator at the same spatial point is approximately proportional to the identity. The approximation depends on the taming parameters and can be made arbitrarily good by increasing $K$, $p\_S$, and $n\_T$ while keeping their ratios suitably small.

To precisely state this fact about the decomposition of $\A$, define the Fourier transforms
\algns{\label{def k}
  \varphi_x \equiv \frac1{\sqrt N} \sum_{k \in \Pbb} \varphi_k \, \e^{\frac{2\pi\i}N kx}, \quad
  \pi_x \equiv \frac1{\sqrt N} \sum_{k \in \Pbb} \pi_k \, \e^{\frac{2\pi\i}N kx},
}
with $\Pbb = \{-\frac N2, \ldots, \frac N2 - 1\}$. From \eqref{comm rel bos} it follows that
\bel{\label{comm rel bos mom}
  [\varphi_k, \pi_l] \approx \i \delta_{k,\, -l}.
}
If the canonical commutation relation were replaced by a commutator of the form $[\varphi_x, \pi_y] = \O_x \delta_{x y}$, the momentum space fields would obey $[\varphi_k, \pi_l] = \O_{k + l}$. For instance, accounting for the first $\varphi\_T$ correction gives
\bel{
  [\varphi_k, \pi_l] \approx \i \delta_{k,\, -l} + \i (\varphi^2)_{k + l}
}
when acting on tame wavefunctionals. This way higher-momentum operators arise out of products of lower-momentum ones, meaning that operators at different momenta are not linearly independent from each other. This observation is particularly important when considering high powers of operators, e.g.\ $\varphi_x^{p\_S}$, in which $1/p\_S$ corrections may become significant.


\subsection{The clock model: dynamics}

The starting microscopic Hamiltonian in this paper is  the nonchiral clock model,
\bel{\label{def H micro}
  H = \frac{g^2}{2 (\d\phi)^2} \sum_{x = 1}^N \left(2 - X_x - X_x^{-1} \right) + \frac1{2g^2}\sum_{x = 1}^N \left(2 - Z_x Z_{x + 1}^{-1} - Z_x^{-1} Z_{x + 1} \right),
}
with $x + N \equiv x$. The basic phase structure of this model is well studied \cite{Jose:1977gm, Elitzur:1979uv, Alcaraz:1980sa, Alcaraz:1980bb, Cardy:1980, Frohlich:1981yn}. At extremely large couplings, which here means $g^2 \gg \d\phi$, the unique ground state is completely disordered in the clock eigenbasis, with all clocks decoupled. At extremely small couplings, $g^2 \ll \d\phi$, there are $K$ ordered ground states in which all clocks point in the same direction. For $K \geq 5$, these two phases are separated by a region of nonzero width (in coupling space) near $g^2 \sim \d\phi$. For each $g$ deep inside this region, the theory is in the BKT universality class \cite{Berezinsky:1970fr, Kosterlitz:1973xp} and its infrared behavior can be described by a free scalar cQFT when $N$ is large.

The taming procedure allows a detailed study of the correspondence between the above lattice model and the scalar cQFT that emerges from it. It is not obvious that this Hamiltonian has eigenstates that remain pure upon site-by-site taming. By analogy with the harmonic oscillator in $(0+1)$D \cite{Radicevic:1D}, and in keeping with BKT lore, it is reasonable to \emph{conjecture} that there exists a window of couplings $g\_{min} \leq g \leq g\_{max}$ for which one can define parameters $p\_S$ and $n\_T$ such that enough --- e.g.\ more than $(2n\_T)^N$ --- energy eigenstates remain pure upon taming. The consistency of this conjecture will be further affirmed in Subsection \ref{subsec T dual}.

\newpage

If the coupling is chosen to lie in this ``tamable'' window, the effective Hamiltonian for the tame wavefunctionals can be obtained by simply projecting the microscopic Hamiltonian \eqref{def H micro} onto the algebra $\A$.  When acting on tame wavefunctionals, this projection is
\bel{\label{def HT}
  H\_T \approx \frac{g^2}{2} \sum_{x \in \Mbb} \pi_x^2  + \frac1{2g^2} \sum_{x \in \Mbb} (\del \varphi_x)^2, \quad \del\varphi_x \equiv \varphi_{x + 1} - \varphi_x.
}
Higher-order corrections in $\varphi\_T \sim n\_T/p\_S$ and $p\_S/K$ can be readily computed. They need to be taken into account when considering high temperatures, long times, or more generally correlation functions involving many copies of $H\_T$.

The Fourier transform \eqref{def k} can be used to bring  $H\_T$ closer to the form \eqref{def H general}. It gives
\bel{\label{HG k space}
  H\_T \approx \frac{g^2}2 \sum_{k \in \Pbb} \pi\+_k \pi_k + \frac1{2g^2} \sum_{k \in \Pbb} 4 \sin^2 \frac{\pi k}N \, \varphi\+_k \varphi_k.
}
Since $\pi_x$ and $\varphi_x$ are Hermitian, their Fourier transforms obey $\pi_k\+ = \pi_{-k}$ and $\varphi_k\+ = \varphi_{-k}$. The momentum modes obey $[\pi_k\+, \pi_l] = [\varphi_k\+, \varphi_l] = 0$ in addition to the approximate relation \eqref{comm rel bos mom}.

For each momentum $0 < k < \frac N2$, the Hamiltonian contains a term built out of operators $\pi_k$ and $\varphi_k$ that is equivalent to the Hamiltonian of a simple harmonic oscillator of frequency $\omega_k \equiv |2 \sin\frac{\pi k}N|$ and mass $1/g^2$. This makes it natural to define ladder operators
\bel{\label{def ak}
  a_k \equiv \frac1{\sqrt 2} \left(\frac{\sqrt{\omega_k}}g \varphi_k +  \frac{\i g}{\sqrt{\omega_k}} \pi_k \right).
}
They obey
\bel{\label{comm rel bos ladder}
  [a_k\+, a_l] = \frac12 \left(-\i [\pi_{-k}, \varphi_l] + \i [\varphi_{-k}, \pi_l] \right) \approx - \delta_{k, \, l},
}
and so they behave like ordinary SHO ladder operators to first order in the taming parameters.

Due to \eqref{comm rel bos ladder}, ladder operators of a precontinuum basis can be defined as
\bel{\label{def ck}
  c_k \equiv \trm P_k\+\, a_k\, \trm P_k , \quad k \neq 0,
}
where $\trm P_k$ is a projector onto the subspace spanned by states $\qvec{0_k}$, $a_k\+ \qvec{0_k}$, \ldots, $(a_k\+)^{J_k - 1}\qvec{0_k}$. Here $\qvec{0_k}$ is the tame null vector of $a_k$, and $J_k \gg 1$ is chosen so that all the above states are eigenstates of $a_k\+ a_k$ with approximately integer eigenvalues. This ensures that the various correction terms will not accumulate and prevent particle numbers from being integers. This definition also implies that $(c_k\+)^{J_k} = 0$, as required of precontinuum ladder operators. For given $p\_S$ and $n\_T$, there is some freedom in choosing the $J_k$'s, and thus the precontinuum basis, but every choice yields the same cQFT at leading order in $1/J_k$ and other small parameters. Heuristically, one can think of each $J_k$ as being $O(n\_T)$.

The case $k = 0$ is subtle because the harmonic potential in $H\_T$ vanishes and the remaining term, $\frac{g^2}2 \pi_0^2$, seems to  describe a free particle whose position is measured by the operator $\varphi_0$. It is tempting to interpret $\pi_0$ as the ``center-of-mass'' momentum along the target circle in the microscopic theory. Unfortunately, this interpretation would lead to an inconsistency. To see why, recall that a free particle on a circle does not have tame eigenstates \cite{Radicevic:1D}. Therefore, if $H$ contained a free particle at $k = 0$, then taming it would not give a unitary theory --- and so $H\_T$ would not be trustworthy in the $k = 0$ sector, leading to a contradiction.

However, the operator $\pi_0^2$ does not  \emph{necessarily} come from taming a free particle. It may appear in the low-energy Hamiltonian even if the microscopic theory had no free particle behavior at $k = 0$. Indeed, it is not difficult to explicity calculate some matrix elements of the microscopic Hamiltonian in the $k = 0$ sector and to verify that this subtheory is interacting.

To do this, it is easiest to work in the $X_x$ eigenbasis, where states on site $x$ are labeled by an integer $-\frac K2 \leq p_x < \frac K2$. Here the term $\frac1{2g^2} V \equiv \frac1{2g^2} \sum_x (Z_x Z\+_{x + 1} + Z_x\+ Z_{x + 1})$ in the Hamiltonian is interpreted as an interaction potential that creates particle-antiparticle pairs. The aim is to check that this interaction is nontrivial. Starting with the trivial state $\qvec \varnothing$ with $p_x = 0$ at each point, the potential acts as
\algns{
  V\qvec \varnothing &= \sqrt N \big( \qvec{1, \trm -1} + \qvec{\trm -1, 1} \big), \\
  V \qvec{1, \trm -1} &= \qvec{2, \trm -2} + 2 \qvec{1, 0, \trm -1} + \ldots + \qvec{1, \trm -1, 0, \ldots, 0, 1, \trm -1} + \\
  & \qquad + \sqrt N \qvec \varnothing + \qvec{1, \trm -2, 1} + \qvec{\trm -1, 2, \trm -1} + \ldots + \qvec{1, \trm -1, 0, \ldots, 0, \trm -1, 1}.
}
The entries denote the smallest intervals that contain all nonzero eigenvalues of $p_x$; each state consists of a normalized sum over all possible translations of the shown configuration, so e.g.\
\bel{
  \qvec{1, \trm -1} \equiv \frac1{\sqrt N} \sum_{x = 1}^{N} \qvec{p_1 = 0, \ldots, p_x = 1, p_{x + 1} = -1, \ldots, p_N = 0}, \quad p_{N + 1} \equiv p_1.
}
Translation-invariant eigenstates of $X_x$ like $\qvec\varnothing$ are thus decidedly \emph{not} energy eigenstates. The potential $V$ gives rise to an interaction that can, in principle, cause some low-energy $k = 0$ eigenstates to be tame.

Proving that any of these eigenstates \emph{are} tame is beyond this paper. However, Subsection \ref{subsec T dual} will demonstrate the nontrivial self-consistency of the \emph{conjecture} that there exist tame low-energy $k = 0$ eigenstates, and that their spectrum is discretized in a way similar to the spectrum of the free particle on a circle. In fact, this conjecture will be shown to be broadly consistent with the conjecture that low-energy $k \neq 0$ states are tame when $g\_{min} \leq g \leq g\_{max}$ for suitably chosen $g\_{min/max}$. For the time being, the existence of tame $k = 0$ eigenstates and the validity of the $\frac{g^2}2 \pi_0^2$ term in $H\_T$ will be merely assumed.

\newpage

There exists another set of prospective low-energy eigenstates. These are states that do not remain pure upon ordinary taming, but instead remain pure after taming is performed around a position-dependent background $\varphi\^{cl}_x$ at each site. In principle, $\varphi\^{cl}_x$ can describe any profile. In practice, it is sufficient to focus only on configurations $\varphi\^{cl}_x$ that vary slowly, e.g.\ by not changing by more than $2\varphi\_T$ over $N/2k\_S$ sites.

This still leaves a considerable number of possible taming backgrounds $\varphi\^{cl}_x$. A further reasonable \emph{conjecture} is that the only backgrounds featuring in low-energy eigenstates will be minima of the potential term in the Hamiltonian. (The \emph{potential} can be defined as the part of the Hamiltonian that depends only on fields featured in spatial gradients; the notion is ill defined if the Hamiltonian contains products of clock and shift operators.)

In the clock model, the potential is $\frac1{g^2} \sum_x (1 - \cos \del\varphi_x)$, and if $\varphi_x$ varies slowly it can be approximated with $\frac1{2g^2} \sum_x (\del\varphi_x \, \trm{mod} \, 2\pi)^2$. It is minimized for taming backgrounds given by
\bel{\label{taming backgrounds}
  \varphi_x\^{cl} = \varphi\^{cl}\_{const} + \frac{2\pi w}N x,
}
where the \emph{winding number}  $w$ must be an integer, as $x \equiv x + N$. The background offset $\varphi\_{const}\^{cl}$ must be an integer multiple of $2\varphi\_T$ to avoid counting the same state within two sectors.

Requiring that $\varphi_x\^{cl}$ vary slowly, as explained above, limits $w$ to
\bel{\label{bound w}
  |w| \lesssim \varphi\_T k\_S.
}
In particular, if $\varphi\_T k\_S \ll 1$, there are no smooth fields with nontrivial winding around the spatial circle that can be included in the analysis. This is the first example of a relation between taming and spatial smoothing parameters that is needed to get familiar cQFT results. One could also require $\frac{2\pi w}N$ to be an integer multiple of $\d\phi$, so that $w \in \frac NK \Z$. However, here it will just be assumed that $\frac{2\pi w}N x$ stands for the nearest integer multiple of $\d\phi$.

The space of low-energy states thus splits into sectors labeled by taming backgrounds. In addition to the $k = 0$ one, there is a sector for each $\varphi\^{cl}\_{const}$ and $w$, subject to the bound \eqref{bound w}. For any $\varphi_x\^{cl}$, the tame subspace contains wavefunctionals supported on configurations
\bel{
  \varphi_x = \varphi_x\^{cl} + \delta\varphi_x, \quad -\varphi\_T \leq \delta \varphi_x < \varphi\_T.
}
Each of the nonzero-mode sectors will be governed by
\bel{\label{def HT(w)}
  H\_T(w) = H\_T \big|_{\varphi_x \mapsto \delta\varphi_x} + \frac{2\pi^2}{g^2 N} w^2.
}
At weak coupling, states with nontrivial winding are energetically suppressed. They dominate the physics at $g^2 \gg \d\phi$, and must be taken into account at the Kosterlitz-Thouless point.

Note that operators that change the winding sector must be built out of high powers of the shift operator at many different points. They are of the form
\bel{
  \prod_{x \in \Mbb} X_x^{m(x)}
}
for $m(x) \in \Z$ that must exceed $n\_T$ at many points $x$. No such operators exist in the algebra $\A$, so the continuum theory as it stands cannot be used to describe changes in winding.

\subsection{The basic and standard noncompact scalar cQFTs} \label{subsec noncomp scalar}

The Hamiltonian \eqref{def HT}, sans zero-modes and in the sector with taming background $\varphi_x\^{cl} = 0$, can be written in terms of SHO ladder operators as
\bel{
  H_0 \approx \frac12 \sum_{k \in \Pbb}  \omega_k \left( a_k\+ a_k + a_{-k} a_{-k}\+ \right) \approx  \sum_{k \in \Pbb} \omega_k \left(a_k\+ a_k + \frac12 \right).
}
The manipulations that lead to this form are not defined at $k = 0$, but this term does not contribute to the above sum, so all is still well. As stressed already, this expression is only valid when acting on tame wavefunctionals. When the states of interest are further limited to tame wavefunctionals with no more than $J_k$ excitations per mode\footnote{There may exist tame states that have more than $J_k$ particles at mode $k$. Both $H_0$ or $H\_T$ can act on them nontrivially, just as they also act nontrivially on the untame states. The important fact is that $H_0$ approximately preserves the particle numbers and will hence never create more than $J_k$ particles per mode.}, $H_0$ has a simple expression as a free theory in the precontinuum basis discussed around \eqref{def ck}, namely
\bel{\label{def H0}
  H_0 \approx  \sum_{k \in \Pbb} \omega_k \left(c_k\+ c_k + \frac12 \right).
}
This sets the stage for smoothing out the theory and defining the \emph{noncompact scalar cQFT}.

As described in Subsection \ref{subsec definition}, the smooth algebra $\A\_S$ is generated by operators
\bel{\label{def AS}
  \{c_k, c_k\+\}_{-k\_S \leq k < 0} \cup \{c_k, c_k\+\}_{0 < k \leq k\_S} \cup \{c_k\+ c_k \}_{k\_S < k < -k\_S} \equiv \{c_k, c_k\+\}_{k \in \Pbb\_S \backslash \{0\}} \cup \{c_k\+ c_k\}_{k \notin \Pbb\_S},
}
where the smooth momentum space is $\Pbb\_S = \{0, \pm 1, \ldots, \pm k\_S\}$. As usual, the (integer-eigenvalued) particle number is
\bel{
  n_k \equiv c_k\+ c_k.
}
Approximately the same smooth algebra is also generated by $\{\varphi_k, \pi_k\}_{k \in \Pbb\_S \backslash \{0\}}\cup \{n_k\}_{k \notin \Pbb\_S}$. Modes of the same sign of $k$ have the same chirality.

It is possible to define a slightly larger smooth algebra $\A\_S^0$, generated by
\bel{\label{alg basis noncomp w zero}
  \{\varphi_k, \pi_k\}_{k \in \Pbb\_S} \cup \{n_k \}_{k \notin \Pbb\_S}.
}
This algebra differs from $\A\_S$ by the addition of the Hermitian zero-mode generators, $\varphi_0$ and $\pi_0$. These operators commute with all operators at $k \neq 0$, and hence they are symmetries of $H_0$. One often says that $\pi_0$ generates the \emph{shift symmetry}, as $\e^{\i m \pi_0 \d\phi} \approx \left[\prod_{x \in \Mbb} X_x^m\right]\_T$ generates shifts $\varphi_x \mapsto \varphi_x + m \d\phi$ for small enough $m$. In the noncompact scalar cQFT, this symmetry is rather trivial, since $\pi_0$ does not act on any degree of freedom in the $k \neq 0$ space. The noncompact theory without these zero-modes will be called \emph{basic}; with them, it will be called \emph{standard}. Much more about the shift symmetry will be said in Subsection \ref{subsec shift}.

In a basic noncompact scalar cQFT, the position space is the same as the original space, though certain subtleties regarding the zero modes remain. The position-space fields $\varphi_x$ and $\pi_x$, after smoothing, become the operators
\bel{\label{def phi(x)}
  \varphi(x) \equiv \frac1{\sqrt N} \sum_{k \in \Pbb\_S \backslash \{0\}} \varphi_k \, \e^{\frac{2\pi\i}{N}kx}, \quad
  \pi(x) \equiv \frac1{\sqrt N} \sum_{k \in \Pbb\_S \backslash \{0\}} \pi_k \, \e^{\frac{2\pi\i}{N}kx}.
}
Thus these fields satisfy the constraints
\bel{\label{zero mode constraints}
  \sum_{x \in \Mbb} \varphi(x) = \sum_{x \in \Mbb} \pi(x) = 0,
}
in addition to the usual smoothness constraints shown in \eqref{smooth constr general}, e.g.
\bel{\label{smooth constr scalars}
  \varphi(x + 1) = \varphi(x) + \hat\del \varphi(x) + O\left(k\_S^2/N^2\right).
}

Importantly, the two operators in \eqref{zero mode constraints} do not commute with each other. It is therefore wrong to interpret these as na\"ive constraints on states. The fact that both sums are zero is only possible because neither of these operators is in the algebra; there are no states on which either operator acts nontrivially. To avoid confusion, always remember that the Hilbert space in the basic noncompact scalar cQFT is the direct product of $2k\_S$ Hilbert spaces associated to \emph{nonzero} momenta between $-k\_S$ and $k\_S$.

The ground state of $H_0$ has $\avg{n_k} = 0$ for all $k \neq 0$. It is thus natural to choose the (free, massless) basic noncompact scalar cQFT to lie in the superselection sector labeled by $n_k = 0$ for all $k \notin \Pbb\_S$. The basic noncompact scalar does not include any $k = 0$ or $w \neq 0$ modes in its Hilbert space, and so it does not fully capture the low-energy limit of \eqref{def H general}. As long as translation invariance is maintained, however, no perturbation of the microscopic theory will couple the zero- and winding modes to the rest of the system.

As the theory is approximately free, it is trivial to calculate any desired correlator of few operators to leading order in the taming parameters. More interestingly, it is possible to define and calculate the operator product expansions in this lattice theory, the same way this was done for fermions \cite{Radicevic:2019jfe, Radicevic:2019mle}, as briefly explained in \eqref{OPE general}. Define the OPE as
\bel{
  \O_x \times \~\O_y \equiv \O\~\O(x, y) - \O(x) \~\O(y).
}
To find the OPE of two scalar fields, it is sufficient to find the terms in $\varphi_x \varphi_y$ that do not get projected out by the smoothing. Only the terms with $k = -l$ in the momentum expansion of this product will contribute to the OPE, and so
\bel{\label{OPE scalar sum}
  \varphi_x \times \varphi_y \approx \frac1N \sum_{k \notin \Pbb\_S} \frac{g^2}{2\omega_k} \e^{\frac{2\pi\i}N k (x - y)} \left[c_k c_k\+ + c_{-k}\+ c_{-k} \right] \approx \frac{g^2}N \sum_{k = k\_S + 1}^{\frac N2 - 1} \frac{\cos\frac{2\pi}N k (x - y)}{2 \sin\frac \pi N k}.
}
The above calculation used the fact that $n_k = 0$ holds in the superselection sector of interest. If $|x - y| \ll N$, this can be written as an integral over $\kappa \equiv \frac kN$ between $\kappa\_S \equiv \frac{k\_S}N \ll 1$ and $\frac12$:
\algns{
  \frac{g^2}2 \int_{\kappa\_S}^{\frac12} \d\kappa\, \frac{\cos 2\pi \kappa (x - y)} {\sin \pi \kappa}
  &= \frac{g^2}{4\pi} \left[ \greek B_{\e^{2\pi \i \kappa\_S}} \left(\tfrac12 - x + y, 0\right) + \greek B_{\e^{2\pi \i \kappa\_S}} \left(\tfrac12 + x - y, 0 \right) - \i \pi\right].
}
This expression with incomplete Beta functions is not terribly illuminating. To make progress, expand in powers of $\kappa\_S$ to get an expression involving harmonic numbers that can be readily approximated further,
\algns{\label{OPE scalar}
  \varphi_x \times \varphi_y
  &= -\frac{g^2}{4\pi} \left[ H \left(x - y - \tfrac12 \right) + H\left(y - x - \tfrac12\right) + 2\log(2\pi \kappa\_S) \right] + O(\kappa\_S^2) \\
  &= -\frac{g^2}{2\pi} \left[\log\left(\frac{2\pi k\_S}N |x - y|\right) + \gamma \right] + O\left(\frac 1{(x - y)^2}\right) + O(\kappa\_S^2),
}
where $\gamma \approx 0.577$ is the Euler-Mascheroni constant. This is the familiar OPE of two scalar fields in $(1+1)$D.\footnote{In string theory, $1/g^2$ is called the string tension, and one typically expresses formul\ae\ in terms of the Regge slope $\alpha' \equiv g^2/2\pi$. In this notation, the leading position-dependent part of the scalar field OPE \eqref{OPE scalar} is $\varphi_x \times \varphi_y \supset - \alpha' \log |x - y|$, which is of course in agreement with the standard CFT result that mentions no lattices at all \cite{Polchinski:1998rq}. What is more commonly called a string scale is defined as $\ell\_{string} \equiv \sqrt{\alpha'}$ and has nothing to do with the field-theoretic ``string scale,'' $\ell\_S$, that is defined here. These scales are similar because each controls a derivative expansion: $\ell\_S$ for the effective cQFT on the worldsheet, and $\ell\_{string}$ for the effective cQFT in the ``spacetime,'' i.e.\ in the target space.} The length scale in the logarithm,
\bel{\label{def ell S}
  \ell\_S \equiv \frac N{2k\_S},
}
is the smearing length or the ``string scale'' of this cQFT.

When studying the OPE of two \emph{elementary} fields, i.e.~of two linear combinations of ladder operators $c_k$, the product here denoted by $\O \~\O(x,y)$ is typically written as $\red{\O(x) \~\O(y)}$, while the product here denoted $\O(x) \~\O(y)$ is typically written as a normal-ordered product $\red{\nord{\O(x) \~\O(y)}}$. (The red color serves to highlight notation that is \emph{\color{red} not} used in this paper.) To illustrate this, consider the product of $n$ fields $\varphi$ at points $x_1, \ldots, x_n$. The correspondence between the standard conventions (e.g.\ as used in \cite{Polchinski:1998rq} and shown here on the l.h.s.\ in red) and the present notation (shown on the r.h.s.\ in black) is
\gathl{\label{Polchinski map}
  \red{\varphi(x_1) \cdots \varphi(x_n)} \quad \longleftrightarrow\quad \varphi^n(x_1, \ldots, x_n),\\
  \red{\nord{\varphi(x_1) \cdots \varphi(x_n)}} \quad \longleftrightarrow\quad  \varphi(x_1) \cdots \varphi(x_n).
}

However, when studying the OPE of \emph{composite} operators,  which are defined as linear combinations of products of multiple ladder operators, the correspondence between the present notation and the customary one is a bit trickier. For example, one has
\bel{
  \red{\nord{\varphi(x_1) \cdots \varphi(x_n)} \, \nord{\varphi(y_1) \cdots \varphi(y_m)}} \ \longleftrightarrow\, \varphi(x_1) \cdots \varphi (x_n) \varphi(y_1) \cdots \varphi(y_m) + \parbox{4.5em}{\centering \scriptsize external\\  contractions}. 
}
This means that the product of two normal-ordered operators is not customarily normal-ordered itself, but is defined to include those OPE singularities that arise when elementary fields from different operators get to within a smearing length $\ell\_S$ of each other (``external contractions''). Notably, OPE singularities from elementary fields belonging to the same composite operator (``internal contractions'') are not included.

Finally, note that the OPE \eqref{OPE scalar} depends solely on high-momentum values of $n_k$ in the superselection sector that contains the ground state. (These are the same quantities that contribute to the two-point function of scalars; hence the OPE necessarily encodes the same singularities as the correlation function.) Including a zero-mode in the definition of $\varphi(x)$ would not affect this computation in any way.

\subsection{Vertex operators}

\emph{Vertex operators} in a scalar cQFT are typically distinguished in the literature by virtue of being conformal primaries. However, conformal symmetry is actually not needed to see that these operators play a special r$\hat{\trm o}$le in both basic and standard noncompact scalar cQFT. This Subsection  will show that vertex operators are ``eigenstates'' of the smoothing operation, which causes them to obey a special kind of product structure \cite{Polchinski:1998rq, DiFrancesco:1997nk},  presented here in eq.\ \eqref{VO prod}. This is an old result derived in a way that reveals all its microscopic underpinnings. The present lattice-centric view will also show how the space of vertex operators naturally acquires boundaries set by the taming parameters $n\_T$ and $p\_S$.

Vertex operators can be defined as powers of the original clock operators $Z_x$ to which appropriate smoothing has been applied. However, to avoid technical complications due to fractional powers of $Z_x$, it is better to define a vertex operator as the element of $\A\_S$ given by
\bel{
  V^p(x) \equiv \N \e^{\i p  \varphi(x)}, \quad p \in \R,
}
for some normalization $\N$ that will be fixed later. The remarkable fact about vertex operators is that $\N$ can be chosen so that $V^p(x)$ agrees with the smoothing $Z^p(x)$ of the microscopic operator $Z_x^p$ --- at least when acting on tame wavefunctionals with few excitations.

This definition deserves some comments. The exponential is defined by the power series
\bel{
  V^p(x) = \N \sum_{n = 0}^\infty \frac{(\i p)^n}{n!} \varphi(x)^n,
}
where it is, as usual, crucial  to remember that operators at the same site $x$ are multiplied first, and tamed second; the fact that spatially smooth operators are being multiplied does not change this multiplication rule. When $n$ is small, this comment is not relevant. However, when $n$ becomes $O(1/\varphi\_T)$ or $O(p\_S)$, various small by-products of taming must be taken into account, and the behavior of $\varphi(x)^n$ is no longer simply related to the behavior of $\varphi(x)$. This qualitative change in the behavior of $\varphi(x)^n$ will not affect $V^p(x)$ if $p$ is small enough so that large-$n$ terms are all parametrically suppressed. This means that the properties of vertex operators derived here cannot hold at arbitrarily large values of $p$.

In a similar vein, it is often said that a free scalar cQFT has a continuous spectrum of vertex operators --- a different operator for each $p \in \R$. Such statements must be amended in the presence of a microscopic lattice description. The operator $\varphi(x)$ acts on a finite-dimensional Hilbert space (its dimension is, roughly speaking, $2n\_T$; the details depend on the exact values of the $J_k$'s chosen when building the precontinuum basis). This means that there can be at most $O(n\_T)$ linearly independent vertex operators per taming background.

Finally, note that $V^p(x)$ does \emph{not} act on zero modes, by definition \eqref{def phi(x)}. This operator is not charged under the shift symmetry, and does not create momentum in the target space. 

In the game of smoothing, the real fun always comes from microscopic operators such as $\e^{\i p \varphi_x}$. Such an operator is nontrivially related to $V^p(x)$ by smoothing. Consider the expansion
\bel{
  V^p_x \equiv \e^{\i p \varphi_x} = \sum_{n = 0}^\infty \frac{(\i p)^n}{n!} \varphi_x^n.
}
This series can be smoothed term by term, using
\bel{
  \varphi^n(x) = \varphi(x)^n + \trm{contractions}.
}
Here a contraction means the replacement of two $\varphi(x)$'s with the OPE
\bel{
  C \equiv \varphi_x \times \varphi_x = \frac{g^2}{2N} \sum_{k = k\_S + 1}^{\frac N2 - 1} \frac1{\sin\frac{\pi}N k} \approx \frac{g^2}2 \int_{\kappa\_S}^{\frac12} \frac{\d\kappa}{\sin \pi \kappa} = -\frac{g^2}{2\pi} \log\tan \frac{\pi k\_S}{2N} \approx \frac{g^2}{2\pi} \log \frac{2N}{\pi k\_S}.
}
(Smoothing of composite operators is governed by contractions because the only high-mo\-men\-tum factors that are retained are products, or ``contractions,'' of pairs of ladder operators that combine into a number operator. This is explained in more detail in \cite{Radicevic:2019mle}.) The sum over contractions is then
\algns{
  \varphi^n(x)
  &= \varphi(x)^n + \binom n 2 C \varphi(x)^{n - 2} + 3!! \binom n 4 C^2 \varphi(x)^{n - 4} + \ldots = \exp\left\{\frac C2 \hat\del_{\varphi(x)}^2 \right\} \varphi(x)^n.
}
This exponential of a formal derivative is an efficient way to encode all contractions as a linear operation on the space of operators \cite{Polchinski:1998rq}. Thus the smoothed version of $V^p_x$ is
\bel{
  \e^{\i p \varphi}(x) = \exp\left\{\frac C2 \hat\del_{\varphi(x)}^2 \right\} \e^{\i p \varphi(x)} = \e^{- \frac C2 p^2} \e^{\i p \varphi(x)} \approx \left( \frac{2N}{\pi k\_S} \right)^{-g^2 p^2/4\pi} \e^{\i p \varphi(x)}.
}
In other words, the smoothing of an exponential is proportional to the exponential of the smoothing. In particular, choosing $\N \equiv (2N/\pi k\_S)^{-g^2p^2/4\pi}$ gives the advertised relation
\bel{\label{VO relation}
  V^p(x) = \e^{\i p \varphi}(x).
}

Interesting structure arises when studying OPEs of vertex operators. Consider
\bel{
  V^{p_1}_x \times V^{p_2}_y  \equiv \e^{\i p_1 \varphi_x} \times  \e^{\i p_2 \varphi_y} = \e^{\i p_1 \varphi} \e^{\i p_2 \varphi}(x, y) - \e^{\i p_1 \varphi}(x) \, \e^{\i p_2 \varphi}(y).
}
The first term can be calculated using the above trick with exponentiated formal derivatives, but now one must account for there being both ``internal'' and ``external'' contractions, getting
\bel{
  \e^{\i p_1 \varphi} \e^{\i p_2 \varphi}(x, y) = \exp\left\{\frac C2 \hat\del_{\varphi(x)}^2  + \frac C2 \hat\del_{\varphi(y)}^2 + C_{xy} \hat\del_{\varphi(x)} \hat\del_{\varphi(y)} \right\} \e^{\i p_1 \varphi(x)} \e^{\i p_2 \varphi(y)},
}
with the OPE at different points given by \eqref{OPE scalar}
\bel{
  C_{xy} \equiv \varphi_x \times \varphi_y \approx -\frac{g^2}{2\pi} \log \left( \frac{2\pi \e^\gamma k\_S}N |x - y|\right),
}
assuming $|x - y| \gg 1$.

Performing the formal derivatives gives
\algns{
  \e^{\i p_1 \varphi} \e^{\i p_2 \varphi}(x, y)
  &\approx \left( \frac{2N}{\pi k\_S} \right)^{-\tfrac{g^2}{4\pi} (p_1^2 + p_2^2)} \left(\frac{\e^\gamma \pi |x - y|}{\ell\_S}\right)^{\tfrac{g^2}{2\pi} p_1 p_2} \e^{\i p_1 \varphi(x)} \e^{\i p_2 \varphi(y)} \\
  &= \left(\frac{\e^\gamma \pi |x - y|}{\ell\_S}\right)^{\tfrac{g^2}{2\pi} p_1 p_2} \e^{\i p_1 \varphi}(x)\, \e^{\i p_2 \varphi}(y).
}
Thus the OPE is
\bel{
  V^{p_1}_x \times V^{p_2}_y \approx \left[ \left(\frac{\e^\gamma \pi |x - y|}{\ell\_S}\right)^{\tfrac{g^2}{2\pi} p_1 p_2} - 1\right] V^{p_1}(x) V^{p_2}(y).
}
The interesting thing here is that it is the product of vertex operators that exhibits simple behavior, without needing to subtract the product of individually smoothed operators,
\bel{\label{VO prod}
  V^{p_1} V^{p_2}(x, y) \approx \left(\frac{\e^\gamma \pi |x - y|}{\ell\_S}\right)^{\tfrac{g^2}{2\pi} p_1 p_2}  V^{p_1}(x) V^{p_2}(y).
}
This relation is often referred to in the literature as the OPE of two vertex operators. From the present point of view, this is a slight abuse of the term ``OPE.''

It is also possible to define ``standard'' vertex operators that include the zero mode,
\bel{\label{def vertex op}
  \V^p(x) \equiv \N \e^{\i p \left[\varphi(x) + \varphi_0/\sqrt N \right]}.
}
As with the OPE of scalar fields \eqref{OPE scalar}, the zero mode has no effect on the OPEs of vertex operators, so $\V^p(x)$ also obeys the structure \eqref{VO prod}. The zero mode becomes important when calculating correlation functions, however. To see this, consider the vacuum expectation value
\bel{
  \avg{\V^{p_1} \V^{p_2}(x, y)} = \avg{V^{p_1} V^{p_2}(x, y)}_{k \neq 0} \avg{\e^{\i (p_1 + p_2) \varphi_0/\sqrt N}}_{k = 0}.
}
The correlator in the nonzero-momentum space can be evaluated using the same approach that led to \eqref{VO prod}. In the zero-mode space, the situation is simple: since the ground state is an eigenstate of $\pi_0^2$, the only way for the above correlator to be nonzero is to have $p_1 + p_2 = 0$, known as the ``neutrality condition'' for vertex operators. This gives the final answer
\bel{\label{VO correlator}
  \avg{\V^{p_1} \V^{p_2}(x, y)} \approx \frac{\delta_{p_1, \, -p_2}}{(\e^\gamma \pi|x - y|/\ell\_S)^{g^2 p_1^2/2\pi}},
}
which indicates that standard vertex operators $\V^p(x)$, unlike the basic ones $V^p(x)$, after rescaling by $\big( \frac{\e^\gamma\pi}{\ell\_S} \big)^{g^2p^2/4\pi}$ behave like conformal primaries of dimension $g^2 p^2/4\pi$.

\newpage

As in the previous Subsection, the notation used here deserves a careful comparison to the more customary one. As before, red will denote the standard notation in the literature, e.g.\ as found in \cite{Polchinski:1998rq}. Per \eqref{Polchinski map}, vertex operators themselves map between notations as
\bel{
  \red{\nord{\e^{\i p \varphi(x)}}} \quad \longleftrightarrow \quad \e^{\i p \varphi(x)},
}
and the remarkable relation \eqref{VO relation} has the dual notations
\bel{\label{VO nord comparison}
  \red{\e^{\i p \varphi(x)}  = \N \nord{\e^{\i p \varphi(x)}}} \quad \longleftrightarrow\quad V^p(x) = \e^{\i p\varphi}(x) = \N \e^{\i p \varphi(x)}.
}
Note, however, that the literature very often drops the normal-ordering signs, making it difficult to distinguish the two sides of the equality on the l.h.s.~of the above equation. It is also customary to avoid expressions in which the normalization factor $\N$ appears.

Now consider the comparison between notations for products of vertex operators. By \eqref{VO nord comparison}, the product of two normal-ordered operators would be, in traditional notation,
\bel{\label{VO prod comparison}
  \red{\nord{\e^{\i p_1 \varphi(x)}} \, \nord{\e^{\i p_2 \varphi(y)}}} \ = \red{\frac1{\N_1 \N_2} \e^{\i p_1 \varphi(x)} \e^{\i p_2 \varphi(y)}}.
}
The way operator products are traditionally defined, the l.h.s.\ above is (in both notations)
\bel{
  \red{ \exp\left\{ C_{xy}\hat\del_{\varphi(x)} \hat\del_{\varphi(y)}  \right\} \nord{\e^{\i p_1\varphi(x)} \e^{\i p_2\varphi(y)} } }
  \quad \longleftrightarrow\quad   \exp\left\{ C_{xy}\hat\del_{\varphi(x)} \hat\del_{\varphi(y)}\right\} \left(\e^{\i p_1\varphi(x)} \e^{\i p_2\varphi(y)}\right).
}
The r.h.s.\ is, according to the first line in \eqref{Polchinski map},
\bel{
  \red{\frac1{\N_1 \N_2} \e^{\i p_1 \varphi(x)} \e^{\i p_2 \varphi(y)}} \quad \longleftrightarrow\quad \frac1{\N_1 \N_2} \e^{\i p_1\varphi} \e^{\i p_2 \varphi} (x, y),
}
In this paper's notation, after using the product relation \eqref{VO prod} and the r.h.s.\ expressions in the above two lines, the relation \eqref{VO prod comparison} is thus
\bel{
  \exp\left\{ C_{xy}\hat\del_{\varphi(x)} \hat\del_{\varphi(y)}\right\} \left( \e^{\i p_1\varphi(x)} \e^{\i p_2 \varphi(y)} \right) = \frac1{\N_1 \N_2} \exp\left\{ C_{xy}\hat\del_{\varphi(x)} \hat\del_{\varphi(y)}\right\} \left( \e^{\i p_1\varphi}(x) \e^{\i p_2 \varphi} (y) \right).
}
This equality indeed holds, as ensured by \eqref{VO relation}. When in doubt, this exercise can be used to verify the consistency of the notation used here and to compare it to the more standard one. The notation developed in this paper is conceptually simpler than the standard one because it does not rely on the distinction between external and internal contractions when defining products of operators. This notation will be used throughout this series without much further apology.

\subsection{T-duality} \label{subsec T dual}

Any clock model on a lattice admits an exact duality that maps it to another clock model (possibly with gauge constraints) on the dual lattice. In $(1+1)$D, the backbone of this \emph{Kramers-Wannier duality} \cite{Kramers:1941kn} is the map between algebra generators
\bel{\label{KW}
  Z_x\+ Z_{x + 1} = X_{x + \frac12}^\vee, \quad X_x = \big(Z_{x - \frac12}^\vee \big)\+ Z^\vee_{x + \frac12}.
}
The dual lattice in this case is a ring $\Mbb^\vee$ of $N$ sites that correspond to links of the original lattice $\Mbb$; the site of $\Mbb^\vee$ that lies between sites $x$ and $x + 1$ of $\Mbb$ will be denoted $x + \frac12$. The checked operators in \eqref{KW} are defined on sites of $\Mbb^\vee$ and they generate the algebra of another $\Z_K$ clock model.

Kramers-Wannier duality is ``singlet-singlet.'' This means that taking the product of duality relations \eqref{KW} over all $x \in \Mbb$ gives  singlet constraints for the shift symmetries,
\bel{\label{singlet constr}
  \prod_{x \in \Mbb} X_x = \1 \quad \trm{and} \quad \prod_{x + \frac12 \in \Mbb^\vee} X_{x + \frac12}^\vee = \1,
}
that must hold in all states for which the duality is consistent. Another way to say this is that no operators charged under the shift symmetry map under the duality, and in particular individual operators $Z_x$ and $Z_{x + \frac12}^\vee$ have no well defined maps. In order to define maps of these operators it is necessary to ``twist'' the dualities by adding topological degrees of freedom on either or both sides \cite{Radicevic:2018okd}; gauge-fixing these degrees of freedom gives rise to familiar nonlocal dualities involving strings of operators \cite{Kadanoff:1970kz}.

In the dual basis, the microscopic theory \eqref{def H micro} is
\gathl{
  H = \frac{g^2}{2 (\d\phi)^2} \sum_{x = 1}^N \left[2 - \big(Z_{x - \frac12}^\vee\big)\+ Z^\vee_{x + \frac12} - \big( Z_{x + \frac12}^\vee \big)\+ Z^\vee_{x - \frac12} \right] + \frac1{2g^2} \sum_{x = 1}^N \left[ 2 - X_{x + \frac12}^\vee - \big(X_{x + \frac12}^\vee \big)\+ \right].
}
Up to an additive constant, this Hamiltonian describes the same theory as the original one, except with coupling
\bel{\label{KW g}
  g^\vee = \frac{\d\phi}{g}.
}
Kramers-Wannier thus exchanges the strong and weak coupling regimes of the theory \eqref{def H micro}. In particular, the theory is self-dual at coupling $g_\star = \sqrt{\d\phi}$. In order for both the theory and its dual to have a continuum description, one must have
\bel{
  g\_{min} g\_{max} = \d\phi.
}

The same smoothing and taming procedures accomplish different reductions in different duality frames. Consider, first, the case of target space smoothing \cite{Radicevic:1D}. In the original theory, this is a projection that leaves each $X_x$ invariant, while removing all off-diagonal elements in $Z_x$ that do not lie in a $2p\_S \times 2p\_S$ block (when expressed in the momentum basis). In the dual basis, this projection leaves all clock operators invariant while restricting the shift operators $X_{x + \frac12}^\vee$. This means that target space smoothing dualizes to target space compactification: the dual clocks are limited to $2p\_S$ consecutive sites in their $K$-site targets. (The fact that the duality only relates singlets of shift symmetry in the two theories makes it meaningless to ask \emph{which} $2p\_S$ positions are picked out.) In radians, the dual of a target-smoothed theory has clock variables restricted to the window $2p\_S \d\phi \sim p\_S/K$.

Now consider compactifying the smoothed target spaces in the original theory. This results in a tame operator algebra of a theory with a $2n\_T$-dimensional target space per site. On the dual side, this compactification dualizes to smoothing, with wavefunctionals obeying smoothness relations like \eqref{smooth constr general} in which derivatives are suppressed by $n\_T/p\_S$.

In short, the Kramers-Wannier dual of a tamed theory with target-space derivatives suppressed by $p\_S/K$ and maximal angle $2\varphi\_T \sim n\_T/p\_S$ is a tame theory with target-space derivatives suppressed by $n\_T/p\_S$ and maximal angle $\sim p\_S/K$. The duality exchanges the tameness parameters. In particular, this means that a dual of a tame state remains tame.

This conclusion makes it possible to express the duality \eqref{KW} in a simple way that is only meaningful when both sides act on tame states in the appropriate theory. Assuming that gradients between clock positions are much smaller than $\varphi\_T$ or $p\_S/K$ on the appropriate sides of the duality, one can write the map as
\bel{
  \frac1{\d\phi} \del\varphi_x \approx  \pi^\vee_{x + \frac12}, \quad \pi_x \approx  \frac1{\d\phi}  \del\varphi_{x - \frac12}^\vee.
}
The duality of the noncompact scalar cQFT that exchanges $\del \varphi_x$ and $\pi_x$ is usually called \emph{T-duality}. The analysis so far shows precisely under which conditions T-duality emerges from Kramers-Wannier duality.

The singlet constraints \eqref{singlet constr} after taming become the na\"ive zero-mode constraints
\bel{
  \sum_{x \in \Mbb} \pi_x = 0 \quad \trm{and} \quad \sum_{x \in \Mbb^\vee} \pi^\vee_{x + \frac12} = 0.
}
This shows that zero-modes do not map under T-duality in a tame sector. These constraints are na\"ive because they are only valid in states with a few excitations. More generally, as will be discussed in Subsection \ref{subsec shift}, the microscopic singlet constraints will be fulfilled whenever
\bel{\label{pi 0 quantization}
  \sum_{x \in \Mbb} \pi_x, \sum_{x \in \Mbb^\vee} \pi^\vee_{x + \frac12} \in K \Z.
}

T-duality between two ordinary tame sectors does not map zero modes of the momentum operators. Speaking slightly imprecisely, these modes enter the story when T-duality is applied to sectors with nontrivial winding. Consider the taming of Kramers-Wannier duals that brings the original theory to a sector of winding $w$. In this case T-duality would take the form
\bel{
  \frac1{\d\phi} \left(\frac{2\pi w}N + \del\varphi_x\right) \approx  \pi^\vee_{x + \frac12}, \quad \pi_x \approx  \frac1{\d\phi}  \del\varphi_{x - \frac12}^\vee,
}
and the zero-mode constraints that follow from summing over $x$ would be
\bel{\label{T dual wind mom}
  \sum_{x \in \Mbb} \pi_x = 0, \quad \sum_{x \in \Mbb^\vee} \pi^\vee_{x + \frac12} = K w.
}
In other words, a sector with nontrivial winding but with no zero-mode $\pi_0$ is T-dual to a sector with no winding but with a nontrivial zero-mode $\pi_0^\vee$.

This duality between zero-modes of the target momentum and the winding sectors was imposed by consistency of the microscopic theory. It can now be used to provide the promised consistency check on the assumption that zero-modes indeed enter the low-energy theory with the tame term $\frac{g^2}2 \pi_0^2$ in the Hamiltonian.

If the original theory is in the winding sector $w$ and in zero-mode sector $\pi_0 = 0$, its low-energy Hamiltonian \eqref{def HT(w)} is
\bel{
  H_0 + \frac{(2\pi)^2}{2g^2N} w^2,
}
where $H_0$ describes the noncompact scalar that has no zero-modes and therefore maps without issue under T-duality. On the other hand, a dual theory with zero winding and with zero-mode $\pi_0^\vee \equiv \frac1{\sqrt N} \sum_{x \in \Mbb^\vee} \pi^\vee_{x + \frac12} \neq 0$ has Hamiltonian
\bel{
  H_0 + \frac{(g^\vee)^2}2 \left(\pi_0^\vee \right)^2.
}
By using \eqref{T dual wind mom} and the duality between couplings \eqref{KW g}, the energy contributed by the winding sector is found to be equal to the energy carried by zero-modes in the dual theory.

This equality of energies justifies keeping the zero-mode term obtained by na\"ive taming of the original Hamiltonian. Remarkably, the equality (and hence T-duality) says more: overall consistency requires the eigenvalues of $\pi_0^\vee$ to be integer multiples of $K/\sqrt N$, as per \eqref{pi 0 quantization}, subject to the bound \eqref{bound w}. This is a nontrivial constraint that does not follow from any simple analysis of the zero-modes in the microscopic theory.

Nothing so far says how T-duality acts on the zero-mode $\varphi_0$ or the background $\varphi\_{const}\^{cl}$. To understand this it is necessary to twist the lattice duality \eqref{KW} and make the above zero-mode analysis much more precise. This will be done in great detail in \cite{Radicevic:3D}.

\subsection{The shift symmetry and the many compact scalar cQFTs} \label{subsec shift}

The microscopic theory \eqref{def H micro} has a shift symmetry generated by
\bel{\label{def micro shift}
  Q \equiv \prod_{x \in \Mbb} X_x.
}
This symmetry has been weaving in and out of the entire narrative so far. Now it is time to confront it head-on.

The generator $Q$ does not belong to the tame algebra $\A$, regardless of the choice of taming background $\varphi_x\^{cl}$. In fact, no energy eigenstates in any tame subspace can be eigenstates of $Q$. On the other hand, tame eigenstates with few excitations are \emph{approximate} eigenstates of $Q$. In these states the shift generator acts as
\bel{
  Q\_T = \prod_{x \in \Mbb} \left( \e^{\i\, \d\phi \, \pi_x} + O\left(p\_S^2/K^2\right) \right) \approx \e^{\i \, \d\phi \sqrt N \, \pi_0},
}
simply by definition of $\pi_x$ in \eqref{def phi pi}. The approximation is valid if $N \gg \frac K{p\_S}$. The shift symmetry preserves tame states ($Q\_T \approx \1)$ if
\bel{\label{pi 0 quantization redux}
  \pi_0 \in \frac{K}{\sqrt N} \Z,
}
which is precisely the condition imposed by T-duality in \eqref{pi 0 quantization}. Note that the number of different eigenvalues of $\pi_0$ that may appear is bounded by the fact that target space momenta must be below $p\_S$, so there cannot be more than $k\_S \frac{p\_S}K$ zero-mode eigenstates that are also spatially smooth. This bound is the T-dual of the winding number bound in \eqref{bound w}.

It is only in this loose sense that the low-energy eigenstates of the scalar cQFT encountered so far can be considered as shift symmetry eigenstates.  The illusion falls apart when one considers shifts by large angles in the target space. For instance, shifting by $2\varphi\_T = 2\pi \frac{n\_T}{p\_S}$, which is enacted by the operator $S \equiv Q^{n\_T K/p\_S}$, changes the taming background by
\bel{
  \varphi_x\^{cl} \mapsto \varphi_x\^{cl} + 2\varphi\_T.
}
This means that for any specific choice of taming background, taming $S$ gives
\bel{
  S\_T = 0.
}
From the point of view of the cQFT, as the power $n$ of the operators $Q^n$ increases from $O(1)$ to $O(n\_T K/p\_S)$, this operator will create so many excitations that it is no longer possible to approximate each $X_x$ with $\1 + \i \, \d\phi\, \pi_x$. In other words, the operators $\pi_x$ stop being canonical in such highly excited states.

\newpage

To construct proper low-energy eigenstates of this ``untame'' shift symmetry, one must consider superpositions of tame states from different superselection sectors, e.g.
\bel{\label{shift eigenstate}
  \sqrt{\frac{\varphi\_T}\pi} \sum_{n = 1}^{\pi/\varphi\_T} \e^{2 \i n q \varphi\_T} \qvec{\psi; \varphi_x\^{cl} + 2 n \varphi\_T}.
}
Here $\psi$ labels an eigenstate of $H_0$ with few excitations, and the field after the semicolon is the superselection sector label. States constructed this way are eigenstates of both $H$ and the shift operator $S$.

Note that shifts enacted by $S$ are still infinitesimal, in the sense that they only shift the clocks by $2\varphi\_T \ll 1$. This is the kind of ``continuous'' shift symmetry that can be simply discussed within a tame context. The charge under this shift symmetry --- labeled by $q$ in \eqref{shift eigenstate} --- does \emph{not} enter the microscopic Hamiltonian at leading order in taming parameters.

Asking about shifts by angles smaller than $2\varphi\_T$ is well defined from the microscopic point of view, but is tricky from a cQFT perspective. On the one hand, the shift by $\d\phi$ was above shown to approximately \emph{preserve} all states of the form \eqref{shift eigenstate}. On the other hand, shifting by angles $\Delta \phi$ that satisfy $\d\phi \ll \Delta \phi < \varphi\_T$ is more troublesome. Doing so to a particular state $\qvec{\psi; \varphi_x\^{cl}}$ creates a lot of excitations in two superselection sectors --- in the starting one and in the one labeled by $\varphi_x\^{cl} + 2\varphi\_T$. Neither of the resulting states in the two sectors is an eigenstate of $H_0$ with few excitations.

This subtlety of shifting sheds more light on the ``quantization'' \eqref{pi 0 quantization redux} of the momentum zero-mode $\pi_0$, which was seen in the previous Subsection to rather mysteriously follow from T-duality. From the current perspective, this quantization ensures that tame states with few excitations are approximate eigenstates of the microscopic shift symmetry. The eigenvalues of $\pi_0$ can be \emph{very loosely} interpreted as labels for superselection sectors of the microscopic shift symmetry that were missed by restricting to shifts by integer multiples of $2\varphi\_T$.

This discussion leads to several definitions of a \emph{compact scalar cQFT}. Recall that recent literature distinguishes two notions of a compact scalar \cite{Seiberg:2016gmd}: the ``high energy (HE) compact scalar'' is any theory in which the target space is compact, while the ``condensed-matter (CM) compact scalar'' is a theory in which the Hamiltonian contains the operators that change winding sectors. As it stands, this distinction is too coarse once one demands that a compact scalar come from a finite microscopic theory. There are two reasons for saying this. First, from a finitary point of view \emph{every} bosonic theory has a compact target space. In this sense, the HE definition is merely a truism. Second, there are multiple labels for a sector within which a noncompact scalar cQFT is defined --- beside the winding $w$, one must specify the constant background $\varphi\^{cl}\_{const}$ and the eigenvalue of the tame momentum zero-mode $\pi_0$. Thus the CM definition does not allow the Hamiltonian to contain some operators that reflect the target compactness, like $S$, just because they do not change the winding number.

It is thus necessary to be more precise about the definition of a ``compact scalar theory.'' In general, in the presence of a symmetry in the microscopic theory, one can define different subtheories by choosing different answers to the following questions:
\begin{enumerate}
  \item Which eigenstates of the symmetry generator belong to the Hilbert space of the subtheory?
  \item Assume that more than one eigenstate of the symmetry is allowed in the subtheory. Are operators charged under the symmetry allowed in the operator algebra of the subtheory?  Equivalently, are superpositions between different symmetry eigenstates allowed, or are there superselection rules?
  \item If the charged operators are in the algebra, are they included in the Hamiltonian of the subtheory? Equivalently, is the symmetry used to define the subspace explicitly broken by adding operators to the Hamiltonian of the subtheory?
\end{enumerate}

As an example, consider how the different noncompact scalar cQFTs of Subsection \ref{subsec noncomp scalar} fit into this framework. There the symmetry generator of interest was the tame momentum zero-mode $\pi_0$:
\begin{itemize}
  \item The basic noncompact scalar is the subtheory that contains only the $\pi_0 = 0$ eigensector. Since only one sector is allowed, the remaining two questions are moot.
  \item The standard noncompact scalar is the subtheory that contains all eigensectors of $\pi_0$. On historical grounds, these tame states may be called the \emph{Anderson tower of states} \cite{Anderson:1952}. Its operator algebra, but not the Hamiltonian, includes the operators $\varphi_0$ conjugate to $\pi_0$. These were crucial in the definition of vertex operators $\V^p(x)$ in \eqref{def vertex op}.
\end{itemize}
Both noncompact scalars were subtheories in a single $\varphi\^{cl}_x$ sector.

Another example comes from the particle numbers $n_k$. The smoothing procedure discards all operators charged under the particle number symmetries at $k \notin \Pbb\_S$. The continuum subtheory is conventionally chosen to lie in just one superselection sector of these high-momentum symmetries --- the sector that contains the ground state. This is the case in any cQFT, as per the definition in Section \ref{sec definitions}.

As for the particle number symmetries for $k \in \Pbb\_S$, here every cQFT is a subtheory of the lattice model that retains all the relevant superselection sectors. In addition, all the charged (ladder) operators remain in the algebra. In the free cQFT, the Hamiltonian does not contain operators that violate these particle number symmetries. There may also exist interacting cQFTs with Hamiltonians of form \eqref{def H general} in which the $n_k$'s are still symmetries. However, interacting cQFTs whose Hamiltonians contain ladder operators at $k \in \Pbb\_S$ are also readily constructed.

Compact scalar theories can now be generally \emph{defined} as those subtheories of the clock model whose state space contains states that are tame relative to any taming background $\varphi\^{cl}_x$ given by \eqref{taming backgrounds}. There is a plethora of compact theories one may now define by choosing different answers to the remaining questions in the ``decision tree'' on the previous page.

All compact scalar theories considered here will have algebras whose generating set consists of, at least, the set \eqref{alg basis noncomp w zero} and the operator $S \equiv Q^{n\_T K/p\_S}$ that generates shifts by $2\varphi\_T$, defined based on \eqref{def micro shift}. In other words, all compact theories will have a Hilbert space that includes superpositions of states with different offsets $\varphi\^{cl}\_{const}$. One such state is \eqref{shift eigenstate}.

Any compact theory that is defined in a \emph{single} winding sector, i.e.\ at fixed $w$,  will be referred to as the \emph{basic compact scalar theory}. This is not a theory that is typically discussed in the literature --- usually either all backgrounds \eqref{taming backgrounds} are included, or none are. Nevertheless, keeping track of this theory is conceptually useful, as it separates the degenerate sectors of the microscopic theory labeled by $\varphi\^{cl}\_{const}$ from the gapped sectors labeled by $w$.

An interacting basic compact scalar Hamiltonian may contain terms involving the operator $S$. Such terms explicitly break the shift symmetry $\varphi\^{cl}\_{const} \mapsto \varphi\^{cl}\_{const} + 2\varphi\_T$. This paper will not consider such perturbations. A ``basic compact scalar theory'' will here always have a degenerate ground state due to these shifts.

A compact theory whose Hilbert space includes \emph{all} winding sectors will be called the \emph{high energy (HE) compact scalar}. Here it may be assumed that the algebra of this subtheory contains operators
\bel{\label{def vortex op}
  \prod_{x \in \Mbb} X_x^{m(x)}, \quad m(x) = \left[\frac{K}N  x\right].
}
These are \emph{vortex operators} --- no relation to vertex operators. They change the winding number of a given background $\varphi\^{cl}_x$ by $w \mapsto w + 1$. Including them in the algebra means that the Hilbert space contains superpositions of states of different windings. This aligns the present terminology with that of \cite{Seiberg:2016gmd}.

Note, by the way, that the vortex operators \eqref{def vortex op} were defined, for convenience, using the floor function and without regards to spatial smoothing. It is possible to define related operators in which $m(x)$ is constant over a ``string length'' $\ell\_S$ and then jumps by $[\ell\_S K/N] \sim [K/k\_S]$. The microscopic choices that go into the definition of vortex operators should not matter at leading order in the various smoothing parameters.

A HE compact scalar that comes from the clock model has an approximately conserved winding number. In particular, no tame operators can change $w$. However, it is possible to consider perturbations of the HE compact scalar that \emph{do} include vortex operators in the Hamiltonian. Any such interacting theory will be called the \emph{condensed matter (CM) compact scalar}. CM compact scalars will not be studied in this paper.


\newpage

\subsection{Dimensionful variables and energy scales} \label{subsec dimensionful}

The entire analysis so far has been done in terms of dimensionless numbers or matrices. This was the case even for cQFT variables. Such a description of the continuum is very unusual. When trying to pass from the lattice to the continuum, the usual approach \cite{Wilson:1974sk} is to introduce a lattice spacing $a \propto 1/N$, to rescale all fields and couplings by powers of $a$ (these exponents are called \emph{engineering dimensions}), and to aver that the continuum theory is obtained by sending $a \rar 0$ and keeping whatever part of the Hamiltonian remains finite in this limit. This is a convenient heuristic, but as shown in this paper, there is more to the continuum limit than just taking $N \rar \infty$. In fact, the preceding Subsections show that all the hallmarks of a cQFT appear without ever rescaling the operators or even introducing a separate quantity called the lattice spacing.

It is nevertheless instructive to carry out the usual motions and rewrite the various cQFTs developed so far in terms of dimensionful variables. Define the \emph{lattice spacing}
\bel{\label{def a}
  a \equiv \frac L N, \quad L \in \R.
}
The system size $L$ is an \emph{arbitrary} number that will never enter any correlation functions except through the combination $L/a$. However, conceptually it is simplest to think of $L$ as a positive $O(1)$ number that is meaningful on its own. This causes the spacing $a$ to be a small, $O(1/N)$, quantity when the lattice contains many sites.

The spatial lattice coordinates $x$ can now be rescaled to define the continuum coordinates
\bel{
  x\^c \equiv x a.
}
Their allowed values range from $x\^c = a$ to $x\^c = L$ in ``small'' steps $a$. Similarly, the lattice derivatives can be given continuum counterparts via
\bel{
  \del\_c f(x) \equiv \frac1a \del f(x) = \frac{f(x + 1) - f(x)}a.
}

It is also natural to define new fields that are functions of continuum coordinates, not lattice ones. These continuum fields can be defined as rescalings of smoothed operators. In the scalar cQFT, this means defining
\bel{
  \varphi\_c(x\^c) \equiv a^{-\Delta_\varphi} \varphi(x), \quad \pi\_c(x\^c) \equiv a^{-\Delta_\pi} \pi(x).
}
The engineering dimensions will turn out to have natural values $\Delta_\varphi = 0$ and $\Delta_\pi = 1$.

\newpage

These natural values of field dimension are found in the traditional way, as follows. The dimensionful Hamiltonian of the standard noncompact scalar can be defined as
\algns{
  H\_c
  \equiv \frac1a H\_T
  &= \frac1a \sum_{x = 1}^N \left[\frac{g^2}2 \pi^2(x) + \frac1{2g^2} (\del\varphi)^2(x) \right] \\
  &= \frac1{a^2} \sum_{x^c = a}^L a \left[a^{2\Delta_\pi} \frac{g^2}2 \pi\_c^2(x\^c) + a^{2 + 2\Delta_\varphi} \frac1{2g^2} (\del\_c \varphi\_c)^2(x\^c) \right] \\
  & \equiv  \int_{a}^L \d x\^c \, \left[\frac{g^2}2 \pi\_c^2(x\^c) + \frac1{2g^2} (\del\_c \varphi\_c)^2(x\^c) \right].
}
In the last line the sum was merely written as an integral, and the engineering dimensions were chosen to remove any explicit prefactors of $a$. The lower bound of the integral can be replaced with $0$ at the expense of adding an $O(1/N)$ correction to $H\_c$.

Note that the above expressions involve smoothed products of microscopic fields $\pi$ and $\del \varphi$. In this paper's notation, it would be incorrect to write e.g.\ $\pi(x)^2$ instead of $\pi^2(x)$. The product $\pi(x)^2$ is normal-ordered, in traditional nomenclature. It does not depend on momentum modes above $k\_S$; if this term were used in the Hamiltonian, the high-momentum ($|k| > k\_S$) modes would all have exactly zero energy.

The choice of engineering dimensions could have been altered by defining a continuum coupling
\bel{
  g\_c \equiv a^{-\Delta_g} g.
}
To find a natural value for this dimension, another requirement can be added: the commutation relation between smooth fields should give a continuum $\delta$-function,
\bel{
  [\varphi\_c, \pi\_c](x\^c, y\^c) \approx \i\, \delta\_c(x^c - y\^c) \equiv \frac\i a \delta_{x, y}.
}
The commutator \eqref{comm rel bos} then implies $\Delta_\varphi + \Delta_\pi = 1$. This justifies the above-quoted values of $\Delta_\varphi$ and $\Delta_\pi$ and fixes $\Delta_g = 0$.

Continuum momenta can be defined as
\bel{
  k\^c \equiv \frac{2\pi}L k \in \left[-\frac\pi a, \frac\pi a\right).
}
This means that the smoothness cutoff in continuum momentum space is $k\_S\^c \equiv \frac{2\pi}{L} k\_S$. The ``string scale'' or smoothing length can also be given a dimensionful analogue,
\bel{
  \ell\_S\^c \equiv a \ell\_S = \frac{L}{2k\_S} = \frac\pi{k\_S\^c}.
}

The rescaling of the Hamiltonian means that the energy spectrum also has its dimensionful version. The $k \neq 0$ excitations in the basic noncompact cQFT have gap $\omega_{k = 1} \sim 1/N$, whose continuum analogue is
\bel{\label{gap basic}
  \E\_{basic}\^c \equiv \frac1a \omega_{k = 1} \sim \frac1L.
}
Excitations at momenta above $k\_S$ have gap $\omega_{k = k\_S} \sim k\_S/N$, or
\bel{\label{gap nonsmooth}
  \E\_S\^c \equiv \frac1a \omega_{k = k\_S} \sim \frac{2k\_S}L = \frac1{\ell\_S\^c}.
}
Moving away from the basic cQFT, excitations of the tame momentum zero-mode have gap
\bel{\label{gap zero modes}
  \E\_{momentum}\^c \sim \frac{g^2 K^2}L \sim \frac1{(g^\vee)^2 L},
}
which is $\sim K/L$ at the self-dual coupling $g_\star = \sqrt{\d\phi}$. Next, entering the realm of the compact scalar cQFTs, states of nontrivial winding have gap
\bel{\label{gap winding}
  \E\_{winding}\^c \sim \frac1{g^2 L},
}
which is $\sim K/L$ at the self-dual coupling --- the same as $\E\_{momentum}\^c$, in accord with T-duality. And finally, excitations associated to ``untamed'' shifts $\varphi\_{const}\^{cl} \mapsto \varphi\_{const}\^{cl} + 2\varphi\_T$ are gapless (or, more precisely, have a gap that is so small that it is not detectable by the tame Hamiltonian at leading order in the taming parameters).

Unsurprisingly, the smooth scale is always much greater than the scale of low-momentum excitations, $\E\_S\^c \gg \E\_{basic}\^c$. The other two scales are more interesting.  At the self-dual point, both zero-modes and winding sectors are at high energies compared to $\E\_{basic}\^c$. However, as the coupling is dialed away from this point, one of these scales will be lowered until it becomes comparable to $\E\_{basic}\^c$. This signals the imminent breakdown of the tameness assumption: as windings or zero-mode excitations become energetically favorable, the low-energy spectrum will start including states that are not smooth or compact in the target space.

There is one further rescaling that is often used when working with a scalar cQFT. Let
\bel{\label{def radius}
  R^2 \equiv \frac{2g^2}{\d\phi}, \quad \varphi_R(x\^c) \equiv \frac{\varphi\_c(x\^c)}R, \quad \pi_R(x\^c) \equiv R\, \pi\_c(x\^c).
}
The coupling $R$ is usually called the \emph{radius} of the compact boson, and the self-dual point is at $R_\star = \sqrt 2$. The continuum Hamiltonian takes a form that will later lead to a particularly simple action,
\bel{
  H\_c = \int_a^L \d x\^c \, \left[ \frac{\d\phi}4 \pi^2_R(x\^c) + \frac1{\d\phi} (\del\_c \varphi_R)^2(x\^c) \right].
}

\subsection{Spontaneous breaking of the shift symmetry}

The notion of symmetry and its breaking has played a central role in QFT for decades now. When viewed through this lens, the phase structure of the nonchiral clock model \eqref{def H micro} can be described as follows:
\begin{enumerate}
  \item At $g \rar \infty$, the $\Z_K$ shift symmetry, generated by $Q$ from \eqref{def micro shift}, is unbroken.
  \item At $g \rar 0$, the $\Z_K$ shift symmetry is spontaneously broken, and the ground state has a $K$-fold degeneracy.
  \item For $K < 5$, these two phases extend to finite values of $g$, and a second-order phase transition separates them.  In particular, for $K = 2$ the onset of the spontaneous breaking of the $\Z_2$ symmetry is described by the Ising conformal theory.
  \item For $K \geq 5$, the situation is more complicated. The intact and broken $\Z_K$ phases are separated by a line of critical points in $g$-space where the fate of the $\Z_K$ symmetry is less obvious.
\end{enumerate}

A careful reader of Subsection \ref{subsec shift} will now immediately notice that its analysis of tame zero-modes and ``untamable'' shifts translates to a rather precise statement about the fate of the $\Z_K$ symmetry in the middle of the BKT region in $g$-space, when $K \gg 1$:
\begin{enumerate}
  \item[5.] In the vicinity of the self-dual point $g_\star = \sqrt{\d\phi}$, the shift symmetry is \emph{partially broken}. More precisely, only a subgroup $\Z_{K'} \subset \Z_K$  is spontaneously broken. This subgroup corresponds to shifts by $K/K'$ sites in the target space. It is consistent to assume that the taming parameters $n\_T$ and $p\_S$ can be chosen in such a way that this $\Z_{K'}$ is generated by the shift operator $S = Q^{n\_T K/p\_S}$, meaning that $K' = \pi/\varphi\_T = p\_S/n\_T$, and that $\Z_{K'}$ is broken by a specific choice of $\varphi\_{const}\^{cl}$. The remainder of the symmetry, corresponding to the coset $\Z_K/\Z_{K'}$, is approximately generated by the tame zero-mode $\pi_0$.
\end{enumerate}

Such a ``consistent'' choice of taming parameters ensures that a nonzero gap \eqref{gap zero modes} is associated to tame zero-modes. This implies that there is no ground state degeneracy in the $k = 0$ sector beyond what is generated by $S$. As $g$ is decreased, it will become impossible to choose taming parameters such that the tame zero-modes do \emph{not} have small gaps; this signals the transition towards a phase in which the full $\Z_K$ is spontaneously broken.

More generally, as $g$ is dialed from $O(1/K)$ to $O(K^0)$, the spontaneously broken group should change from $\Z_K$ to $\Z_1$. This is the scenario depicted on Fig.\ \ref{fig phases}. This means that the BKT regime can be viewed as a spontaneous breakdown of $\Z_K$ staggered over an $O(1)$ interval of parameter space. Such a crossover stands in contrast to ordinary critical points, where all gapless modes are gapped out simultaneously as $g$ sweeps over an interval whose size vanishes as $K, N \rar \infty$.

\newpage

All this talk of regimes with broken symmetry appears to run counter to a famous piece of lore, the Coleman-Hohenberg-Mermin-Wagner (CHMW) theorem: in one spatial dimension, a continuous symmetry cannot be spontaneously broken \cite{Mermin:1966fe, Hohenberg:1967zz, Coleman:1973ci}. At first glance, it may appear that this theorem does not apply, as $\Z_K$ is not continuous. However, this paper is interested precisely in the $K \gg 1$ regime in which $\Z_K$ can, for all intents and purposes, be identified with U(1). In fact, it is easy to use the taming formalism to give a simple proof of the CHMW theorem for the clock model. This proof will also highlight how, for arbitrarily large $K$, the theorem stops applying at small but still $O(K^0)$ couplings.

Consider the basic noncompact scalar cQFT with Hamiltonian \eqref{def H0}. Taming is only consistent if the inequality
\bel{\label{CHMW bound origin}
  \avg{\varphi_x^2} < \varphi\_T^2
}
is satisfied for all $x$. This expectation value can be calculated by the same approach as \eqref{OPE scalar sum},
\bel{
  \avg{\varphi_x^2} \approx \frac{g^2}N \sum_{k = 1}^{\frac N2 - 1} \frac1{2 \sin\frac \pi N k} \approx \frac{g^2}{2\pi} \log N.
}
The CHMW conclusion is that, for $N \gg 1$, this expectation diverges and violates the bound \eqref{CHMW bound origin}. This argument goes through for any $K$ that is large enough to allow a definition of the taming parameter $\varphi\_T \gg \d\phi = 2\pi/K$; no further requirement on $K$ is assumed.

This derivation also makes it clear that the CHMW theorem does \emph{not} hold when the coupling satisfies
\bel{
  g \ll \frac{\varphi\_T}{\sqrt{\log N}}.
}
In fact, unless the spatial system size is taken to be more than exponentially larger than the target space size, the meat of this bound is captured by the weaker constraint
\bel{
  g \ll \varphi\_T.
}
This is certainly true in the ordered phase, when $g \leq g\_{KT}^\vee \sim 1/K$. Therefore there is no contradiction in stating that the $\Z_K$ symmetry is broken at extremely small couplings.

It is also not difficult to imagine that for large enough $K$ and $N$ one can choose $\varphi\_T$ such that the CHMW theorem fails in the vicinity of the self-dual point $g_\star \sim 1/\sqrt K$. In fact, this paper tacitly assumes this is the case throughout. The corresponding noncompact cQFT degrees of freedom, associated to the free boson CFT, can then be viewed as the Nambu-Goldstone bosons of the spontaneously broken $\Z_{K'}$ symmetry for $K' = \pi/\varphi\_T$.

\newpage

\section{Continuum path integrals for scalars} \label{sec bos path int}

\subsection{Preliminaries} \label{subsec path int prelims}

Section \ref{sec scalars} identified, among others, four continuum theories based on the nonchiral clock model with coupling $g$ close to the self-dual point $g_\star = \sqrt{\d\phi}$:
\begin{enumerate}
  \item The basic noncompact scalar, with three nondynamical labels of note: the tame zero-mode $\pi_0$, the constant piece of the background $\varphi\^{cl}\_{const}$, and the winding number $w$.
  \item The standard noncompact scalar (the basic noncompact scalar with a dynamical $\pi_0$).
  \item The basic compact scalar (the standard noncompact scalar with a dynamical $\varphi\^{cl}\_{const}$).
  \item The HE compact scalar (the basic compact scalar with a dynamical $w$).
\end{enumerate}
In each theory, the nondynamical variables label different superselection sectors. Making each variable dynamical means expanding the algebra to include operators that change that variable, or expanding the Hilbert space to include superpositions of different variable values.

Each theory in turn leads to a different kind of path integral. The framework is as follows. Given a theory whose superselection sectors are labeled by $\greek s$, with a state space denoted by
\bel{
  \H = \bigoplus_{\greek s} \H_{\greek s} \equiv \bigoplus_{\greek s} \trm{span}\left\{ \qvec{ \greek f; \greek s} \right\},
}
the path integral is obtained by inserting a decomposition of unity
\bel{\label{decomposition of unity}
  \1 = \sum_{\greek s,\, \greek f} \qproj{\greek f; \greek s}{\greek f; \greek s}
}
at each time step, i.e.\ between every two operators in the desired correlation function. The fact that $\greek s$ labels superselection sectors means that no operator can change $\greek s$, so correlation functions get expressed as
\bel{\label{correlator decompositions}
  \qmat{\greek f\_f; \greek s\_f}{\prod_{n = 1}^{N_0} \O_n}{\greek f\_i; \greek s\_i} = \delta_{\greek s\_i, \, \greek s\_f} \sum_{\{\greek f_n\}} \prod_{n = 1}^{N_0 + 1} \qmat{\greek f_{n + 1}; \greek s\_i}{\O_n}{\greek f_n; \greek s\_i}, \quad \greek f_1 \equiv \greek f\_i, \ \greek f_{N_0 + 1} \equiv \greek f\_f.
}

In the basic noncompact scalar, the labels $\greek f$ will denote linearly independent, tame, spatially smooth states; $\greek s$ will denote $k \notin \Pbb\_S$ occupation numbers, zero-modes, and taming backgrounds. As more variables are made dynamical in the progression of cQFTs towards the HE compact scalar, the corresponding superselection sector labels are moved from $\greek s$ to $\greek f$, and it is understood that the operators $\O_n$ may now change these new dynamical labels.

To make things concrete, consider the thermal partition function of the microscopic theory,
\bel{
  \Zf \equiv \Tr\, \e^{-\beta H}, \quad \beta \in \R^+.
}
The usual way of expressing $\Zf$ as a path integral is to write
\bel{
  \e^{-\beta H} = \prod_{\tau = \d\tau}^{\beta} \e^{-\d\tau H}, \quad \d\tau \equiv \frac\beta{N_0},
}
and to insert a decomposition of unity between every pair of sequential operators $\e^{-\d\tau H}$. To get an exact answer, it is necessary to insert a complete set of basis states --- i.e.\ a sum over $D = K^N$ states --- at each time step.

A \emph{continuum path integral} is obtained if the standard procedure is modified as follows:
\begin{itemize}
  \item The states $\{\qvec{\greek f; \greek s}\}$ are chosen to form an \emph{undercomplete} basis; any state that is not tame w.r.t.\ some background $\varphi_x\^{cl}$ is simply dropped from the decomposition \eqref{decomposition of unity}.
  \item Instead of performing the sum over superselection sectors, $\greek s$ is fixed to a specific value.
\end{itemize}
If the chosen sector captures the low-energy spectrum of the microscopic theory, the corresponding continuum path integral can be a good approximation to the exact answer $\Zf$ as long as the temperature $1/\beta$ is much smaller than the highest energy accessed by this sector.\footnote{\label{foot roughening}The precise meaning of ``much smaller'' depends on the detailed spectrum of the microscopic eigenstates that are not captured by the cQFT. If $\E$ is the highest energy of a tame state, for example, then the na\"ive temperature bound for the validity of the continuum path integral is $1/\beta \ll \E$. An infinitesimally less na\"ive bound is $1/\beta \ll \E/\log D$, which takes into account that the temperature must be low enough so that the \emph{sum} of all states at energy $\gtrsim \E$ is negligible in the partition function. When temperatures are between $\E/\log D$ and $\E$, the exact path integral experiences a ``roughening'' as it gradually starts discovering the untame and nonsmooth states in the spectrum, which causes it to deviate from the continuum path integral. This ``roughening transition'' was discussed in some detail in \cite{Radicevic:1D}; in that QM context, only the untame states contributed to the roughening, while here it is possible to distinguish between spatial and target roughenings.}


The simplest continuum path integral is associated to the basic noncompact scalar.  Here the label $\greek f$ used in \eqref{correlator decompositions} collects the nontrivial eigenvalues of all spatially smooth position operators $\widehat\varphi(x)$ at $2k\_S$ different sites $x$. It is natural to take these points to be a ``string length'' apart,
\bel{\label{def xi}
  x = \ell\_S\, \xi, \quad \xi \in \{ 1, \ldots, 2k\_S\}.
}
The superselection sector label, on the other hand, contains particle numbers of spatially nonsmooth but tame states, as well as data on the tame zero-modes and taming backgrounds,
\bel{\label{def s}
  \greek s = \left(\{n_k\}_{k \notin \Pbb\_S}, \pi_0,  \varphi\_{const}\^{cl}, w\right).
}
By definition, all of these labels will be fixed in the continuum path integral.

To understand how well this continuum path integral can approximate the exact answer, consider the energy scales associated to each superselection sector. The dimensionful energy scales are shown in eqs.\ \eqref{gap basic}--\eqref{gap winding}. This Section will mostly work with dimensionless energies, which are obtained from the dimensionful ones by simply multiplying them by $a$.

The first consequence of these energy estimates is that, if $1/\beta \ll \E\_{basic} \sim 1/N$, the basic cQFT modes will fail to contribute to the partition function. The resulting path integral will not depend on local degrees of freedom. Such situations will be considered in \cite{Radicevic:3D}, but for now, assume that $1/\beta \gg 1/N$. In both lattice and continuum notational conventions, this means
\bel{
  \beta \ll N \quad \Leftrightarrow \quad \beta\^c \equiv a\beta \ll L.
}

Next, if the temperature is low enough, it may be possible to ignore some superselection sectors entirely. Roughly speaking (see footnote \ref{foot roughening}), if $1/\beta$ is much smaller than $k\_S/N$, $1/g^2 N$, and $1/(g^\vee)^2 N$, then one may disregard nonsmooth states, winding sectors, and tame momentum modes, respectively. In this regime it is appropriate to assume that $\e^{-\beta H}$ will not change the superselection labels $\{n_k\}_{k \notin \Pbb\_S}$, $w$, and $\pi_0$, and that one may simply set these labels to zero and hence minimize the total energy. In other words, at sufficiently low temperatures, meaning
\bel{
  \beta \gg \ell\_S, \ g^2 N, \ (g^\vee)^2 N \quad \Leftrightarrow \quad \beta\^c \gg \ell\^c\_S, \ g^2 L, \ (g^\vee)^2 L,
}
one can get a good approximation to $\Zf$ by computing the continuum path integral in just one superselection sector associated to these three types of labels.

Finally, there appears to be no temperature that is small enough to justify excluding the shift eigenstates \eqref{shift eigenstate} from the partition function, which is what working with a noncompact cQFT path integral does. However, the tame Hamiltonian $H\_T$ can never change $\varphi\_{const}\^{cl}$, and it is independent of its value. Thus, even though a continuum path integral based on a noncompact scalar cQFT cannot reproduce the correct partition function $\Zf$, the sum over superselection sectors labeled by $\varphi\_{const}\^{cl}$ will simply end up multiplying the single-sector result by $\pi/\varphi\_T = p\_S/n\_T$. Such a prefactor merely shifts the free energy by a constant; it is important when calculating the Casimir energy, but does not influence any thermal correlators.

This conclusion is only valid as long as only tame excitations contribute to $\Zf$. The relevant energy scale $\E\_T$ above which untame states dominate the partition function is difficult to estimate. A na\"ive guess, $\E\_T \sim n\_T/N$, comes from assuming that each momentum mode has a $\sim n\_T$-dimensional state space.  However, this guess is inconsistent: since $n\_T \ll K$, this value of $\E\_T$ would be much lower than $\E\_{electric}$ or $\E\_{magnetic}$ near the self-dual point. This paper will not attempt to calculate $\E\_T$. Instead, it will be assumed that it exists and satisfies
\bel{\label{gap tameness}
  \E\_T \gg \E\_S.
}

There are thus four scales to keep in mind when working with continuum path integrals.  Two of them, $\E\_T$ and $\E\_S$, control the validity of excluding untame states and smoothing sectors that do not contain the ground state. To get the traditional path integrals, in addition to \eqref{gap tameness}, $\E\_S$ must be assumed to be much greater than all other scales.

The remaining two energy scales, $\E\_{momentum}$ and $\E\_{winding}$, control the validity of excluding tame zero-modes and winding sectors. They determine which of the four continuum path integrals is sufficient to compute the correct partition function $\Zf$ at a given temperature.
\begin{enumerate}
  \item The basic noncompact scalar continuum path integral is (almost) sufficient when
      \bel{
        \beta \gg g^2 N, \ (g^\vee)^2 N \quad \Leftrightarrow \quad  \beta\^c \gg g^2 L, \ (g^\vee)^2 L.
      }
      In this regime $\Zf$ can be calculated, up to a $p\_S/n\_T$ prefactor, by setting all the superselection labels in \eqref{def s} to $\pi_0 = w = n_k = \varphi\_{const}\^{cl} = 0$. Theories in this class can be made interacting by weakly perturbing the Hamiltonian with smooth operators (arbitrary products of modes $c_k$ and $c_k\+$ at $k \in \Pbb\_S \backslash \{0\}$).
  \item The standard noncompact scalar continuum path integral may be used when
      \bel{
        \beta \lesssim (g^\vee)^2 N, \quad \beta \gg g^2 N \quad \Leftrightarrow \quad \beta\^c \lesssim (g^\vee)^2 L, \quad \beta\^c \gg g^2 L.
      }
      In this case the eigenvalue of $\pi_0$ is summed over, but it remains valid (again, up to a $p\_S/n\_T$ prefactor) to fix $w = n_k = \varphi\_{const}\^{cl} = 0$. The interactions in this case may also include operators $\pi_0$ and $\varphi_0$. Note that this regime is only possible when $g^\vee \gg g$, or
      \bel{
        g \ll \sqrt{\d\phi}.
      }
      This is an extreme weak-coupling limit in which windings can be ignored.
  \item The basic compact scalar continuum path integral may be used in the same parameter regime as above. Now the summation over $\varphi\_{const}\^{cl}$ gives the correct constant prefactor. Interactions can include the shift operator $S$ and a potential $V(\varphi\_{const}\^{cl})$ for the taming background.
  \item The HE compact scalar cQFT must be used when
      \bel{
        \beta \lesssim g^2 N, \ (g^\vee)^2 N \quad \Leftrightarrow \quad  \beta\^c \lesssim g^2 L, \ (g^\vee)^2 L.
      }
      The only fixed superselection labels are now $n_k = 0$ at $k \notin \Pbb\_S$, and the interaction terms may include vortex operators \eqref{def vortex op}. In the absence of such interactions, the sum over winding numbers is weighted by the potential shown in \eqref{def HT(w)}.
\end{enumerate}

Note that it is also possible to consider a regime in which
\bel{
  \beta \gg (g^\vee)^2 N, \quad \beta \lesssim g^2 N \quad \Leftrightarrow \quad \beta\^c \gg (g^\vee)^2 L, \quad \beta\^c \lesssim g^2 L.
}
This is the T-dual of a standard noncompact scalar in which winding sectors must be summed over while the tame zero-modes are kept nondynamical with a fixed value of $\pi_0 = 0$. This is necessarily a strong-coupling regime, available only when $g \gg \sqrt{\d\phi}$. It will not feature prominently in this paper as it has the same physics as the standard noncompact scalar.

At the self-dual point, where $g = g^\vee = \sqrt{\d\phi}$, only the basic noncompact scalar and the HE compact scalar can be used to build a continuum path integral. Which one is used depends on the temperature. The noncompact theory is applicable when
\bel{
  g^2 = \frac{2\pi}K \ll \frac \beta N = \frac{\beta\^c}L.
}
Otherwise, 
one must sum over both windings and zero-mode states, as in the compact theory. This last statement can be expressed in terms of the boson radius $R = O(1)$, defined in \eqref{def radius}, as
\bel{
  \beta\^c \lesssim \hbar R^2 L.
}
The label $\hbar \equiv \d\phi$ will be justified below, cf.\ \eqref{def S nc cont R}.


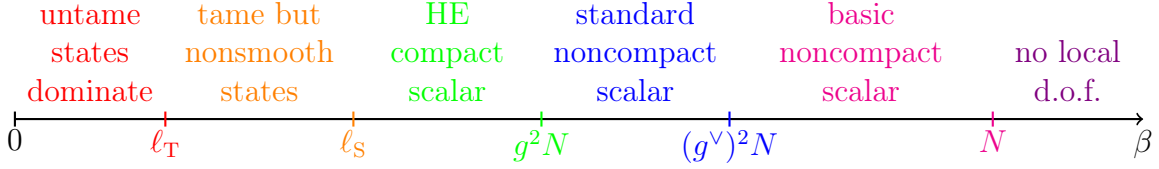
\begin{figure}
\begin{center}
\begin{tikzpicture}[scale = 1]
  \contourlength{1pt}


  \draw[->, thick] (0, 0) -- (15, 0);
  \draw (15, 0) node[below] {$\beta$};
  \draw[thick] (0, -0.1) -- (0, 0.1);
  \draw (0, 0) node[below] {0};

  \draw[thick, red] (2, -0.1) -- (2, 0.1);
  \draw[red] (2, 0) node[below] {$\ell\_T$};
  \draw[red] (1, 0.1) node[above] {\parbox{5em}{\centering untame \\ states \\ dominate}};

  \draw[thick, orange] (4.5, -0.1) -- (4.5, 0.1);
  \draw[orange] (4.5, 0) node[below] {$\ell\_S$};
  \draw[orange] (3.25, 0.1) node[above] {\parbox{5em}{\centering tame but\\ nonsmooth \\ states}};

  \draw[thick, green] (7, -0.1) -- (7, 0.1);
  \draw[green] (7, 0) node[below] {$g^2 N$};
  \draw[green] (5.75, 0.1) node[above] {\parbox{5em}{\centering HE compact \\ scalar}};

  \draw[thick, blue] (9.5, -0.1) -- (9.5, 0.1);
  \draw[blue] (9.5, 0) node[below] {$(g^\vee)^2 N$};
  \draw[blue] (8.25, 0.1) node[above] {\parbox{6em}{\centering standard\\ noncompact \\ scalar}};

  \draw[magenta] (11.25, 0.1) node[above] {\parbox{6em}{\centering basic\\ noncompact \\ scalar}};

  \draw[thick, magenta] (13, -0.1) -- (13, 0.1);
  \draw[magenta] (13, 0) node[below] {$N$};
  \draw[violet] (14, 0.1) node[above] {\parbox{5em}{\centering no local\\ d.o.f.}};


\end{tikzpicture}
\end{center}
\caption{\small Which states contribute nontrivially to the partition function as the inverse temperature is varied? This is the schematic answer for couplings near the self-dual point, $g , g^\vee \sim g_\star = \sqrt{2\pi/K}$, with $1/\sqrt{k\_S} \ll g < g^\vee \ll 1$. The microscopic partition function has at least six distinct phases. \\
\hspace*{1ex} Increasing temperature corresponds to moving from right to left. As each colored point is crossed, new degrees of freedom must be included in the continuum path integral in order for it to reproduce the correct result (if it can ever do so). The two most familiar continuum path integrals arise when $N/\sqrt K \ll \beta \ll N$ and when $\ell\_S \ll \beta \sim N/\sqrt K$. The former is the basic noncompact scalar, for which computing $\Zf$ only involves Gaussian integrals at nonzero momenta. The latter is the HE compact scalar, where computing $\Zf$ involves summing winding and momentum zero-modes in addition to doing Gaussian integrals. \\
\hspace*{1ex} As $\beta$ is dialed below the ``string scale'' $\ell\_S$, spatially nonsmooth states become important. This is the origin of the roughening transition. Here it is assumed that the coupling is chosen such that a smoothing scale exists. For any $g$, there will be a minimal possible value of $\ell\_S(g)$, and its lower bound is the scale $\ell\_T(g)$ at which low-energy states can no longer be described as tame w.r.t.\ any taming parameters $1 \ll n\_T \ll p\_S \ll K$.}
\label{fig energies}
\end{figure}

\newpage

\subsection{The basic noncompact scalar} \label{subsec path int bnc}

The goal of this Subsection is to explicitly construct the continuum path integral $\Zf\_{bnc}$ associated to the basic noncompact scalar cQFT. For simplicity, let the temperature be low enough so that tame zero-modes can be ignored and only the $k \neq 0$ tame/smooth excitations matter,
\bel{\label{beta bound noncomp path int}
  \beta \gg g^2N, \ (g^\vee)^2 N, \ \frac N{k\_S}.
}
As explained in Subsection \ref{subsec path int prelims}, this ensures that $\frac{p\_S}{n\_T} \Zf\_{bnc}$ is a good approximation to $\Zf$.

The states $\qvec{\greek f}$ that will be used in decompositions \eqref{decomposition of unity} will be chosen to be nontrivial eigenvectors of the smooth position operators $\widehat\varphi(x)$ at a set $\Mbb\_S$ of $2k\_S$ equidistant points $x$, as shown in eq.\ \eqref{def xi}.\footnote{Of course, since the free noncompact scalar is solvable, one can choose to use the set of states $\qvec{\{n_k\}_{k \in \Pbb\_S \backslash \{0\}}}$ which exactly diagonalize the Hamiltonian. But the goal here is to pretend that the energy eigenstates are not available, and to develop the path integrals in terms of original/position space variables. These path integrals will have straightforward generalizations to local interacting theories.} It may be instructive to describe this space of states in some detail.

In the clock eigenbasis, the Hilbert space of the microscopic clock model is spanned by states
\bel{
  \qvec{\e^{\i\phi}} = \bigotimes_{x \in \Mbb} \qvec{\e^{\i\phi_x}}.
}
In a trivial taming background, the tame clock eigenstates are
\bel{
  \qvec{\e^{\i\varphi}} = \bigotimes_{x \in \Mbb} \qvec{\e^{\i\varphi_x}},
}
with each $\qvec{\e^{\i\varphi_x}}$ obtained by smearing the original clock states along target directions, as described in \eqref{def varphi}. Any state $\qvec{\e^{\i\varphi}}$ is an eigenstate of the tame position operators $\widehat\varphi_x$ for all $x \in \Mbb$. Smooth position operators $\widehat\varphi(x)$ are projections of $\widehat\varphi_x$ onto the algebra $\A\_S$ generated by \eqref{def AS}. They can be understood as smearings of microscopic position operators along the original lattice,
\bel{
  \widehat\varphi(x) = \sum_{y \in \Mbb} f_{xy}\^S \widehat\varphi_y, \quad f_{xy}\^S \equiv \frac1N \sum_{k \in \Pbb\_S} \e^{\frac{2\pi\i}N k(x - y)}.
}
\emph{All} states $\qvec{\e^{\i\varphi}}$ are eigenstates of $\widehat\varphi(x)$, but most of them are (approximate) null states. Only states in which the field configuration $\varphi_x$ varies smoothly have nontrivial eigenvalues. For an example of how $\widehat\varphi(x)$ projects out nonsmooth states, consider its action on a field configuration $\varphi = \theta \delta^{(y)}$ with $\theta = n\d\varphi \neq 0$:
\bel{
  \widehat\varphi(x) \qvec{\e^{\i\theta \delta^{(y)}}} = \sum_{z \in \Mbb} f\^S_{xz} \widehat\varphi_z \qvec{\e^{\i\theta \delta^{(y)}}} \approx \sum_{z \in \Mbb} \theta \delta^{(y)}_z f\^S_{xz} \qvec{\e^{\i\theta \delta^{(y)}}} = \theta f_{xy}\^S \qvec{\e^{\i\theta \delta^{(y)}}} = \theta \, O(k\_S/N).
}

Thus it makes sense to define nontrivial eigenstates of $\widehat \varphi(x)$ to be those tame states $\qvec{\e^{\i\varphi}}$ that satisfy the same spatial smoothness constraints \eqref{smooth constr scalars} as the field operators, namely
\bel{\label{smooth constr fields}
  \varphi_{x + 1} = \varphi_x + \hat\del \varphi_x + O\left(k\_S^2/N^2\right).
}
The corresponding states will be denoted $\qvec{\varphi}$, in contrast to the not-necessarily-smooth states $\qvec{\e^{\i\varphi}}$. Strictly speaking, the dimension of the space spanned by the set $\{\qvec\varphi\}$ depends on the choice of the $J_k$'s, but heuristically it can be thought to be $(2n\_T)^{2k\_S}$. (The least smooth states will be suppressed at temperatures of interest, so this sloppiness will not be of immediate concern.) The exponent $2k\_S$ reflects the fact that all linearly independent functions $\varphi_x$ satisfying \eqref{smooth constr fields} are described by specifying the value of the function at $2k\_S$ different points $x$. A natural choice for this set of $2k\_S$ points is the coarser lattice $\Mbb\_S$ from \eqref{def xi}.

States given by
\bel{
  \varphi_x = \theta, \quad x \in \Mbb
}
are certainly smooth, but they are annihilated by the operators $\widehat\varphi(x)$. This follows from the definition \eqref{def phi(x)}, where the zero-mode is excluded from the sum over $k \in \Pbb\_S$.

The continuum path integral is easy to construct now that the structure of the states $\qvec{\greek f}$ has been explicated. It is given by
\bel{\label{def Zf nc}
  \Zf\_{bnc} = \e^{-\beta \E(\greek s_0)} \sum_{\{\varphi\}} \prod_{\tau = \d\tau}^\beta \qmat{\varphi_{\tau + \d\tau}}{\e^{-\d\tau H_0}}{\varphi_\tau},
}
where $\qvec{\varphi_\tau}$ denotes the state $\qvec{\varphi}$ that enters the decomposition of unity inserted at time $\tau$, and $\E(\greek s_0)$ is the energy of the superselection sector that contains the ground state. If $\d\tau$ is small enough, even at the highest energies probed by $H_0$ it will be possible to write
\bel{
  \e^{-\d\tau H_0} \approx \1 - \d\tau H_0.
}
These highest energies are achieved when all the $2k\_S$ modes are excited to energies $\sim n\_T/N$, and so the constraint of interest is
\bel{
  \d\tau \ll \frac N{n\_T k\_S} \sim \frac{\ell\_S}{n\_T}.
}
Since $\d\tau = \beta/N_0$, the bounds \eqref{beta bound noncomp path int} imply that the number of time steps in the path integral must satisfy
\bel{\label{N0 constraints}
  N_0 \gg k\_S, \, g^2 k\_S n\_T, \, (g^\vee)^2 k\_S n\_T.
}
It will always be assumed that $N_0$ is chosen large enough to obey these bounds.  Unlike in QM \cite{Radicevic:1D}, due to \eqref{beta bound noncomp path int} here it is not possible to choose $\beta$ so small that $N_0$ can be arbitrary.

The Hamiltonian \eqref{def H0} can be written as
\bel{
  H_0 = \frac{g^2}2 \sum_{x \in \Mbb} \widehat\pi^2_x + \frac1{2g^2} \sum_{x \in \Mbb} (\del \widehat\varphi)^2_x
}
when acting on tame and spatially smooth states $\qvec\varphi$. When $\d\tau$ is small enough so that \eqref{N0 constraints} is valid, the path integrand is given by a product of matrix elements
\algns{
  &\qmat{\varphi_{\tau + \d\tau}}{\left[\1 - \d\tau H_0 + O \left( (\d\tau)^2 \, \tfrac{n\_T^2}{\ell\_S^2}\right) \right]}{\varphi_\tau} \\
  &\qquad \qquad  \approx \frac1{(2p\_S)^N} \sum_{\{p\}} \e^{\sum_{x \in \Mbb} \left[\i p_x (\varphi_{x, \tau + \d\tau} - \varphi_{x, \tau}) - \big(\frac{g^2}2 p^2_x + \frac1{2g^2} (\varphi_{x + 1, \tau} - \varphi_{x, \tau})^2 \big) \d\tau \right]} \\
  &\qquad \qquad \approx \frac1{(2\pi \alpha^2)^{k\_S}}
  \e^{- \frac1{2g^2} \sum_{x \in \Mbb}
  \big[(\del_\tau \varphi)^2_{x, \tau} + (\del_x \varphi)^2_{x, \tau} \big] \d\tau},
}
where $\alpha^2 \equiv g^2 \d\tau/(\d\varphi)^2$ must \emph{not} be small in order to justify using a Gaussian integral to calculate the sum over target momenta $p$ in the second line. The size of this parameter is not controlled by the relations \eqref{N0 constraints}, and so it should be further assumed that $\d\tau$ is chosen large enough to justify $\alpha^2 \gtrsim 1$. The rather familiar steps of the above derivation have been spelled out in detail in \cite{Radicevic:1D}, and so they will not be repeated here.

Taking the product over $\tau$ as in \eqref{def Zf nc} and setting $\E(\greek s_0) = 0$ now gives
\bel{\label{def Zf nc 2}
  \Zf\_{bnc} \approx \frac 1{(2\pi \alpha^2)^{N_0k\_S}} \sum_{\{\varphi\}} \e^{-S[\varphi]},
}
where the action is
\bel{\label{def S nc}
  S[\varphi] \equiv  \sum_{x \in \Mbb} \sum_{\tau = \d\tau}^\beta \L(\varphi_{x,\tau}, \del_\mu \varphi_{x, \tau}) \, \d\tau  \equiv \frac1{2g^2} \sum_{x \in \Mbb} \sum_{\tau \in \Sbb}  \left((\del_\tau \varphi)^2_{x, \tau} + (\del_x \varphi)^2_{x, \tau} \right)\d\tau.
}
The field configurations $\varphi_{x, \tau}$ that enter the sum \eqref{def S nc} take on a precise set of values:
\begin{itemize}
  \item At each spacetime point $(x, \tau)$, $\varphi_{x, \tau}$ can take $2n\_T$ different values, spaced $\d\varphi$ apart between $-\varphi\_T$ and $\varphi\_T$. This is the consequence of target tameness.
  \item The field configurations are further constrained to vary slowly along the $x$-direction, reflecting spatial smoothness \eqref{smooth constr fields}. Furthermore, spatially constant configurations are excluded. These constraints can be simply imposed by requiring that the Fourier transform $\varphi_{k, \tau}$ only receive contributions from $k \in \Pbb\_S \backslash \{0\}$.
  \item At this point, there are no constraints like \eqref{smooth constr fields} governing smoothness in the temporal direction. These must be introduced by hand in order to rewrite $\Zf\_{bnc}$ in a doable form.
\end{itemize}

\subsection{The high-energy compact scalar}

The partition function $\Zf\_{bnc}$ in \eqref{def Zf nc 2} can approximate the microscopic answer (up to a $p\_S/n\_T$ prefactor) when the temperatures are low enough to justify excluding momentum and winding modes, cf.\ Fig.\ \ref{fig energies}. The goal of this Subsection is to derive the continuum path integral that approximates the microscopic answer at temperatures $\beta \sim g^2 N$ for couplings near the self-dual point, where $g^2 \sim 1/K$. This is the regime in which the low-energy physics is described by the HE compact scalar cQFT, and both winding and momentum modes must be included.

Consider first a continuum path integral that includes only tame momentum zero-modes. This corresponds to the standard noncompact scalar cQFT. The corresponding partition function is
\bel{\label{def Zf snc}
  \Zf\_{snc}
    =
  \Zf\_{bnc}
  \sum_{\pi_0}
    \e^{-\frac{\beta g^2}2 \pi^2_0},
}
where the sum runs over all eigenvalues of $\pi_0$. By \eqref{pi 0 quantization redux}, these are given by integer multiples of $K/\sqrt N$, with the maximal eigenvalue bounded by $k\_S p\_S / \sqrt N$, as follows from smoothness and tameness considerations.

Practically speaking, this is the most convenient expression for $\Zf\_{snc}$. Let $g^2 = R^2 \d\phi/2$ like in \eqref{def radius}. When $R = O(1)$ and $\beta \sim g^2 N \sim N/K$, the sum over $\pi_0$ can be approximated by a Gaussian integral, giving
\bel{
  \Zf\_{snc} \approx \frac1R \sqrt{\frac{2N}{\beta K}} \Zf\_{bnc}.
}
This completely solves the path integral over tame momentum zero-modes.

However, \ae sthetically speaking, a better answer includes the sum over $\pi_0$ into the sum over the other $\varphi$'s that appears in \eqref{def Zf nc 2}. There is a simple way to do this. First, split
\bel{
  \e^{- \frac{\beta g^2}2 \widehat\pi_0^2} = \prod_{\tau = \d\tau}^\beta \e^{- \frac{g^2}2 \widehat\pi_0^2 \d\tau},
}
and then insert a set of $\widehat \pi_0$ eigenstates between each factor. This way the sum over $\pi_0$ in \eqref{def Zf snc} becomes
\bel{
  \sum_{\{\pi_{0, \tau}\}} \prod_{\tau = \d\tau}^\beta \e^{-\frac{g^2}2 \pi_{0, \tau}^2 \d\tau} \delta_{\pi_{0, \tau}, \, \pi_{0, \tau + \d\tau}}.
}
The Kronecker delta can now be implemented by a Lagrange multiplier $\varphi_{0, \tau}$, giving
\bel{
  \frac1{M^{N_0}} \sum_{\{\pi_{0, \tau}, \, \varphi_{0, \tau} \}} \prod_{\tau = \d\tau}^\beta \e^{-\frac{g^2}2 \pi_{0, \tau}^2 \d\tau + \i \varphi_{0, \tau} \del_\tau \pi_{0, \tau} \, \d\tau}.
}

It is important to be careful about the range of $\varphi_{0, \tau}$. At each $\tau$, this variable takes $M$ different values given by integer multiples of
\bel{\label{def d varphi0}
  \d\varphi_0 \equiv \frac{2\pi}M \frac{\sqrt N} K.
}
The number $M$ must be chosen to be much bigger than $k\_S p\_S/K$, the number of different values of $\pi_0$. With this convention, the sum over $\pi_{0, \tau}$ is again approximated by a Gaussian integral, so
\bel{\label{def Zf snc 2}
  \Zf\_{snc}
    \approx
  \Zf\_{bnc}
  \left(\frac{2\pi N}{g^2 K^2 M^2 \d\tau }\right)^{N_0/2}
  \sum_{\{\varphi_{0,\d\tau}\}}
    \e^{
      -\frac1{2g^2}
      \sum_{\tau = \d\tau}^\beta
        (\del_\tau \varphi_0)^2 \d\tau
    }
}
Thus the $\pi_0$ sum precisely contributes a $k = 0$ mode to the action \eqref{def S nc}.

Notice the logic here. The continuum path integral for the standard noncompact scalar was not obtained by starting with a path integral that includes a $k = 0$ mode. Such a construction would not have known that $\d\varphi_0$ should be given by \eqref{def d varphi0}. Instead, this special discretization of $\varphi_0$ was derived using Hamilotonian methods, as in Subsection \ref{subsec T dual}, and it was then manually inserted into the foundation of the path integral.

The path integral for the HE compact scalar is obtained after including taming backgrounds \eqref{taming backgrounds} into the sum appearing in $\Zf\_{snc}$. This is not difficult. It is sufficient to replace
\bel{
  \varphi_{x, \tau} \mapsto \varphi_{x, \tau} + \varphi\^{cl}_x
}
in the action \eqref{def S nc}. Of course, the sum over appropriate taming backgrounds must be added to the sum over tame field configurations $\varphi_{x, \tau}$ appearing in \eqref{def Zf nc 2}. The resulting partition function, $\Zf\_{hec}$, thus includes the sum over tame fields $\varphi_{x, \tau}$, Lagrange multipliers $\varphi_{0, \tau}$, and taming backgrounds $\varphi\^{cl}_x$. When evaluated at appropriate temperatures, $\Zf\_{hec}$ approximates the microscopic action and even correctly includes the $p\_S/n\_T$ prefactor.

Note that the taming backgrounds are all time-independent. This causes a certain asymmetry between space and time. It is easy to envision a path integral in which backgrounds that describe winding along the thermal circle are also included in the sum. In fact, such configurations would have entered the path integral constructed directly from \eqref{def H micro} by summing over all possible fields $\phi_{x, \tau}$.
The problem is that this path integral would also not have had the nice action \eqref{def S nc}, and it would have known nothing about smooth and tame states.

There are thus two conceptually different ways of studying path integrals. The first, advocated in this paper, explicitly uses only tame/smooth configurations and the appropriate backgrounds, at the cost of having less symmetry in the resulting path integral. The second way merely sums over all configurations $\phi_{x, \tau}$. In this context it is a nontrivial dynamical question to prove that the path integral is dominated by tame and smooth configurations.

\newpage

\subsection{Temporal smoothing} \label{subsec temp smoothing}

The usual way to evaluate the path integral \eqref{def S nc} is by performing a Fourier transform in spacetime. Let
\bel{
  \varphi_{x, \tau} \equiv \frac1{\sqrt{N_0N}} \sum_{k \in \Pbb\_S} \sum_{n \in \Fbb} \varphi_{k, n} \, \e^{\frac{2\pi\i}N kx + \i \omega_n \tau},
}
where the Matsubara frequencies are
\bel{
  \omega_n \equiv \frac{2\pi}\beta n, \quad n \in \Fbb \equiv \left\{-\frac{N_0}2, \ldots, \frac{N_0}2 - 1\right\}.
}
Note the asymmetry between the spatial and temporal transforms. The difference is not only notational, with $x$ being an integer while $\tau$ has its own ``lattice spacing'' $\d\tau$. The actually significant difference is that the sum over frequencies goes all the way to $\pm N_0/2$, while the sum over momenta terminates at $\pm k\_S$. As stressed in Subsection \ref{subsec path int bnc}, this is because the integration variables $\varphi_{x, \tau}$ are required to be smooth only along the spatial directions. This asymmetry becomes evident when the action is expressed as
\bel{
  S = \frac1{2g^2} \sum_{k \in \Pbb\_S} \sum_{n \in \Fbb} \varphi_{k, n}\+ \varphi_{k,n} \left[ 4\sin^2\frac{\omega_n \d\tau}2 + \left(\frac{2\pi k}{N}\right)^2 + O\left( \frac{k\_S^4}{N^4} \right) \right].
}

It is tempting to discard high-frequency terms (say, at $|n| > n\_S$) and to approximate $4\sin^2\frac{\omega_n\d\tau}2 \approx \omega_n^2 (\d\tau)^2$, thereby bringing $n$ and $k$ to an equal footing. Indeed, this is part of standard operating procedures in all textbooks. However, \emph{dropping high-frequency terms is not justified}:  this truncation leads to a quantity $\~\Zf\_{bnc}$ that in no way approximates $\Zf\_{bnc}$. This was described in detail in the QM context in \cite{Radicevic:1D}, and the same conclusions hold here. What saves the day is \emph{universality}: in well behaved theories, a multiplicative factor of $\~\Zf\_{bnc}$ --- independent of $N_0$ and $n\_S$ --- will agree with the corresponding part of $\Zf\_{bnc}$, up to a choice of finite \emph{counterterms} that are included by fiat when the high-frequency modes are dropped.\footnote{Very roughly, theories are ``well behaved'' if their continuum couplings are $O(1)$ when compared to scales set by all the cutoffs. A precise lattice formulation of universality criteria, and the associated continuum notions of ``renormalizability'' and ``naturalness,'' will not be explored in this paper.}

Dropping the sum over high frequencies amounts to restricting the frequency space $\Fbb$ to a smaller set
\bel{
  \Fbb\_S \equiv \{-n\_S, \ldots, n\_S - 1\}.
}
This defines a projection of position space fields that will be called \emph{temporal smoothing}:
\bel{
  \varphi_{x, \tau} \mapsto \varphi(x, \tau) \equiv  \frac1{\sqrt{N_0N}} \sum_{k \in \Pbb\_S} \sum_{n \in \Fbb\_S} \varphi_{k, n} \, \e^{\frac{2\pi\i}N kx + \i \omega_n \tau}.
}

The path integral variables $\varphi(x, \tau)$ are thus smooth in both spatial and temporal directions. The temporally smoothed continuum path integral is then defined as
\bel{
  \~\Zf\_{bnc} \equiv \frac{\e^{-\beta \E(\greek s_0)}}{(2\pi \alpha^2)^{N_0k\_S}} \sum_{\{\varphi\}} \e^{- \~S[\varphi] - S\_{ct}[\varphi]},
}
where the truncated action is
\bel{\label{def S nc tilde}
  \~S[\varphi] \equiv \frac1{2g^2} \sum_{(x,\tau) \in \Ebb}  \left[ \big(\del_\tau \varphi(x, \tau)\big)^2 + \big(\del_x \varphi(x, \tau)\big)^2 \right]\d\tau,
}
with the spacetime lattice
\bel{
  \Ebb \equiv \Mbb\times \Sbb.
}
Here $S\_{ct}[\varphi]$ is a functional of fields $\varphi(x, \tau)$ that contains all the counterterms --- terms inserted by hand in order to make the universal part of $\~\Zf\_{bnc}$ match the universal part of $\Zf\_{bnc}$.

As in Subsection \ref{subsec dimensionful}, it is possible to rewrite this action using rescaled, dimensionful variables. The Euclidean time coordinate must also be rescaled, so define
\bel{
  \tau\^c \equiv a \tau , \quad \d\tau\^c \equiv a \d\tau .
}
Taking into account that $\Delta_\varphi = \Delta_g = 0$, the continuum presentation of the action is
\algns{\label{def S nc cont}
  \~S[\varphi\_c]
  &= \frac1{2g\_c^2} \sum_{(x,\tau) \in \Ebb}  a \d\tau\^c \left[ \big(\del_{\tau\^c} \varphi_c(x\^c, \tau\^c)\big)^2 + \big(\del_{x\^c} \varphi_c(x\^c, \tau\^c)\big)^2 \right] \\
  &\approx \frac1{2g\_c^2} \int_0^L \d x\^c \int_0^{\beta\^c} \d\tau\^c \left[ \big(\del_{\tau\^c} \varphi\_c\big)^2 + \big(\del_{x\^c} \varphi\_c\big)^2 \right].
}
Finally, after the boson radius is introduced as in \eqref{def radius}, the action becomes
\bel{\label{def S nc cont R}
  \~S[\varphi_R] \approx \frac1{\d\phi} \int_0^L \d x\^c \int_0^{\beta\^c} \d\tau\^c \left[ \big(\del_{\tau\^c} \varphi_R\big)^2 + \big(\del_{x\^c} \varphi_R\big)^2 \right].
}
This rewriting makes it clear that $\d\phi$ is precisely the Planck constant $\hbar$.

Actions like \eqref{def S nc tilde} are the starting point of many continuum analyses. While they may look simple, they are an extremely long way away from a microscopically defined theory like \eqref{def H micro} (around thirty pages in the current font). A major aim of this paper is to stress that actions like $\~S[\varphi_R]$, far from being fundamental objects, in reality conceal a rich pattern of ideas that is needed to derive them as effective descriptions of a finite theory. Their validity necessarily fails if $R$ is dialed to become $O(\hbar)$ or $O(1/\hbar)$, or if $\beta\^c$ violates any of the (dimensionful versions of) bounds \eqref{beta bound noncomp path int} --- and each of these failure modes is different.


\newpage

\section{Continuum path integrals for fermions} \label{sec ferm path int}

Fermionic cQFTs in $(1+1)$D were constructed using the canonical formalism in \cite{Radicevic:2019jfe, Radicevic:2019mle}. The goal of this Section is to derive continuum path integrals for one of these theories, the free Dirac fermion, in analogy with the scalar integrals from Section \ref{sec bos path int}. Fermionic cQFTs are simpler than bosonic ones, as they do not require target space taming, and the corresponding path integrals will be used to study symmetries of smooth actions.

The Hamiltonian of the free Dirac fermion on a lattice was given in eq.\ \eqref{def H Dirac},
\bel{
  H = \i \sum_{v = 1}^{2N} \left(\psi_v\+ \psi_{v + 1} - \psi_{v + 1}\+ \psi_v  \right) = 2 \sum_{k = -N}^{N - 1} n_k \, \sin\frac{\pi k}N.
}
The dispersion relation has two nodes, at $k = 0$ and $k = N$, as discussed in Subsection \ref{subsec chiral}. The modes in the vicinity of each node have different chiralities. Modes of opposite chiralities can be collected into a two-component object, a Dirac spinor
\bel{
  \Psi_k = (\Psi_k^+, \Psi^-_k) \equiv (\psi_k, \psi_{k + N}), \quad k \in \Pbb \equiv \left\{-\frac N2, \ldots, \frac N2 - 1\right\}.
}
The corresponding particle numbers are $n_k^\alpha \equiv (\Psi^\alpha_k)\+ \Psi^\alpha_k$, and the Hamiltonian is
\bel{\label{def Luttinger H}
  H = \sum_{k \in \Pbb} \left(n_k^+ - n_k^-\right) \omega_k, \quad \omega_k \equiv  2\sin\frac{\pi k}N.
}

As before, the goal is to derive a path integral expression for
\bel{
  \Zf = \Tr\, \e^{-\beta H}.
}
The path integral philosophy differs from the scalar case. Instead of inserting an undercomplete set of position eigenstates at each time step $\tau \in \Sbb = \{\d\tau, 2\d\tau, \ldots, \beta\}$, here it is straightforward to split $\Zf$ into a product of partition functions at individual momenta $k \in \Pbb$, and then to write a Berezin integral for each mode. This was explained in \cite{Radicevic:1D} for a single mode, and the generalization to many modes is trivial. This gives
\gathl{\label{def Zf ferm UV}
  \Zf
  = \int [\d\eta^+ \d\bar\eta^+  \d\eta^- \d\bar\eta^-]\, \e^{-S\_{UV}}, \\
  S\_{UV}
  \equiv \sum_{\tau \in \Sbb} \sum_{k \in \Pbb} \d\tau \left[ - \left(\bar \eta^+_{k, \tau} (\del_\tau \eta^+_k)_{\tau - \d\tau} + \bar \eta^-_{k, \tau} (\del_\tau\eta^-_k)_{\tau - \d\tau}\right) +  \omega_k \left(\bar \eta^+_{k, \tau} \eta^+_{k,\tau} - \bar \eta^-_{k, \tau} \eta^-_{k, \tau}\right) \right].
}
Each $\eta^\alpha_{k, \tau}$ and $\bar\eta^\alpha_{k, \tau}$ is an independent Grassmann variable. Neither complex nor Hermitian conjugation relates $\eta$'s to $\bar\eta$'s!

To Fourier-transform along the thermal direction, define
\bel{
  \eta_{k, \tau} \equiv \frac1{\sqrt{N_0}} \sum_{n \in \Fbb} \eta_{k, n}\, \e^{\i \omega_n \tau}, \quad
  \bar \eta_{k, \tau} \equiv \frac1{\sqrt{N_0}} \sum_{n \in \Fbb} \bar \eta_{k, n}\, \e^{-\i\omega_n \tau},
}
with the frequency space $\Fbb = \left\{-\frac{N_0}2, \ldots, \frac{N_0}2 - 1 \right\}$ and fermionic Matsubara frequencies
\bel{
  \omega_n \equiv \frac{2\pi}\beta \left(n + \frac12 \right).
}
(Note that $\omega_n$ and $\omega_k$ are different functions!) This results in the action
\bel{\label{def SE ferm UV}
  S\_{UV} = \sum_{n \in \Fbb} \sum_{k \in \Pbb} \left[ \bar\eta^+_{k, n} \eta^+_{k, n} \left(\e^{-\i \omega_n \d\tau} - 1 + \omega_k \d\tau \right) + \bar\eta^-_{k, n} \eta^-_{k, n} \left(\e^{-\i \omega_n \d\tau} - 1 - \omega_k \d\tau \right) \right].
}

Once again, this formula awakens a strong urge to expand the exponentials and put $k$ and $n$ on an equal footing. This can be done by restricting to modes with $n \in \Fbb\_S \equiv \{-n\_S, \ldots, n\_S - 1\}$ for $1 \ll n\_S \ll N_0$. The action for these low-frequency modes is
\bel{\label{def S UV tilde ferm}
  \~S\_{UV} \approx \sum_{n \in \Fbb\_S} \sum_{k \in \Pbb} \d\tau  \left[ \bar\eta^+_{k, n} \eta^+_{k, n} \left(- \i \omega_n + \omega_k \right) + \bar\eta^-_{k, n} \eta^-_{k, n} \left(-\i \omega_n - \omega_k \right) \right].
}

Subsection \ref{subsec temp smoothing} has stressed that restricting to low Matsubara frequencies is not justified. The cutoff $n\_S$ is unphysical, and
\bel{
  \int [\d\eta^+ \d\bar\eta^+ \d\eta^- \d\bar\eta^-]\, \e^{-\~S\_{UV}}
}
is not even approximately equal to $\Zf$. As in the scalar case, one must include counterterms $S\_{ct}$ in order to ensure that universal ($N_0$- and $n\_S$-independent) parts of the above expression match those of $\Zf$. These counterterms were explicitly computed for a single (spinless) fermionic mode in \cite{Radicevic:1D}, where it was shown that the only counterterm needed was a constant whose finite part was proportional to $\omega_k$. When the path integral includes the same number of modes of either chirality, the needed finite counterterms will cancel out, so $S\_{ct} = 0$ holds.

The restriction to low frequencies implements temporal smoothing. Let
\bel{
  \eta_k^\alpha(\tau) \equiv \frac1{\sqrt{N_0}} \sum_{n \in \Fbb\_S} \eta_{k,n}^\alpha \, \e^{\i\omega_n \tau}, \quad  \bar \eta_k^\alpha(\tau) \equiv \frac1{\sqrt{N_0}} \sum_{n \in \Fbb\_S} \bar \eta_{k,n}^\alpha \, \e^{-\i\omega_n \tau}.
}
These variables satisfy the usual kind of smoothness relation,
\bel{
  \eta_k^\alpha(\tau + \d\tau) = \eta_k^\alpha(\tau) + \d\tau\, \hat\del_\tau \eta_k^\alpha(\tau) + O\left(n\_S^2/N_0^2 \right).
}

While the action $\~S\_{UV}$ treats the frequencies $\omega_n$ and $\omega_k$ on the same footing, the latter is still a nonlinear function of $k$. This disparity is fixed by spatial smoothing, which \emph{does} have a canonical formulation \cite{Radicevic:2019jfe}. In the path integral language, the idea is analogous to how spatial smoothing was implemented for scalars: all modes at momenta $k \notin \Pbb\_S$ are ``frozen'' --- their occupation numbers $n_k^\alpha$ are treated as fixed superselection sector labels, and their contribution to the Hamiltonian is resummed into a quantity $\E(\greek s_0)$ and taken outside the continuum path integral. This leads to the final continuum path integral expression,
\bel{\label{def Zf ferm}
  \~\Zf \equiv \e^{-\beta \E(\greek s_0)} \int [\d\eta^+ \d\bar\eta^+ \d\eta^- \d\bar\eta^-]\, \e^{-\~S - S\_{ct}},
}
where the temporally and spatially smoothed action is
\algns{\label{def SE ferm smooth}
  \~S
  &= \sum_{n \in \Fbb\_S} \sum_{k \in \Pbb\_S} \d\tau  \left[ \bar\eta^+_{k, n} \eta^+_{k, n} \left(- \i \omega_n + \omega_k \right) + \bar\eta^-_{k, n} \eta^-_{k, n} \left(-\i \omega_n - \omega_k \right) \right] \\
  &= - \sum_{\tau \in \Sbb} \sum_{x \in \Mbb^\star} \d\tau \left[\bar \eta^+(x,\tau) \left(\del_\tau  + \i \del_x\right) \eta^+(x, \tau) + \bar \eta^-(x, \tau) \left(\del_\tau  - \i \del_x\right) \eta^-(x, \tau) \right],
}
with
\bel{
  \omega_k \approx \frac{2\pi}N k \ll 1.
}
Here the spatially smoothed Grassmann fields are defined in the now-familiar way,
\bel{
  \eta^\alpha(x, \tau) \equiv \frac1{\sqrt{N}} \sum_{k \in \Pbb\_S} \eta_{k}^\alpha(\tau) \, \e^{\frac{2\pi\i}N k x}, \quad
  \bar \eta^\alpha(x, \tau) \equiv \frac1{\sqrt{N}} \sum_{n \in \Fbb\_S} \bar \eta_{k,n}^\alpha \, \e^{-\frac{2\pi\i}N k x}.
}
Note that the position space $\Mbb^\star$ contains $N$ sites (labeled by $x$), unlike the original space $\Mbb$ which in this case contains $2N$ sites (labeled by $v$). As explained in Subsection \ref{subsec spaces}, this is the simplest example in which the two spaces are not the same. Henceforth, let $\Ebb \equiv \Sbb \times \Mbb^\star$.

The ground state of the Hamiltonian \eqref{def Luttinger H} is fourfold degenerate, with a ``Dirac sea'' structure
\bel{
  \avg{n_k^+} = \theta(-k), \quad \avg{n_k^-} = \theta(k), \quad \trm{for} \quad k \in \Pbb \backslash \{0\}.
}
Freezing out fluctuations outside $\Pbb\_S$ is acceptable only when, roughly speaking (footnote \ref{foot roughening}),
\bel{\label{gap nonsmooth ferm}
  \beta \gg \frac{N}{k\_S}.
}
This is the only constraint on the applicability of the path integral \eqref{def Zf ferm}. The energy contribution from these frozen modes is
\bel{
  \E(\greek s_0) = \sum_{k \notin \Pbb\_S} \big(\theta(-k) - \theta(k)\big) \omega_k = \sum_{k \notin \Pbb\_S} |\omega_k| \approx \frac{4N}\pi.
}

The action \eqref{def SE ferm smooth} already looks familiar. Only a few simple rescalings are needed to bring it into a completely standard continuum form. Before introducing a lattice spacing, it is instructive to rewrite \eqref{def SE ferm smooth} using complex coordinates that are explicitly defined on the lattice $\Ebb$. Let
\gathl{
  z^0 \equiv \tau - \frac12\beta, \quad z^1 \equiv x - \frac12N,\\
  z \equiv z^0 + \i z^1, \quad \bar z \equiv z^0 - \i z^1.
}
One can think of the map $(z^0, z^1) \mapsto (z, \bar z)$  as a coordinate transformation
\bel{
  \bcol z {\bar z} = \bmat 1 \i 1 {-\i} \bcol{z^0}{z^1}.
}
Complex derivatives can also be uniquely specified by their actions on $z$ and $\bar z$,
\gathl{
  \del_z z = 1, \ \del_{\bar z} z = 0, \
  \del_z \bar z = 0, \ \del_{\bar z} \bar z = 1
  \implies \del_z \equiv \frac12 \left(\del_\tau - \i \del_x \right), \ \del_{\bar z} \equiv \frac12 \left(\del_\tau + \i \del_x \right).
}

The path integral variables in \eqref{def SE ferm smooth} can be written as $\eta^\alpha(z, \bar z)$.  The smoothness constraints can be expressed as Taylor expansions for parameters $|\eps^0| \ll N_0/n\_S$, $|\eps^1| \ll N/k\_S$,
\bel{\label{smooth constr complex ferm}
  \eta^\alpha(z + \eps, \bar z + \bar \eps) \approx \eta^\alpha(z, \bar z) + \eps \, \del_z \eta^\alpha(z, \bar z) + \bar\eps \, \del_{\bar z} \eta^\alpha(z, \bar z) \equiv \eta^\alpha(z, \bar z) + \eps^\mu \, \del_\mu \eta^\alpha(z, \bar z),
}
where $\eps \equiv \eps^z \equiv \eps^0 + \i \eps^1$ and $\bar \eps \equiv \eps^{\bar z} \equiv \eps^0 - \i \eps^1$. Formally, one can define the derivative as $\del_z \eta^\alpha(z, \bar z) = \frac1\eps [\eta^\alpha(z + \eps, \bar z) - \eta^\alpha(z, \bar z)]$ if $\eta^\alpha(z, \bar z)$ is holomorphic, $\del_{\bar z} \eta^\alpha(z, \bar z) = 0$. However, there is a snag here. It is not meaningful to think of $\eta^\alpha(z, \bar z)$ as a complex function, holomorphic or otherwise. As explained in \cite{Radicevic:1D}, each $\eta^\alpha(z, \bar z)$ can be thought of as a specific operator acting on an auxiliary fermionic system. As such, the Berezin integral does \emph{not} sum over different values of $\eta^\alpha(z, \bar z)$ at all possible points $z$; it is better thought of as a trace over a large matrix given by products of various fixed matrices $\eta_{k, n}^\alpha$. It is therefore also impossible to isolate a specific set of ``functions'' $\eta^\alpha(z, \bar z)$ that are holomorphic.\footnote{Said another way, there is no way to pick out a subset of ``holomorphic Grassmann numbers'' out of the set $\{\eta^\alpha_{k, n}\}$ for a fixed chirality $\alpha$ and $(k, n) \in \Pbb\_S \times \Fbb\_S$. Said in yet another way, the constraints \eqref{smooth constr complex ferm} cannot be further restricted to ``holomorphic'' constraints; they merely reflect the fact that not all $\eta^\alpha(z, \bar z)$'s are linearly independent.} The notion of holomorphy in a fermionic path integral \emph{only} makes sense in a much weaker context, as a statement about the vanishing of certain correlation functions involving e.g.\ $\del_{\bar z} \eta^z(z, \bar z)$.

To understand the sense in which Grassmann fields are holomorphic, the path integral variables $\eta^\alpha(z, \bar z)$ should be contrasted with canonical operators $\Psi^\alpha(x, \tau)$. These are defined without any reference to smoothing along the thermal circle, namely
\bel{
  \Psi^\alpha(x, \tau) \equiv \e^{-H \tau} \Psi^\alpha(x) \e^{H\tau}.
}

In the free theory, the operators $\Psi^\alpha(x, \tau)$ obey the equation of motion
\bel{
  \left(\del_\tau + \i \alpha \del_x \right) \Psi^\alpha(x, \tau) = 0.
}
In complex notation, this takes the simple form
\bel{\label{eom fermions}
  \del_{\bar z} \Psi^+(z, \bar z) = 0, \quad \del_z \Psi^-(z, \bar z) = 0.
}
This is why the chiral components of the Dirac spinor can be called holomorphic and antiholomorphic. This can be specially denoted by writing these fields as $\Psi^+(z)$ and $\Psi^-(\bar z)$.

Now consider a time-ordered thermal correlation function of multiple operators $\Psi^{\alpha_i}(z_i, \bar z_i)$. Due to important technicalities involving contact terms, discussed in detail in \cite{Radicevic:1D}, such a correlation function can be calculated by inserting Grassmann variables $\eta^{\alpha_i}(z_i, \bar z_i)$ into the path integral \eqref{def SE ferm smooth}, \emph{provided that all the insertion points $z_i$ are over a smearing length apart}. Under these conditions, taking a derivative of the correlation functions and applying the equation of motion \eqref{eom fermions} gives
\bel{\label{fermion holomorphy}
  \avg{\del_{\bar z_i} \eta^+(z_i, \bar z_i) \cdots}_\beta = 0, \quad  \avg{\del_{z_i} \eta^-(z_i, \bar z_i) \cdots}_\beta = 0,
}
regardless of the choice of operators inserted in place of the ellipsis. It is only in this sense that these path integral variables are holomorphic.

After this aside on holomorphy, it is time to return to massaging the action \eqref{def SE ferm smooth} into a standard continuum form. In complex coordinates, this action is
\bel{\label{def SE ferm complex}
  \~S = - 2 \sum_{z \in \Ebb} \d\tau \left[\bar \eta^+(z, \bar z) \, \del_{\bar z} \eta^+(z, \bar z) + \bar \eta^-(z, \bar z)\, \del_z \eta^-(z, \bar z) \right].
}
Further, define the two-component objects
\bel{
  \eta \equiv \bcol{\eta^+}{\eta^-}, \quad \bar\eta \equiv \bcol{\bar\eta^+}{\bar\eta^-},
}
and then, using $\gamma^0 \equiv \sigma^x = \bmat0110$ and $\gamma^1 \equiv \sigma^y = \bmat0{-\i}\i0$, write the action as
\bel{\label{def gamma z}
  \~S = -\sum_{z \in \Ebb} \d\tau \, \bar \eta^\alpha (\gamma^0\gamma^\mu)_{\alpha\beta} \del_\mu \eta^\beta, \quad  \gamma^{\bar z} \equiv \gamma^0 - \i \gamma^1 = \bmat0020, \ \gamma^z \equiv \gamma^0 + \i \gamma^1 = \bmat0200.
}
(As in \eqref{smooth constr complex ferm}, repeated indices $\mu$ and $\alpha$ are summed, with $\mu \in \{z, \bar z\}$. It should always be clear from the context whether $\beta$ is the inverse temperature or a spinor index.)

The Lagrangian can be written telegraphically, with \verb+\overline+s instead of \verb+\bar+s denoting the inclusion of $\gamma^0$ to the right of $\bar \eta$, and with $\gamma^\mu \del_\mu \equiv \c\del$:
\bel{\label{def L tilde}
  \~\L \equiv - \bar \eta^\alpha (\gamma^0\gamma^\mu)_{\alpha\beta} \del_\mu \eta^\beta \equiv - \overline{\eta}\, \c\del \eta.
}
The minus sign comes from the particular convention chosen when defining the Berezin integrals. Apart from this sign, the above result has a completely familiar Lagrangian form.  Note that the points $z$ at which the Grassmann fields are defined are still explicitly sites of a lattice. There is no fermion doubling ``problem'' here because the component fields $\eta^\alpha(z, \bar z)$ were \emph{defined} to each have a single node in the momentum-frequency space $\Pbb\_S \times \Fbb\_S$.

The final --- and completely cosmetic --- step is to define the lattice spacing \eqref{def a} and all the attendant dimensionful quantities.  The dimensionful Grassmann fields are
\bel{
  \eta\_c(z\^c, \bar z\^c) \equiv a^{-\Delta_\eta} \eta(z, \bar z), \quad \bar\eta\_c(z\^c, \bar z\^c) \equiv a^{-\Delta_{\bar\eta}} \bar\eta(z, \bar z).
}
In order to conform to the correspondence between canonical operators and Grassmann variables, one should require $\Delta_\eta = \Delta_{\bar \eta}$. Then the condition that the action has no explicit dependence on $a$ translates into
\bel{
  \Delta_\eta = \frac12.
}
The continuum action is then simply
\bel{\label{def S cont ferm}
  \~S = - \sum_{z \in \Ebb} a\, \d\tau\^c\  \overline{\eta}\_c\, \c\del\_c \eta\_c \approx - \int_0^L \d x\^c \int_0^{\beta\^c} \d\tau\^c \ \overline{\eta}\_c\, \c\del\_c \eta\_c.
}

The condition for the validity of the continuum path integral, \eqref{gap nonsmooth ferm}, now becomes
\bel{\label{gap nonsmooth ferm cont}
  \beta\^c \gg \frac L{2k\_S} = \ell\^c\_S.
}
As described below Fig.\ \ref{fig energies}, if the temperature is so high that this bound is violated, $\~\Zf$ will further deviate from the correct answer because fluctuations at $k\_c \gtrsim 1/\ell\_S\^c$ are not taken into account. It is possible to deepen one's reliance on universality and to plough ahead and compute $\~\Zf$ even at $\beta\^c \ll \ell\^c\_S$: in that situation, further counterterms may be needed, and $\~\Zf$ will only reproduce the spatially universal ($N$- and $k\_S$-independent) parts of $\Zf$. This situation has its appeal, as it treats spatial and temporal universality equitably. In particular, it is then possible to state that universal terms are those that do not depend on $a$ or any other dimensionful energy scale except for $L$ and $\beta\^c$. The downside of this approach, which is standard throughout the literature, is that it obfuscates the existence of multiple cutoff scales, like $n\_S$ and $k\_S$, or $p\_S$ and $n\_T$ in the scalar case.

\newpage

\section{Symmetries of smooth actions} \label{sec symmetries}

\subsection{Quantum symmetries}

Symmetries of the Hamiltonian, which can also be called \emph{quantum symmetries}, are manifested as invariances of the action under \emph{time-independent} changes in the fields. Examples include time translations,
\bel{\label{transf time transl}
  \eta^\alpha(x, \tau) \mapsto \eta^\alpha(x, \tau + v^0), \quad \bar\eta^\alpha(x, \tau) \mapsto \bar\eta^\alpha(x, \tau + v^0),
}
spatial translations,
\bel{\label{transf space transl}
  \eta^\alpha(x, \tau) \mapsto \eta^\alpha(x + v^1, \tau), \quad \bar\eta^\alpha(x, \tau) \mapsto \bar\eta^\alpha(x + v^1, \tau),
}
phase rotations,
\bel{\label{transf phase}
  \eta^\alpha(x, \tau) \mapsto \e^{\i\theta} \eta^\alpha(x, \tau), \quad \bar\eta^\alpha (x, \tau) \mapsto \e^{-\i\theta} \bar\eta^\alpha (x, \tau),
}
and axial transformations,
\bel{\label{transf axial}
  \eta^\alpha(x, \tau) \mapsto \e^{\i\alpha\vartheta} \eta^\alpha(x, \tau), \quad \bar\eta^\alpha (x, \tau) \mapsto \e^{-\i\alpha \vartheta} \bar\eta^\alpha (x, \tau).
}
It is easy to check that none of these formal substitutions change the action \eqref{def SE ferm smooth}. The corresponding parameters $v^\mu$, $\theta$, and $\vartheta$ are arbitrary real numbers. In particular, even though $\eta^\alpha(x, \tau)$ is defined only at lattice sites $(x, \tau) \in \Ebb$, the smoothness relations \eqref{smooth constr complex ferm} and the periodicity requirements $\eta^\alpha(x + N, \tau) = - \eta^\alpha(x, \tau + \beta) = \eta^\alpha(x, \tau)$  allow this definition to be extended to arbitrary $(x, \tau) \equiv z \in \C$.

The canonical operators that generate these symmetries are, respectively, the Hamiltonian
\bel{
  H = \sum_{k \in \Pbb} \left(n_k^+ + n_k^- \right) 2\sin \frac{\pi k}N,
}
the spatial momentum
\bel{
  P \equiv \sum_{k \in \Pbb} \left(n_k^+ + n_k^- \right) \frac{2\pi k}N,
}
the fermion number
\bel{
  N\^F \equiv \sum_{k \in \Pbb} \left(n_k^+ + n_k^- \right),
}
and the axial number
\bel{
  N\^A \equiv \sum_{k \in \Pbb} \left(n_k^+ - n_k^- \right).
}
The four symmetries analogous to \eqref{transf time transl}--\eqref{transf axial} are effected by respectively conjugating $\Psi^\alpha(x, \tau)$ with the operators
\bel{\label{def symmetry ops}
  \e^{v^0 H}, \quad \e^{\i v^1 P}, \quad \e^{\i \theta N\^F}, \quad \e^{\i \vartheta N\^A},
}
for example
\bel{
  \e^{-v^0 H} \Psi^\alpha(x, \tau) \e^{v^0 H}  = \Psi^\alpha(x, \tau + v^0).
}
The Boltzmann factor $\e^{-\beta H}$ is necessarily invariant under such conjugations, and this in turn means that the action must be invariant under the appropriate transformations of the corresponding path integral variables $\eta^\alpha(x, \tau)$.

The parameters of the symmetry transformations \eqref{def symmetry ops} can be arbitrary complex numbers, or they can be restricted to a discrete set of values. It is most meaningful to choose their range so that all resulting operators \eqref{def symmetry ops} are linearly independent. For example, as $N\^F$ only has $2N + 1$ different eigenvalues, it is natural to take $\theta$ to be an integer multiple of $2\pi/(2N + 1)$. Defining symmetry transformations with a finer-grained parameter, for example $\theta \in [0, 2\pi)$ --- or, equivalently, viewing the symmetry transformations as forming a group of order greater than $2N + 1$, for example U(1) --- leads to subtleties that fall under the heading of \emph{geometric anomalies} \cite{Radicevic:2018zsg}. In the example at hand, these are obstructions to gauging the U(1) fermion number symmetry that are detected by the size of the underlying spatial lattice. This issue will be faced in \cite{Radicevic:3D}, but it is for this reason that this paper generally avoids talking about ``symmetry groups'' and instead focuses just on their generators, like $H$ or $N\^F$.

It is also possible to define ``smooth'' symmetry generators in which the sum over momenta is restricted to $\Pbb\_S$. These will be denoted $H\_S$, $P\_S$, $N\^F\_S$, and $N\^A\_S$. They are all exact symmetries of the theory, and their actions on smooth fields are the same as those of their ``original'' versions. In particular, the smooth Hamiltonian is
\bel{
  H\_S \approx \sum_{k \in \Pbb\_S} \left(n_k^+ - n_k^- \right) \frac{2\pi k}N,
}
and as such it differs from the smooth momentum $P\_S$ only by the sign with which the negative-chirality modes enter.

This similarity between $H\_S$ and $P\_S$ motivates the definition of chiral symmetry generators
\bel{\label{def chiral symms}
  N^\alpha \equiv \sum_{k \in \Pbb} n_k^\alpha, \quad P^\alpha \equiv \sum_{k \in \Pbb} \frac{2\pi k}N n_k^\alpha,
}
Their ``smooth'' versions are defined by simply restricting the momentum $k$ to $\Pbb\_S$. This shows that smoothing is crucial in order to define the familiar ``light-cone momenta'' on the lattice, which is done via
\bel{
  P^\pm\_S = \frac12 (P\_S \pm H\_S).
}

The theory has many more quantum symmetries, since each $n_k^\alpha$ is conserved \cite{Kibble:1965}. This makes it reasonable to generalize chiral particle numbers and momenta by defining Hermitian operators
\bel{
  P_{s}^\alpha \equiv \sum_{k \in \Pbb} \left(\frac{2\pi k}N\right)^s n_k^\alpha, \quad 0 \leq s < N.
}
Taking the integer $s$ above $N$ would produce operators $P_s^\alpha$ that are linearly dependent on the lower-$s$ ones. The corresponding symmetry transformations can be implemented by
\bel{
  \e^{\i^s v_s P^\alpha_s}.
}
(No summation over $s$ is implied.) Conjugation by these enacts the transformations
\bel{\label{def hs symms}
  \Psi^\alpha(x, \tau) \mapsto \Psi^\alpha(x, \tau) + v_s\,  \hat\del^s_x \Psi^\alpha(x, \tau) + O\left(k\_S^{2s}/N^{2s}\right).
}
For $s = 1$, this is simply the first term in the expansion \eqref{smooth constr complex ferm}, corresponding to translations. The transformations for $s \geq 2$ are \emph{higher-spin symmetries}. They are broken by interactions that do not preserve the sum of $s$'th powers of the momentum at each interaction vertex.

There remain three discrete quantum symmetries that have not been mentioned so far, even in the Hamiltonian formalism. In the clock model \eqref{def H micro}, \emph{charge conjugation} is generated by an operator $\Csf$ via
\bel{
  \Csf X \Csf = X\+, \quad \Csf Z \Csf = Z\+,
}
or simply
\bel{
  \Csf \qvec{\e^{\i\phi}} = \qvec{\e^{-\i\phi}}.
}
In the clock basis, $\Csf$ is block-diagonal: one block is of size $1\times 1$ and corresponds to the state $\phi = 0$, while the other block has size $(K - 1)\times (K - 1)$ and has unit entries on the antidiagonal. This matrix is not simply expressed in terms of the shift and clock operators.

The operator $\Csf$ satisfies $\Csf^2 = \1$, and hence it generates a $\Z_2$ symmetry. When $K \rar \infty$, charge conjugation and shift symmetry generators together form the group O(2).

Within the path integral of the basic noncompact scalar, in which the smoothed action is \eqref{def S nc tilde}, charge conjugation is manifested by the invariance of the action under
\bel{\label{def C}
  \varphi(x, \tau) \mapsto - \varphi(x, \tau).
}
If sectors with nontrivial taming offsets, winding numbers, and tame momenta are included, $\Csf$ reverses the signs on all of their labels.  In the fermion path integral, one possible $\Csf$ map is equal to a chiral rotation of the $-$ component of the Dirac field by $\theta = \pi$,
\bel{
  \eta^\alpha(z, \bar z) \mapsto \alpha \eta^\alpha(z, \bar z), \quad \bar \eta^\alpha(z, \bar z) \mapsto \alpha \bar \eta^\alpha(z, \bar z).
} 

The basic noncompact scalar Hamiltonian \eqref{def H0} is also evidently invariant under \emph{parity}, which acts as
\bel{
  \Psf c_k \Psf = c_{-k}, \quad \Psf c_k\+ \Psf = c_{-k}\+.
}
or, in position space,
\bel{
  \Psf \varphi_x \Psf = \varphi_{N - x}, \quad \Psf \pi_x \Psf = \pi_{N - x}.
}
Parity is again a $\Z_2$ symmetry. With translations, it forms another O(2) group when $N \gg 1$. In the path integral, it acts as
\bel{
  \varphi(x, \tau) \mapsto \varphi(N - x, \tau) \equiv \varphi(-x, \tau).
}

Parity is a symmetry of the Dirac fermion if it also exchanges chiralities:
\bel{
  \Psf \Psi^\pm_k \Psf = \Psi^\mp_{- k}, \quad \trm{or} \quad \Psf \Psi_k \Psf = \gamma^0 \Psi_{- k}
}
using conventions of \eqref{def gamma z}. In the path integral, the corresponding map is approximately
\bel{
  \eta(x, \tau) \mapsto \gamma^0 \eta (N - x, \tau) \equiv \gamma^0 \eta (-x, \tau), \quad \bar\eta(x, \tau) \mapsto \gamma^0 \bar\eta (- x, \tau).
}
In complex coordinates, the action of $\Psf$ is given by the simple expression
\bel{
  \eta(z, \bar z) \mapsto \gamma^0 \eta (\bar z, z), \quad \bar\eta^\pm(z, \bar z) \mapsto \gamma^0 \bar \eta (\bar z, z).
}

The last entry in this familiar trifecta of $\Z_2$ symmetries is \emph{time reversal}. In the canonical formalism, it is generated by an antiunitary operator $\Tsf$ that involves complex conjugation. As such, it is not an element of the $\C^{D \times D}$ algebra, which was here taken to be the largest operator algebra of interest. Generalizing the present story to operator algebras that include time reversal is one digression that will not be taken in this paper.

However, in the path integral formalism, time reversal is straightforward. For bosons, it acts as
\bel{
  \varphi(x, \tau) \mapsto \varphi(x, -\tau).
}
For fermions, one natural choice for the action of $\Tsf$ is
\bel{
  \eta(x, \tau) \mapsto \i \gamma^1 \eta(x, -\tau), \quad \bar\eta(x, \tau) \mapsto \i \gamma^1 \bar\eta(x, -\tau).
}
With these choices, the combined transformation $\Csf \Psf \Tsf$ just maps each chirality to itself,
\bel{
  \eta(x, \tau) \mapsto \eta(-x, -\tau), \quad \bar\eta(x, \tau) \mapsto \bar\eta(-x, -\tau).
}

\newpage

There exists a weaker notion of a quantum symmetry transformation that deserves a brief mention here. Certain operators that do not commute with a Hamiltonian may still preserve its spectrum. Such operators map energy eigenstates to each other. As such, they are said to generate a \emph{spectral flow}. They can be understood as ``fractional roots'' of a symmetry, as applying them many times ultimately maps each energy eigenstate to itself. A simple example of such an operator arises after introducing a $\theta$-term into the clock model Hamiltonian. A spectral flow is then generated by operators that change the $\theta$ angle by $\d\phi$ \cite{Gaiotto:2017yup}. Much of what is said about symmetries here can be generalized to spectral flows.

\subsection{Engineering scale invariance} \label{subsec eng scale inv}

Before advancing to other familiar examples of action symmetries, a short interlude is in order. \emph{Engineering scale transformations} are exact invariances of continuum actions that are \emph{not}, on their own, meaningful symmetries in any sense. It is important to logically distinguish them from dilatations, which \emph{are} nontrivial action symmetries, as will be discussed in Subsection \ref{subsec dilatations}.

Recall that dimensionful quantities are introduced by defining an arbitrary lattice spacing $a$ and system size $L$, subject to $N = L/a$. This means that the ``continuum'' actions \eqref{def S nc cont} and \eqref{def S cont ferm} must be invariant under
\bel{\label{def eng scale trafo}
  a \mapsto \lambda a, \quad L \mapsto \lambda L,\quad \lambda \in \R \backslash \{0\}.
}
This transformation causes a rescaling of all objects labeled by ``c.'' For example, an operator $\O\_c$ with engineering dimension $\Delta$ transforms as $\O\_c \mapsto \lambda^{-\Delta} \O\_c$.

The map \eqref{def eng scale trafo} is the announced engineering scale transformation.  It is \emph{not} a quantum symmetry. One reason: if $\O$ has nonzero trace, there is no operator $Q_\lambda$ that can act as $Q_\lambda \O Q_\lambda^{-1} = \lambda \O$ for $\lambda \neq 1$. Therefore the engineering scale change for this operator cannot be implemented in a canonical formalism. Taking $\O \propto \1$, this is simply the statement that c-numbers can never transform under quantum symmetries.

Engineering scale invariance is sometimes expressed by stating that the smooth partition function, viewed as a function of dimensionful quantities $L$ and $\beta\^c$, satisfies
\bel{
  \~\Zf(\beta\^c, L) = \~\Zf(\lambda \beta\^c, \lambda L),
}
and only depends on $\beta\^c/L$. Strictly speaking, this is only valid for the universal part of $\~\Zf$. However, if $\beta\^c \gg \ell\_S\^c$, $\~\Zf$ will correctly capture the factors of $N$ and $k\_S$, cf.\ the discussion below \eqref{gap nonsmooth ferm cont}. Still, even in this temperature regime, $\~\Zf$ will not capture the correct dependence on $N_0$ --- and it will depend on $n\_S$, which does not even have a canonical interpretation.

\newpage

\subsection{Rotations}

A remarkable fact about the path integral formalism is that smooth actions can possess nontrivial (typically approximate) symmetries that are invisible from the canonical formalism. As a first example, consider spacetime rotations. A na\"ive definition of a rotation is the map
\bel{\label{naive rotation}
  \eta^\alpha(z, \bar z) \mapsto \eta^\alpha \left(\e^{\i \theta} z, \e^{-\i\theta} \bar z \right), \quad \bar\eta^\alpha(z, \bar z) \mapsto \bar\eta^\alpha \left(\e^{\i \theta} z, \e^{-\i\theta} \bar z \right).
}
For a generic $\theta \in [0, 2\pi)$, $\e^{\i\theta} z$ does not even belong to the spacetime lattice $\Ebb$. Only within a smooth path integral can $\eta^\alpha(z, \bar z)$ be defined for $z \in \C$ by the smoothness condition \eqref{smooth constr complex ferm}.

Let $z' \equiv \e^{\i\theta} z$. The standard, ``continuum'' way to proceed is to write derivatives of the fields as
\bel{\label{temp rotation of derivatives}
  \del_{\bar z} \eta^\alpha\left(z', \bar z' \right) = \e^{-\i \theta} \del_{\bar z'} \eta^\alpha  \left(z', \bar z' \right), \quad \del_{z} \eta^\alpha\left(z', \bar z' \right) = \e^{\i \theta} \del_{z'} \eta^\alpha \left(z', \bar z' \right).
}
Then one changes the variables of integration in the action \eqref{def SE ferm complex}, finds that the action is not invariant under the transformation \eqref{naive rotation}, and notices that the Grassmann variables must pick up an additional phase factor in order to preserve the action. This is a rather traditional way to detect that fermions must transform covariantly under spacetime rotations.

However, this standard procedure must be handled with care on the lattice. Given any map $z \mapsto z'$, to each $z \in \Ebb$ one can associate at least one point $z_\star$ on the same lattice, such that \bel{\label{complex floor function}
  \left|(z')^0 - (z_\star)^0\right| < \d\tau, \quad
  \left|(z')^1 - (z_\star)^1\right| < 1.
}
In other words, $z' \mapsto z_\star$ acts like a two-dimensional ``floor function.'' There may not be a unique choice $z_\star$, but as all fields vary slowly on lattice scales, any choice $z_\star$ will be acceptable.

To simplify notation in the following lattice derivation of \eqref{temp rotation of derivatives}, $\eta(z)$ will be written instead of $\eta^\alpha(z, \bar z)$. The derivatives in the action map under na\"ive rotations as
\bel{\label{temp deriv trafo}
  \del_z \eta(z) = \frac12 \left(\del_\tau - \i \del_x\right) \eta(z) \mapsto \del_z \eta(z'(z)) = \frac{\eta\big((z + \d \tau)'\big) - \eta\big(z'\big)}{2 \d \tau} + \frac{\eta\big((z + \i)'\big) - \eta\big(z'\big)}{2\i}.
}
This is \emph{not} equal to $\frac12(\del_\tau - \i \del_x) \eta(z') = \frac{\eta(z' + \d\tau) - \eta(z')}{2\d\tau} + \frac{\eta(z' + \i) - \eta(z')}{2\i}$. This distinction is crucial! Instead, use the smoothness condition \eqref{smooth constr complex ferm} to write
\gathl{
  \eta\big((z + \d \tau)'\big)
  \approx \eta(z') + \big[(z + \d \tau)' - z'\big]^0 \del_\tau \eta(z') + \big[(z + \d \tau)' - z'\big]^1 \del_x\eta(z'),\\
  \eta\big((z + \i)'\big)
  \approx \eta(z') + \big[(z + \i)' - z'\big]^0 \del_\tau \eta(z') + \big[(z + \i)' - z'\big]^1 \del_x\eta(z').
}
Here it is assumed that $z \mapsto z'$ preserves adjacency, so that e.g.\ $|(z + \d\tau)' - z'|$ is never greater than a smearing length. At leading order in the smoothing parameters, one can further write $\del_\mu \eta(z') \approx \del_\mu \eta(z_\star)$ in the above formul\ae.

Plugging these expansions back into \eqref{temp deriv trafo} gives
\gathl{\label{derivative trafo}
  \del_z \eta\left(z\right) \mapsto \del_z \eta\big(z'(z)\big) \approx \del_z z'\, \del_z \eta(z_\star) + \del_z \bar z'\, \del_{\bar z} \eta(z_\star), \\
  \del_{\bar z} \eta\left(z\right) \mapsto \del_{\bar z} \eta \big(z'(z)\big) \approx \del_{\bar z} z'\, \del_z \eta(z_\star) + \del_{\bar z} \bar z'\, \del_{\bar z} \eta(z_\star).
}
This calculation justifies the naively expected derivative transformation and resolves all attendant ambiguities. To avoid confusion, always remember the difference between $\del_\mu \eta(z')$ and $\del_\mu \eta\big(z'(z)\big)$, and keep in mind that $\del_z \eta(z_\star)$ depends on the Grassmann fields at the lattice points $z_\star$, $z_\star + \d\tau$, and $z_\star + \i$. Meanwhile, an expression like $\del_z z'$ is a c-number given by
\bel{
  \del_z z' = \frac12(\del_\tau - \i \del_x) z' = \frac{(z + \d\tau)' - z'}{2\d\tau} + \frac{(z + \i)' - z'}{2\i}.
}
Finally, $\bar z'$ denotes the complex conjugate of $z'$, and should be distinguished from the transformation $(\bar z)'$ of the point $\bar z$.

Going back to the specific example of rotations, the map of interest is $z' = \e^{\i\theta} z$ and consequently $\bar z' = \e^{-\i\theta} \bar z$. This establishes the transformation property \eqref{temp rotation of derivatives} in a more precise way,
\gathl{\label{rotation of derivatives}
  \del_{\bar z} \eta\big(z'(z) \big) \approx \e^{-\i \theta} \del_{\bar z} \eta  \left(z_\star \right), \\
  \del_{z} \eta\big(z'(z) \big) \approx \e^{\i \theta} \del_z \eta \left(z_\star \right).
}
It is finally possible to conclude that the transformed action \eqref{def SE ferm complex}, e.g.\ for the $+$ chirality, is different from the original one,
\bel{
  - 2 \e^{-\i\theta} \sum_{z \in \Ebb} \d\tau\,  \bar\eta^+(z_\star) \del_{\bar z} \eta^+(z_\star) \neq - 2\sum_{z \in \Ebb} \d\tau \,\bar \eta^+(z) \del_{\bar z} \eta^+(z).
}
(The term $(z' - z_\star)^\mu \del_\mu \overline \eta(z_\star) \c\del \eta(z_\star)$ that comes from the transformation $\bar \eta(z) \mapsto \bar \eta(z')$ is subleading and can be ignored at this precision.) The inequality persists even if it is assumed that the range of $z_\star$ is the same as the range of $z$, a subtlety that will be addressed below.

There is a simple way to fix the na\"ive transformation \eqref{naive rotation} and eliminate the discordant phase $\e^{-\i\theta}$ from the transformed action. Define the rotation transformation to be
\gathl{\label{def rotation}
  \eta^\pm(z, \bar z) \mapsto \e^{\pm\i\theta/2} \eta^\pm\left(\e^{\i\theta} z, \e^{-\i\theta} \bar z\right), \\ \bar\eta^\pm(z, \bar z) \mapsto \e^{\pm\i\theta/2} \bar\eta^\pm\left(\e^{\i\theta} z, \e^{-\i\theta} \bar z\right).
}
With this definition the phases all cancel out in the transformed action. The factor $1/2$ in the phase reflects the spinorial nature of the fields, and in particular it shows that a rotation by $2\pi$ in the complex plane flips the signs of all fermion fields: it acts as fermion parity. Note that, as is familiar from the Lorentzian case, it is $\overline \eta = \bar\eta \gamma^0$ and not $\bar\eta$ that transforms as $\Psi\+$, i.e.~with the opposite phase of the one with which $\eta$ transforms.

As already hinted, however, the above conclusion was coy about the range of the variable $z_\star$ in the transformed action. The lattice $\Ebb$ is rectangular, and hence generically invariant only under rotations by $\pi$. Rotations by other values of $\theta$ can lead to unpleasant situations. For example, assume that $\beta \geq 2N$, and consider the rotation of the point $z = N$ by $\theta = \pi/2$, giving $z' = \i N$. Periodicity in the $x$-direction implies that $z' \equiv 0$, so the conclusion is that this rotation maps $z$ to the origin. This ruins the group structure of rotations, as the origin remains invariant under rotations while two sequential rotations of $z = N$ by $\pi/2$ should rightfully yield $z' = -N$. More seriously, the rotation by $\pi/2$ identifies $\eta(z = 0)$ and $\eta(z = N)$, signifying that there is an ambiguity in what one means by a field $\eta(z = 0)$ after that rotation.

To avoid such issues, rotations should be defined only for a \emph{subset} $\Ebb_R \subset \Ebb$ such that $|z| \leq R$ for $R \leq \trm{min}(N, \beta)$. The other shoe drops immediately: a rotation that only acts on some points in $\Ebb$ does not preserve the notion of adjacency and causes the action to change. This means that rotations are not symmetries of the action \eqref{def SE ferm complex}, and as a result this whole Subsection starts looking a bit pointless. The way to proceed is to again genuflect to universality, and to simply \emph{drop} all fields at $|z| > R$ from the action. The expectation is that doing so (and including appropriate boundary counterterms at $|z| = R$) would result in a partition function, $\~\Zf_R$, whose $R$-independent part matches that of the exact answer $\Zf$.

There is an informal but intuitive way to rephrase the above restriction of the theory to a disk. In continuum notation, instead of talking about $R$-independent parts of the partition function, one can talk about parts that are independent of $L$ and $\beta\^c$. (Recall that parts of the partition function that depend on $a$ and $\d\tau$ were already discarded when agreeing to drop modes at frequencies above $n\_S$ and momenta above $k\_S$ while constructing the smooth continuum path integrals.) This restriction of the path integral to a spacetime disk is the idea implicit in all constructions of a continuum path integral on an infinite spacetime plane, which can be visualized as the limit $\beta\^c, L \rar \infty$.

A more general lesson can be drawn from this discussion. When considering symmetries of an action, the map $z \mapsto z'$ should always be a bijection, so that for any $z \neq w$ one has $z' \neq w'$. Note that this does not preclude the possibility that $z_\star = w_\star$ for some pair of points.

\subsection{Dilatations} \label{subsec dilatations}

Define a \emph{dilatation} of a Grassmann field as the map
\bel{\label{def scale trafo}
  \eta(z, \bar z) \mapsto \lambda^{1/2}\, \eta(\lambda z, \lambda \bar z), \quad \bar\eta(z, \bar z) \mapsto \lambda^{1/2}\, \bar\eta(\lambda z, \lambda \bar z), \quad \lambda \in \R^+.
}
The power of $\lambda$ in front of the $\eta$'s was chosen to equal its engineering dimension. Unlike the engineering scale transformation \eqref{def eng scale trafo}, this map is defined \emph{without} introducing $a$ and $L$.


By the general formula \eqref{derivative trafo}, when $z' = \lambda z$ derivatives map under \eqref{def scale trafo} as
\bel{
  \del_\mu \eta\left(z', \bar z'\right) \approx \lambda\, \del_\mu \eta(z_\star, \bar z_\star).
}
Unlike rotations, however, dilatations do not necessarily preserve adjacency. At sufficiently large $\lambda$, any two neighboring points on $\Ebb$ may end up separated by more than a smoothing length, in which case the ``change of variables'' formula \eqref{derivative trafo} fails. To make sure this does not happen, it is necessary to require that the dilatation parameter is small enough so that
\bel{\label{dilatation max}
  \lambda \ll \frac{N_0}{n\_S}, \ \frac N{k\_S}.
}
Imposing this requirement means that dilatations \emph{cannot} form a group. Nevertheless, they may still define legitimate symmetries.

If $\lambda$ is sufficiently small, it may happen that $z_\star = 0$ for every $z \in \Ebb$. In this case, the action clearly stops being equivalent to the starting one. To avoid collapsing $\Ebb$ into a single smearing length, one must have
\bel{\label{dilatation min}
  \lambda \gg \frac1{n\_S}, \ \frac1{k\_S}.
}

The presence of (anti)periodic boundary conditions complicates the situation and puts dilatations at risk of not being bijective. One way to avoid this is to only perform dilatations that do not shift any point by more than a smoothing scale, e.g.\ via $|\lambda - 1| \ll 1/k\_S, 1/n\_S$. This is extremely restrictive, however.

An alternative is to consider a family of path integrals defined on spacetime disks $\Ebb_R$ at various $R$. This is the choice that leads to the most conventional situation. In this case, the transformed action \eqref{def SE ferm complex} becomes
\bel{
  - \lambda^2 \sum_{z \in \Ebb_R} \d\tau\, \overline \eta(z_\star, \bar z_\star) \c\del \eta(z_\star, \bar z_\star).
}
The sum can be interpreted as going only over those points in $\Ebb_{\lambda R}$ that are images of a point $z \in \Ebb_R$ under the dilatation. To first order in the smoothing parameters, the sum over $z_\star \in \Ebb_R$ can be replaced by a sum over all $z \in \Ebb$, rescaled by a factor of $1/\lambda^2$. (This is just a lattice version of the Jacobian.) In other words, the transformed action can be written as
\bel{
  - \lambda^2 \sum_{z \in \Ebb_R} \d\tau\, \overline \eta(z_\star, \bar z_\star) \c\del \eta(z_\star, \bar z_\star) = - \sum_{z \in \Ebb_{\lambda R}} \d\tau\, \overline \eta(z, \bar z) \c\del \eta(z, \bar z).
}
Dilatations subject to \eqref{dilatation max} and \eqref{dilatation min} thus preserve the action on a spacetime disk, up to rescaling $R \mapsto \lambda R$. In particular, all $R$-independent quantities survive the rescaling \eqref{def scale trafo}.

\newpage

\subsection{Noether currents}

\subsubsection{Noether's theorem}

Consider a general transformation of Grassmann fields
\bel{\label{general symm transf}
  \eta(z, \bar z) \mapsto \eta'(z', \bar z'), \quad \bar \eta(z, \bar z) \mapsto \bar \eta'(z', \bar z').
}
This may be an exact symmetry of the action, like in the case of quantum symmetries, or it may be a symmetry of a universal part of the action, as in the case of spacetime rotations and dilatations. In either case, the existence of the symmetry implies useful constraints on (universal parts of) various correlation functions. \emph{Noether currents} are functions of path integral variables that play a crucial r$\hat{\trm o}$le in the study of these constraints. Their existence, constructively demonstrated by Noether's theorem, is standardly derived for continuous symmetries in a cQFT framework (see e.g.\ \cite{DiFrancesco:1997nk}). In this Subsection, Noether's theorem will be rederived in the finitary setup, taking into account the possibility that the symmetry parameters of interest might not be infinitesimally small.

Noether's theorem is very simple to state in a lattice framework. Instead of applying the map \eqref{general symm transf} at all points $z \in \Ebb$, apply it at a \emph{single} point $z$, and consider the resulting change in the action \eqref{def SE ferm complex},
\bel{
  \~S \mapsto \~S + \delta \~S(z).
}
The proof proceeds by the following three steps:
\begin{enumerate}
  \item Note that the partition function is approximately preserved by this local transformation.
  \item If the transformation is sufficiently close to the identity, one has $\e^{-\delta \~S(z)} \approx 1 - \delta \~S(z)$. The invariance of the partition function then implies that $\avg{\delta \~S(z) \cdots} \approx 0$ for all possible field insertions more than a smoothing length  away from $z$.
  \item The fact that \eqref{general symm transf} is a symmetry guarantees that $\sum_{z \in \Ebb} \delta \~S(z) = 0$, possibly up to nonuniversal corrections. This indicates that it is possible to write $\delta \~S(z) \propto \del_\mu J^\mu(z)$ for some function $J^\mu(z)$ of fields at or near $z$. The previous step then implies that $J^\mu(z)$ is conserved, $\del_\mu \avg{J^\mu(z)\cdots} \approx 0$. The object $J^\mu(z)$ is the Noether current, and its conservation is the most basic constraint on the correlation functions in the theory.
\end{enumerate}

The plan is to apply these steps to the various symmetries discussed in this Section, to derive the corresponding currents, and to precisely understand when a symmetry transformation is ``sufficiently close to the identity'' for Noether's theorem to hold in its conventional sense. The last point is novel, and working it out will explain how Noether's theorem can be generalized to symmetries that are not continuous.

\subsubsection{Phase rotations}

The chiral particle number symmetry, generated e.g.\ by $N^+$ from eqs.\ \eqref{def chiral symms}, provides the simplest setting in which to study the nitty-gritty of Noether's theorem. The corresponding map is
\bel{\label{def chiral phase rotation}
  \eta(z) \mapsto \Lambda\, \eta(z), \quad \bar\eta(z) \mapsto \Lambda^*\, \bar \eta(z),
}
with $\Lambda \equiv \e^{\i\theta}$. For convenience, this is written in ``holomorphic form,'' without $\bar z$ and the chirality label $+$. The Noether trick entails performing \eqref{def chiral phase rotation} at one point $z$ only.

A phase rotation at all $z \in \Ebb$ clearly preserves the action, and hence also the partition function. The fact that a localized phase rotation leaves the partition function approximately invariant is less obvious. In fact, the situation is subtle because such a localized transformation can violate the smoothness constraints \eqref{smooth constr complex ferm}. To precisely state the issue, recall that the variables of the smooth path integral are the Grassmanns $\{\eta_{k, n}^\alpha, \bar\eta_{k, n}^\alpha\}_{k, n, \alpha}$, and the ``measure'' is
\bel{\label{def measure}
  [\d\eta \d\bar\eta] = \prod_{k \in \Pbb\_S} \prod_{n \in \Fbb\_S}  \d\eta_{k, n}^+ \d\bar\eta_{k, n}^+ \d\eta_{k, n}^- \d\bar\eta_{k, n}^-.
}
Although the action is conventionally expressed in terms of position-space fields $\eta^\alpha(z, \bar z)$, this does \emph{not} mean that for each $z \in \Ebb$ the variable $\eta^\alpha(z, \bar z)$ is independent of all the others. 

One rather natural way to proceed is to define the transformed function to be a smoothed phase rotation of the microscopic field $\eta_z$, which is given by a Fourier transform of $\eta_{k, n}$ involving the whole range $\Pbb \times \Fbb$ of momenta and frequencies. This will make sure that the resulting field still obeys the appropriate smoothness conditions.

To make this precise, consider the measure \eqref{def measure} and observe that any mapping
\bel{\label{temp phase rotation}
  \eta_{k, n}^\alpha \mapsto \Lambda \eta_{k, n}^\alpha, \quad \bar\eta_{k, n}^\alpha \mapsto \Lambda^* \bar\eta_{k, n}^\alpha
}
of a single momentum mode will preserve the integral. This is easily seen as follows. The action $\e^{-\~S}$ can be expanded into a Taylor series. The only terms in the expansion that will survive the Berezin integrals contain each Grassmann field exactly once. These terms will have an equal number of $\Lambda$'s and $\Lambda^*$'s after the above map, and hence they will not change.

Another way to state the same thing is that a map $\eta^\alpha_{k, n} \mapsto \Lambda \eta^\alpha_{k, n}$ of any individual mode is equivalent to multiplying the integral by $\Lambda$ while leaving the integrand unchanged. This is the ``change of variables'' point of view that most books advocate. In a similar way, a map
\bel{
  \eta^\alpha_{k,n} \mapsto \sum_{(l, m) \in \Pbb\_S \times \Fbb\_S} U_{k, n}^{l, m} \eta_{l, m}^\alpha, \quad
  \bar\eta^\alpha_{k,n} \mapsto \sum_{(l, m) \in \Pbb\_S \times \Fbb\_S} \left(U_{k, n}^{l, m}\right)^* \bar\eta_{l, m}^\alpha
}
changes the integral by $\det (U\+ U)$, which is unity for any unitary $4n\_S k\_S \times 4n\_S k\_S$ matrix $U$.

The local phase rotation demanded by Noether's theorem is then given by
\bel{\label{def U}
  U_{k,n}^{l,m} = \delta_{nm} \delta_{kl} + \frac{\Lambda - 1}{N_0N} \e^{\i(\omega_m - \omega_n) z^0 + \i (\omega_l - \omega_k) z^1},
}
where $z = z^0 + \i z^1$ is the point at which the phase rotation by $\Lambda$ is enacted. If the range of the momenta and frequencies were the whole space $\Pbb\times \Fbb$, this would have been a unitary transformation that, in position space, acts as $\eta_w \mapsto \eta_w + (\Lambda - 1) \eta_z \delta_{w, z}$. However, upon restricting to $(k, n) \in \Pbb\_S \times \Fbb\_S$, this map becomes nonunitary. In position space, it is given by
\bel{\label{Noether transf}
  \eta(w) \mapsto \eta(w) + (\Lambda - 1) \,\eta(z) \!\! \sum_{(k, n) \in \Pbb\_S \times \Fbb\_S} \frac{\e^{\i \omega_n (w^0 - z^0) + \i \omega_k (w^1 - z^1)}}{N_0 N},
}
where the sum is readily interpreted as a $\delta$-function smeared in both spacetime directions.

If the matrix $U$ deviates from unitarity, so that
\bel{\label{def delta U}
  U\+ U \equiv \e^{-\delta U} \neq \1,
}
then the result of the transformation \eqref{def U} will be to multiply the path integral by $\det \e^{-\delta U}$. If the deviation from unitarity is small, it will be possible to use
\bel{
  \det \, \e^{-\delta U} \approx \det(\1 -\delta U) \approx 1 - \tr(\delta U).
}
This deviation can be calculated explicitly from \eqref{def U} and \eqref{def delta U}, getting
\bel{
  \left(\e^{-\delta U}\right)_{k, n}^{l, m}
  = \delta_{nm} \delta_{kl} + \left(\Lambda + \Lambda^* - 2\right) \left(1 -\frac{4k\_S n\_S}{N_0N}\right) \frac{\e^{\i(\omega_m - \omega_n) z^0 + \i (\omega_l - \omega_k) z^1}}{N_0 N}.
}
The second term is always suppressed by $N_0N$, so $\delta U$ is small for any choice of $\Lambda$, and it can be expressed as
\bel{
  \delta U_{k, n}^{l, m} \approx \frac{4\sin^2(\theta/2)}{N_0 N} \e^{\i(\omega_m - \omega_n) z^0 + \i (\omega_l - \omega_k) z^1}.
}
Its trace is independent of the choice of $z$, and equals
\bel{\label{Noether Zf change}
  \tr(\delta U) = \frac{4 k\_S n\_S}{N_0 N} \ 4\sin^2 \frac\theta2 \equiv \frac{4}{ A\_S} \sin^2\frac\theta 2.
}

The conclusion is that the ``smoothing area'' $A\_S \gg 1$ makes this quantity small for any $\Lambda = \e^{\i\theta}$, and so the partition function is indeed approximately invariant under the local transformation \eqref{def U}. Furthermore, the partition function is exactly invariant if \eqref{def U} is performed at each $z$; this global transformation is implemented by a manifestly unitary matrix $U = \Lambda \1 $. This calculation is valid for either chirality separately.


The second step in Noether's theorem is to calculate the variation in the action under \eqref{Noether transf}, rewritten for simplicity as
\bel{\label{Noether transf simplified}
  \eta(w) \mapsto \eta(w) + \delta\Lambda \, \eta(z) \, f(w - z), \quad \bar\eta(w) \mapsto \bar\eta(w) + \delta\Lambda^* \, \bar\eta(z) \, f(w - z).
}
This variation is
\algns{\label{Noether variation}
  \delta \~S(z)
  &\approx - 2 \sum_{w \in \Ebb} \d\tau\, \big[\delta \Lambda\, \bar \eta(w) \eta(z) \del_{\bar w} f(w - z) + \delta \Lambda^*\, \bar \eta(z) f(w - z) \del_{\bar w} \eta(w)  \big] \\
  &\approx - 2 \sum_{w \in \Ebb} \d\tau\, \big[-\delta \Lambda \, \del_{\bar w} \bar \eta(w) \eta(z) + \delta \Lambda^*\, \bar\eta(z) \del_{\bar w} \eta(w) \big] f(w - z) \\
  &\approx  \frac{2 \d\tau}{A\_S} \, \big[\delta \Lambda \, \del_{\bar z} \bar \eta(z) \eta(z) - \delta \Lambda^*\, \bar\eta(z) \del_{\bar z} \eta(z) \big] \\
  &\approx \frac{2\d\tau}{A\_S} \, \left[ \left(\e^{\i\theta} - 1\right) \, \del_{\bar z} \left( \bar \eta(z) \eta(z)\right) - 4\sin^2\frac\theta2\,  \bar\eta(z) \del_{\bar z} \eta(z) \right].
}
In the first line, the term proportional to $|\delta \Lambda|^2$ was dropped because it was subleading in the smoothing parameters. The second line features a ``summation by parts,'' while in the third one the summation over $w$ treats $f(w - z)$ as a $\delta$-function (proportional to the inverse smoothing area $A\_S^{-1} = 4n\_S k\_S/N_0N$). Finally, the fourth line simply uses the product rule, which is correct to leading order in the smoothing parameters. At no point was it necessary to assume that $\delta\Lambda$ was small.

The upshot of this calculation is that the local transformation \eqref{Noether transf} has two effects: it rescales the Lagrangian at $z$, and it adds to the action a total derivative of a particular combination of the fields. The local rescaling of the Lagrangian precisely matches the change \eqref{Noether Zf change} in the partition function. When $\theta \ll 1$, this effect is not only small on its own, but it is also much smaller than the $O(\theta)$ term in \eqref{Noether variation}. Nevertheless, it is important to note that Noether's theorem actually continues to hold in its familiar form even if $\theta$ is not small.

The physically important result is that the local phase rotation of fields of $+$ chirality changes the action by a divergence of a spacetime current whose $z$-component is zero. A phase rotation of $-$ chirality fields does the same, except the corresponding current has a zero $\bar z$-component. Thus these two chiral components can be assembled into a single two-component object, the Noether current for phase rotations, given by
\bel{\label{def J phase}
  J^{\bar z}(z, \bar z) =  \bar\eta^+(z, \bar z) \eta^+(z, \bar z), \quad J^{z}(z, \bar z) = \bar\eta^-(z, \bar z) \eta^-(z, \bar z).
}
The presence of $A\_S^{-1} \ll 1$ in $\delta \~S(z)$ justifies expanding $\e^{-\delta \~S(z)}$, and this completes Noether's construction. The final result can be recorded in the general form
\bel{
  \avg{\del_{\bar z} J^{\bar z}(z, \bar z) \cdots} = \avg{\del_{z} J^{z}(z, \bar z) \cdots} = 0.
}

\newpage

\subsubsection{Translations}

Recall that translations \eqref{transf time transl} and \eqref{transf space transl} are defined for any $v \in \C$ as
\bel{
  \eta(z, \bar z) \mapsto \eta(z + v, \bar z + \bar v), \quad
  \bar\eta(z, \bar z) \mapsto \bar\eta(z + v, \bar z + \bar v).
}
When $v^\mu$ is smaller than the appropriate smoothing length, the constraints \eqref{smooth constr complex ferm} imply
\bel{
  \eta(z + v, \bar z + \bar v) \approx \eta(z, \bar z) + v^\mu \del_\mu \eta(z, \bar z).
}
Independently of this, note that if $v = n^0 \d\tau + \i n^1$ for integer $n^\mu$, then the action \emph{does not} change, as this translation merely reshuffles terms in the sum over $z$. Thus it is sufficient to only focus on transformations with $-\frac12 \d\tau \leq v^0 < \frac12 \d\tau$ and $-\frac12 \leq v^1 < \frac12$. Here it is natural to pick the ``floor function'' to be $z_\star = z$. The resulting translations still do not truly act ``on-site'' like the phase rotations did, as they mix the fields with their derivatives.

The guaranteed smallness of the translation parameter $v$ makes the Noether procedure simpler than in the case of phase rotations. The reason is that, to first order in the smoothing parameters, there is no obstruction to promoting $v$ to a local parameter $v_z$;\footnote{Keep in mind that $z$ in $v_z$ indicates position, not a covariant index. The two components of $v_z$ are $v^\mu_z$.} the resulting fields remain smooth, as their change is $O(n\_S/N_0)$ or $O(k\_S/N)$. Thus the partition function remains approximately invariant under a translation by an arbitrary local parameter $v_z$. The corresponding variation of the action \eqref{def SE ferm complex} is, with implied summations over $\mu, \nu \in \{0, 1\}$,
\algns{\label{def T}
  \delta \~S[v]
  &\approx -\sum_{z \in \Ebb} \d\tau \left[v_z^\mu\, \del_\mu \overline\eta(z) \c\del \eta(z) + \overline\eta(z) \, \c\del\big(v_z^\mu \del_\mu \eta(z)\big) \right] \\
  &\approx -\sum_{z \in \Ebb} \d\tau \left[ v_z^\mu\, \del_\mu\big(\overline\eta(z) \c\del \eta(z) \big) + \del_\nu v_{z}^\mu\ \overline\eta(z) \gamma^\nu \del_\mu \eta(z)\right] \\
  &\equiv - \sum_{z \in \Ebb} \d\tau\, T_\mu^\nu(z)\, \del_\nu v^\mu_z, \quad \trm{where} \quad T_\mu^\nu(z, \bar z) \equiv \delta^\nu_\mu \, \~\L + \overline\eta(z, \bar z) \gamma^\nu \del_\mu \eta(z, \bar z).
}
Choosing $v_z = \eps \delta_z^{(w)}$ for some point $w \in \Ebb$ then ensures that the corresponding change in the action is
\bel{
  \delta \~S(w) = - \eps \,\d\tau \, \del_\nu T^\nu_\mu(w).
}

The resulting Noether current, the \emph{stress-energy tensor}, is usually reported with both indices lowered. The discussion so far does not motivate the lowering operation, but (merely for completeness) let
\bel{
  T_{\mu\nu}(z, \bar z) \equiv g_{\mu\rho} T^\rho_\nu(z, \bar z), \quad g_{\mu\rho} \equiv \frac12 \bmat0110.
}
The metric $g_{\mu\rho}$ is here defined for $\mu, \rho \in \{z, \bar z\}$. Its effect on indices is to exchange $z \leftrightarrow \bar z$, so a nonstandard but useful notation for the new stress-energy tensor may be
\bel{
  T_{\mu\nu}(z, \bar z) \equiv \frac12 T_\nu^{\bar \mu}(z, \bar z).
}
In components,
\gathl{\label{def T(z)}
  T_{zz}(z, \bar z) = \bar\eta^+(z, \bar z) \, \del_{z} \eta^+(z, \bar z), \qquad T_{z \bar z}(z, \bar z) = -\bar\eta^-(z, \bar z)\, \del_z \eta^-(z, \bar z), \\
  T_{\bar z z}(z, \bar z) = -\bar\eta^+(z, \bar z)\, \del_{\bar z} \eta^+(z, \bar z), \qquad T_{\bar z \bar z}(z, \bar z) = \bar\eta^-(z, \bar z)\, \del_{\bar z} \eta^-(z, \bar z),
}
and the conservation laws are
\gathl{
  \avg{\big(\del_z T_{\bar z z}(z, \bar z) + \del_{\bar z} T_{zz}(z, \bar z) \big) \cdots } = 0,\\
  \avg{\big(\del_z T_{\bar z \bar z}(z, \bar z) + \del_{\bar z} T_{z \bar z}(z, \bar z) \big) \cdots } = 0.
}
This stress tensor can be \emph{improved} to give a traceless object, with $T_\mu^\mu = 0$ or $T_{\mu\bar \mu} = 0$ (summation implied). It can also be made symmetric, which (together with tracelessness) forces the off-diagonal component $T_{z\bar z}$ and $T_{\bar z z}$ to vanish \cite{DiFrancesco:1997nk}. In this case the conservation laws state that the remaining components are (anti)holomorphic, in the same sense as the individual fermion fields \eqref{fermion holomorphy}. This analysis no longer features any subtleties related to the lattice origins of the path integral, and so it will not be further covered here.

\subsubsection{Higher-spin symmetries}

The free Dirac fermion has higher-spin symmetries \eqref{def hs symms} for all $s \leq 2k\_S$. In the path integral, they are expressed as the invariance under
\gathl{\label{def hs symms Grassmann}
  \eta^\alpha(z, \bar z) \mapsto \eta^\alpha(z, \bar z) + v \, \del^s_x \eta^\alpha(z, \bar z), \\
  \bar\eta^\alpha(z, \bar z) \mapsto \bar\eta^\alpha(z, \bar z) + (-1)^{s + 1} v \, \del^s_x \bar\eta^\alpha(z, \bar z).
}
There is a separate higher-spin symmetry for each chirality. To avoid ambiguities, the transformation parameter $v$ should really be written as $v^\alpha_s$.

Note that the smoothness of path integral variables was instrumental in writing these transformations in such a simple position-space form. The same approach was used when discussing translations around eq.\ \eqref{def T}. Only the $s = 0$ case, where no derivatives were used, required the more careful treatment presented in eq.\ \eqref{Noether variation}.

The higher-spin case differs from translations becaise $\del_x^s$ cannot be expressed as a linear combination of complex derivatives. This will make the expressions for the corresponding currents slightly unwieldy. This paper will present only one form for these currents.

Consider the action for the $+$ chirality denoted $\eta(z)$. Its variation under a local version of \eqref{def hs symms Grassmann} is
\algns{
  \delta \~S[v]
  &\approx -2 \sum_{z \in \Ebb}\d\tau \left[(-1)^{s + 1} v_z\, \del^s_x \bar\eta(z)\, \del_{\bar z} \eta(z) + \bar\eta(z) \, \del_{\bar z} \big(v_z\, \del^s_x \eta(z) \big)  \right] \\
  &\approx -2 \sum_{z \in \Ebb}\d\tau \left[ - \bar\eta(z)\, \del^s_x \big(v_z \, \del_{\bar z} \eta(z) \big) + \del_{\bar z} v_z \, \bar\eta(z) \, \del_x^s \eta(z) + v_z\, \bar\eta(z) \, \del_{\bar z} \del^s_x \eta(z)   \right] \\
  &= -2 \sum_{z \in \Ebb}\d\tau \left[ -\sum_{r = 1}^{s} \binom s r \del_x^r v_z \,  \bar\eta(z) \, \del_{\bar z} \del^{s - r}_x \eta(z) + \del_{\bar z} v_z \, \bar\eta(z) \, \del_x^s \eta(z) \right] \\
  &\approx -2 \sum_{z \in \Ebb}\d\tau \left[ \sum_{r = 1}^{s} (-1)^r \binom s r \del_x v_z \,  \del_x^{r - 1} \left(\bar\eta(z) \, \del_{\bar z} \del^{s - r}_x \eta(z) \right) + \del_{\bar z} v_z \, \bar\eta(z) \, \del_x^s \eta(z)  \right].
}
Each term in the sum over $r$'s is a total derivative contribution to the higher-spin current. As with the stress-energy tensor, these terms may be ``improved away.'' Practically, this means that the most important part of the higher-spin current is the remaining, $r$-independent term. Collecting these terms from both chiralities into a single current gives
\bel{
  J_s^{\bar z}(z, \bar z) = \bar \eta^+(z, \bar z) \del_x^s \eta^+ (z, \bar z), \quad
  J_s^{z}(z, \bar z) = \bar \eta^-(z, \bar z) \del_x^s \eta^- (z, \bar z).
}

\subsubsection{Discrete symmetries}

A remarkable fact about the derivation \eqref{Noether variation} was that it was never necessary to assume that the local symmetry transformation was infinitesimal. Requiring the local variation to be smooth, as given by \eqref{Noether transf} or \eqref{Noether transf simplified}, is also enough to make the variation small. This way Noether's theorem for phase rotations was derived by using $k\_S/N$ as the small parameter.

It is thus natural to ask whether this version of the Noether procedure can be applied to discrete symmetries, i.e.\ to symmetries whose parameters \emph{cannot} be made infinitesimal.  This is indeed the case. Consider the charge conjugation symmetry \eqref{def C},
\bel{
  \varphi(x, \tau) \mapsto - \varphi(x, \tau),
}
of the scalar continuum path integral \eqref{def S nc tilde},
\bel{
  \~S[\varphi] \equiv \frac1{2g^2} \sum_{(x,\tau) \in \Ebb}  \left[ \big(\del_\tau \varphi(x, \tau)\big)^2 + \big(\del_x \varphi(x, \tau)\big)^2 \right]\d\tau.
}
In the scalar theory, as opposed to the Dirac fermion, the $\Csf$ transformation is independent of the symmetries generated by particle numbers $n_k$. Thus there is no way to associate a current to $\Csf$ using the standard Noether procedure that relies on infinitesimal transformations.

The local version of charge conjugation is completely analogous to the phase rotation \eqref{Noether transf simplified} by $\Lambda = -1$ in the fermion theory. The transformation of interest is
\bel{\label{Noether transf scalar C}
  \varphi(x, \tau) \mapsto \varphi(x, \tau) - 2 \varphi(x', \tau') \, f(x - x', \tau - \tau'),
}
where $f$ is the same smearing function as in \eqref{Noether transf simplified}. 
The change of the action \eqref{def S nc tilde} under \eqref{Noether transf scalar C} is, with an implied summation over $\mu$,
\algns{
  \delta \~S(x', \tau')
  &\approx - \frac2{g^2} \sum_{(x, \tau) \in \Ebb} \d\tau\ \varphi(x', \tau') \del_\mu \varphi(x, \tau) \del_\mu f(x - x', \tau - \tau') \\
  &\approx \frac2{g^2 A\_S} \left[\del_\mu \big(\varphi(x', \tau') \del_\mu \varphi(x', \tau') \big) - \del_\mu \varphi(x', \tau') \del_\mu \varphi(x', \tau') \right].
}
As long as $g^2 A\_S \gg 1$, this is guaranteed to be a small variation. The current is
\bel{
  J^\mu_\Csf(x, \tau) = \varphi(x, \tau) \del_\mu \varphi(x, \tau).
}
This is a total derivative, and as such the Noether charge will be zero. In fact, the current itself can be ``improved'' to zero. Nevertheless, it is remarkable that it is actually possible to derive a Noether current for a $\Z_2$ symmetry. (See also \cite{Ashton:2008, Seidl:2014} for some earlier proposals.)

A similar procedure can be repeated for $\Psf$ and $\Tsf$ symmetries. However, the corresponding Noether currents will be bilocal. For example, $J^\mu_\Psf(x, \tau)$ will be supported at both $x$ and $-x$.

\subsubsection{Spacetime rotations}

Next off, rotations. They are defined by eq.~\eqref{def rotation}, which can be rewritten as
\gathl{
  \eta^\alpha(z, \bar z) \mapsto \Lambda^{\alpha/2} \, \eta^\alpha (\Lambda z, \Lambda^* \bar z), \quad
  \bar \eta^\alpha(z, \bar z) \mapsto \Lambda^{\alpha/2} \, \bar \eta^\alpha (\Lambda z, \Lambda^* \bar z)
}
for a phase $\Lambda = \e^{\i \theta}$. The transformation of the derivatives is given by \eqref{rotation of derivatives}, and amounts to
\bel{
  \del_\mu \eta^\alpha(z, \bar z) \mapsto \del_\mu(\Lambda z)^\nu  \, \del_\nu \left[\Lambda^{\alpha/2}  \eta^\alpha(z_\star, \bar z_\star)\right], 
}
where $(\Lambda z)^z = \Lambda z$ and $(\Lambda z)^{\bar z} = \Lambda^* \bar z$. (If this is unclear, study eq.\ \eqref{derivative trafo} some more.)

Formally, the Noether trick is clear: generalize from $z' = \Lambda z$ to $z' = \Lambda_z  z$, keep track of the additional terms involving $\del \Lambda$, and at some point specialize to $\Lambda_z = 1 + (\Lambda- 1)\delta^{(w)}_z$ that is nontrivial only at one site $w \in \Ebb$. To make this tractable, constrain the set of allowed functions $\Lambda_z$ by requiring that, for every $z$, $\Lambda z$ and $\Lambda_z z$ have the same modulus and the same ``integer part'' (i.e.~correspond to the same lattice site $z_\star$). This is somewhat analogous to focusing only on translations $v_z$ that moved $z$ by less than half a lattice spacing.

\newpage

These considerations lead to the local generalization of rotations
\bel{
  \Lambda_z z = \e^{\i\theta_{z_\star}} \Lambda z
  \approx \Lambda z + \i \theta_{z_\star} \, z_\star,
}
with the position-dependent deviation $\theta_z$ satisfying $|z| \, |\theta_z| \leq \d\tau$. (Since $\d\tau \ll 1$, this ensures that the local deviation from a vanilla rotation does not move the image of this rotation by more than one lattice spacing in either direction.) In particular, if this rotation-invariant path integral is implemented on a lattice $\Ebb_R$, the local rotation parameter must satisfy $R |\theta_z| \leq \d\tau$.  As with translations, the fact that the local deviation angle $\theta_z$ is small makes it unnecessary to further demand that it also vary smoothly across the lattice, as was needed in the case of phase rotations.

The stage is now set for computing the corresponding Noether current. For simplicity, \emph{replace $z_\star \mapsto z$ in all the arguments of $\theta$ and $\eta$}; do not confuse them with other $z$'s! The transformed action is
\algns{
  -2 \sum_{z \in \Ebb_R} \d\tau \left(\e^{\i \frac{\theta_z}2}\, \bar\eta^+(z)\, \del_{\bar z}\! \left[\e^{\i \theta_z} z\right]^\nu \! \del_\nu \! \left[\e^{\i\frac{\theta_z}2} \eta^+(z) \right] + \e^{-\i \frac{\theta_z}2} \bar\eta^-(z) \del_z\! \left[\e^{\i \theta_z} z\right]^\nu\! \del_\nu\! \left[\e^{-\i\frac{\theta_z}2} \eta^-(z) \right]   \right).
}
The assorted derivatives are
\gathl{
  \del_z \left[\e^{\i \theta_z} z\right]^z = z \, \del_z \e^{\i\theta_z} + \frac12\left(\e^{\i\theta_{z + \d \tau}} + \e^{\i\theta_{z + \i}}\right), \\
  \del_z \left[\e^{\i \theta_z} z\right]^{\bar z} = \del_z \left[\e^{-\i \theta_z} \bar z\right] = \bar z \, \del_z \e^{-\i\theta_z} + \frac12 \left(\e^{-\i\theta_{z + \d\tau}} - \e^{-\i\theta_{z + \i}} \right),\\
  \del_{\bar z} \left[\e^{\i \theta_z} z\right]^z  = z \, \del_{\bar z} \e^{\i\theta_z} + \frac12 \left(\e^{\i\theta_{z + \d\tau}} - \e^{\i\theta_{z + \i}} \right), \\
  \del_{\bar z} \left[\e^{\i \theta_z} z\right]^{\bar z} = \del_{\bar z} \left[\e^{-\i \theta_z} \bar z\right] = \bar z \, \del_{\bar z} \e^{-\i\theta_z} + \frac12 \left(\e^{-\i\theta_{z + \d\tau}} + \e^{-\i\theta_{z + \i}} \right),
}
and
\gathl{
  \del_z \left[\e^{\i \alpha \theta_z/2} \eta^\alpha(z)\right] = \del_z \e^{\i\alpha\theta_z/2}\, \eta^\alpha(z) + \frac12 \e^{\i\alpha\theta_{z + \d \tau}/2}\, \del_\tau \eta^\alpha(z) + \frac1{2\i} \e^{\i\alpha\theta_{z + \i}/2}\, \del_x \eta^\alpha(z), \\
  \del_{\bar z} \left[\e^{\i \alpha \theta_z/2} \eta^\alpha(z)\right] = \del_{\bar z} \e^{\i\alpha\theta_z/2}\, \eta^\alpha(z) + \frac12 \e^{\i\alpha\theta_{z + \d \tau}/2}\, \del_\tau \eta^\alpha(z) - \frac1{2\i} \e^{\i\alpha\theta_{z + \i}/2}\, \del_x \eta^\alpha(z).
}
These expressions all be simplified by expanding to first order in $\theta_z$. This gives
\gathl{
  \del_z \left[\e^{\i \alpha \theta_z/2} \eta^\alpha(z)\right] \approx \del_z \eta^\alpha(z) + \frac{\i\alpha}2  \del_z \theta_z \, \eta^\alpha(z) + \frac{\i\alpha}4 \theta_{z + \d \tau} \, \del_\tau \eta^\alpha(z) + \frac{\alpha}{4} \theta_{z + \i}\, \del_x \eta^\alpha(z), \\
  \del_{\bar z} \left[\e^{\i \alpha \theta_z/2} \eta^\alpha(z)\right] \approx \del_{\bar z} \eta^\alpha(z) +  \frac{\i\alpha}2 \del_{\bar z} \theta_z\, \eta^\alpha(z) + \frac{\i\alpha}4 \theta_{z + \d \tau}\, \del_\tau \eta^\alpha(z) - \frac{\alpha}{4} \theta_{z + \i}\, \del_x \eta^\alpha(z).
}

\newpage

Finally, localize the deviation from a global rotation by setting $\theta_{z_\star} = \theta \, \delta_{z_\star, \, w}$. Then all the terms in the action of the type $\theta_{z + \i} \del_\mu \eta^\alpha(z)$ become approximately $\theta_z\, \del_\mu \eta^\alpha(w)$. This results in more simplifications, and the variation of the action becomes (after relabelling $w \mapsto z$)
\gathl{
  \delta \~S_R
  = -2 \d\tau \Big[ \i z \del_{\bar z} \theta_z \, \bar\eta^+(z)  \del_z \eta^+(z) - \i \bar z \del_{\bar z} \theta_z  \, \bar\eta^+(z) \del_{\bar z} \eta^+(z) + \tfrac\i2 \del_{\bar z} \theta_z \, \bar\eta^+(z) \eta^+(z) + \\
  \quad + \i z \del_z \theta_z \, \bar\eta^-(z) \del_z \eta^- (z) - \tfrac\i2 \del_z \theta_z \, \bar\eta^-(z)\eta^-(z) - \i \bar z \del_z \theta \, \eta^-(z) \del_{\bar z} \eta^-(z) \Big].
}
It is now possible to read off the currents associated to rotations: they are
\gathl{\label{def J rot}
  J\_{rot}^{\bar z}(z, \bar z) = \bar\eta^+(z, \bar z) \left(z \del_z - \bar z \del_{\bar z} + \tfrac12\right) \eta^+(z, \bar z), \\
  J\_{rot}^z(z, \bar z) = \bar\eta^-(z, \bar z) \left(z \del_z - \bar z \del_{\bar z} - \tfrac12 \right) \eta^-(z, \bar z).
}
The half-integer parts are just currents due to chiral phase rotations; these were defined in \eqref{def J phase}. The meat of the rotation currents is in the $z \del_z - \bar z \del_{\bar z}$ operators.

\subsubsection{Dilatations}

Finally, dilatations are defined by \eqref{def scale trafo}, which can be recorded as
\bel{
  \eta(z, \bar z) \mapsto \lambda^\Delta\, \eta(\lambda z, \lambda \bar z), \quad \bar\eta(z, \bar z) \mapsto \lambda^\Delta\, \bar\eta(\lambda z, \lambda \bar z).
}
Recall that $\lambda$ is real and constrained by \eqref{dilatation max} and \eqref{dilatation min}. The derivative transforms as
\bel{
  \del_\mu \eta(z, \bar z) \mapsto \del_\mu(\lambda z)^\nu \del_\nu\left[\lambda^\Delta \eta(z_\star, \bar z_\star)\right].
}
Now consider position-dependent transformations $\lambda_z$ subject to the same kinds of constraints seen with rotations, namely
\bel{
  \lambda_z z = \e^{\epsilon_{z_\star}} \lambda z \approx \lambda z + \epsilon_{z_\star} \, z_\star,
}
for $|\epsilon| < \d\tau/R$. Functionally, dilatations are the same as rotations, except they move things along the radial direction, not perpendicular to it. This is a huge boon, as the same calculation as above gives the dilatation (or scale) currents
\gathl{\label{def J dil}
  J\_{dil}^{\bar z}(z, \bar z) = \bar\eta^+(z, \bar z) \left(z \del_z + \bar z \del_{\bar z} + \Delta \right) \eta^+(z, \bar z), \\
  J\_{dil}^{z}(z, \bar z) = \bar\eta^-(z, \bar z) \left(z \del_z + \bar z \del_{\bar z} + \Delta \right) \eta^-(z, \bar z).
}
The one small issue with this calculation is that, as discussed in Subsection \ref{subsec dilatations}, dilatations are only a symmetry if one ignores the finite size $R$ of the lattice. It is thus plausible to expect that the dilatation conservation law will receive nonuniversal $1/R$ corrections.

\subsubsection{Canonical interpretations of Noether currents}

It is often taken for granted that Noether currents are also canonical operators. The explicit construction presented in this Subsection shows that this intuition is quite problematic. The four currents constructed here --- the fermion number current $J^\mu(z, \bar z)$ in \eqref{def J phase}, the stress-energy tensor $T_{\mu\nu}(z, \bar z)$ in \eqref{def T(z)}, the spacetime rotation current $J\_{rot}^\mu(z, \bar z)$ in \eqref{def J rot}, and the dilatation current $J\_{dil}^\mu(z, \bar z)$ in \eqref{def J dil} --- are all bilinears of smooth spacetime fields $\eta(z, \bar z)$, $\bar\eta(z, \bar z)$, and their derivatives. The temporal smoothness of these fields means that none of them are defined at a single time-slice. In particular, setting $\tau = 0$ (or $z = -\bar z$) does not erase the fact that microscopic path integral fields $\eta_{\tau, x}$ at $\tau = \pm \d\tau$ also contribute to $\eta(z, \bar z)$. Moreover, the fact that each fermionic field in a Noether current is temporally smooth means that the current must contain products of microscopic fields $\bar\eta_{\tau, x} \eta_{\tau', x'}$ for both orderings of $\tau$ and $\tau'$ --- and only one time-ordering has a nice interpretation in terms of canonical operators, as stressed in \cite{Radicevic:1D}. The conclusion is that there is, in general, no natural way to define canonical operators that correspond to Noether currents.

This conclusion contradicts tomes of lore that prescribe rituals, known as quantization rules, for defining canonical operators based on path integral variables. The contradiction is intentional. The point of view of this entire series is that quantum theories are fundamentally defined in a canonical formalism; the pleasant, classical-looking path integrals (and the corresponding classical field theories) arise only from specially chosen quantum theories that are subjected to judicious smoothing procedures. In other words, one should never trust a na\"ively formulated path integral to encode a microscopically well defined quantum theory.

This does not mean that it is always impossible to define canonical versions of Noether currents. Rather obviously, for any quantum symmetry, the temporal component of the Noether current can be associated to the normal-ordered charge density $j^0(x)$, e.g.
\bel{
  (\Psi^+)\+(x) \Psi^+(x) + (\Psi^-)\+(x) \Psi^-(x)
}
for the fermion number symmetry. Spatial components of Noether currents can then also be defined by imposing a conservation equation, e.g.\ $\del_x j^1(x, \tau) \propto \del_\tau j^0(x, \tau)$. This correspondence between operators $j^\mu(x, \tau)$ and Noether currents $J^\mu(z, \bar z)$ is rather loose and should be treated with care, with a clear understanding that it stands a chance of holding only up to the inclusion of contact terms discussed in \cite{Radicevic:1D}.

Such a correspondence is less natural in the case of spacetime symmetries. Here there are no natural canonical counterparts to Noether currents that represent any symmetries of the Hamiltonian. It is still possible to define canonical operators by simply replacing $\eta \mapsto \Psi$ and $\bar\eta \mapsto \Psi\+$. These are still subject to the usual provisos explained above. A more complete analysis of the microscopic properties of these operators is beyond this paper's remit.

\section{Remarks}

This paper has covered a lot of ground, but many ideas remain unexplored. The goal of this final Section is to highlight a few topics of particular interest that did not get sufficient airtime in the bulk of the paper.

\subsection{Fermion doubling}

Fermion doubling is a widely publicized feature (or bug) of lattice theories. The basic conceit is simple \cite{Rothe:1992nt}. If a quadratic, single-derivative Lagrangian in $D$ dimensions is discretized in a way that preserves the antihermiticity of the derivative operator, the resulting action will give rise to propagators with $2^D$ times as many momentum-space poles as expected from the continuum theory. In other words, this na\"ively discretized action will describe $2^D F$ particle types for some $F \in \Nbb$. The $2^D$ different types of particles that result from this doubling are called \emph{tastes}. The main issue here is that generic interactions in the discretized theory will couple different tastes. The resulting theory then has no reason to resemble the desired interacting cQFT, which was only ever supposed to have $F$ coupled types of particles.

Over the years, a huge amount of attention was devoted to formulating lattice actions in which the tastes can be guaranteed to decouple even after interactions are turned on \cite{Ginsparg:1981bj, Kaplan:1992bt, Narayanan:1993ss, Neuberger:1997fp, Neuberger:1998wv, Luscher:1998pqa}. The philosophy of this series is orthogonal to this whole body of work.

It is important to understand this difference. In this series, the Hamiltonian of a lattice theory takes precedence over any other fundamental notion. Doubling in the canonical formalism is slightly different from the story given above. For an example, consider the Dirac Hamiltonian \eqref{def H Dirac},
\bel{
  H = \i \sum_{v, u = 1}^{2N} \psi_v\+ K_{vu} \psi_u, \quad K_{vu} = \delta_{v,\, u - 1} - \delta_{v,\, u + 1}.
}
It is quadratic in fermion fields and involves an antihermitian discrete derivative operator $K_{vu}$. Its dispersion thus has two nodes, at $k = 0$ and $k = N$, and the low-energy excitations around these nodes can be understood to correspond to two different tastes. In $d = 1$, it is perfectly natural to identify the tastes of a spinless fermion with different components of a Dirac spinor \cite{Susskind:1976jm}. In $d > 1$, the situation is a bit subtler, but a similar point of view can be justified. (More on this will be said in the next paper of this series, \cite{Radicevic:3D}.) In any dimension, this ``staggered fermion'' approach makes it possible to write down free Hamiltonians with the desired number of fermion fields, and to then include interactions between just the right components. The idea is thus to give up thinking about generic lattice interactions, and instead to start from a specific lattice Hamiltonian (or a family of them) and to use it to \emph{define} a cQFT using the formalism developed here.

One immediate advantage of this approach is that it eliminates doubling issues associated with the temporal direction in the fermion path integral. For example, consider the Berezin path integral derived from the Dirac Hamiltonian, as given by \eqref{def Zf ferm UV}. The Grassmann fields live in momentum space in this formula, but they can easily be transformed back to position space if desired. Either way, the discrete temporal derivatives emphatically have \emph{no} specific hermiticity. In frequency space, this is reflected by the fact that the kernel $\e^{-\i \omega_n \d\tau} - 1$ in \eqref{def SE ferm UV} is neither real nor imaginary. As mentioned below \eqref{def L tilde}, this means that the spacetime dispersion relation has only one node in frequency space, and so there are no doublers here.

The antihermiticity of the derivative operator in the Lagrangian is thus a red herring. The underlying quantum theory can be perfectly well defined without this requirement. This is a stark example of how seemingly natural requirements in the path integral language are not always justified.

It may seem that this is a dangerous path to go down. After all, if the Lagrangian of a single Grassmann mode has form
\bel{
  \L = \i \bar\eta_\tau K_{\tau,\, \tau'} \eta_{\tau'},
}
then surely $K$ must be antihermitian in order to make $\L$ real, i.e.\ invariant under complex conjugation? The rub lies in the fact that there is no natural action of complex conjugation on $\L$ that sends $\eta \mapsto \bar \eta$. As repeatedly stressed when deriving the Berezin integral in \cite{Radicevic:1D}, and again below \eqref{def Zf ferm UV} in this paper, $\eta$ and $\bar\eta$ are not related by any ordinary Hermitian conjugation. They are totally independent destruction operators that act on an auxiliary fermion system. Thus $\L$ should not be understood as a c-number; it is an operator on an auxiliary Hilbert space, and the Berezin integral is a trace over this space.

The antihermiticity of the analogous derivative operator along spatial directions \emph{does} follow from the Hermiticity of the Hamiltonian. The only doublers that a properly defined path integral should know about are thus the ones coming from spatial directions.

This does not necessarily disqualify numerical results obtained from lattice actions that feature temporal doubling. It is possible that the universal part of a correlation function or free energy computed via Monte Carlo does not care whether the action had temporal doublers that were later removed by explicit ``rooting,'' i.e.\ by manually rescaling the numerically computed free energy by a power of two to eliminate the unwanted tastes. The point here is that lattice theories can be much more than machines for spouting universal terms --- and if they are truly taken seriously, their path integrals must be constructed very carefully.

The point of view described in this short Subsection appears to be rather unorthodox. It is not clear if it can help bring about more efficient Monte Carlo computations in fermionic theories. It certainly does not claim to solve the \emph{sign problem}, which remains the most important numerical bottleneck in these theories (see e.g.\ \cite{deForcrand:2010ys}). Nevertheless, this ``lattice-first'' approach to doubling may prove to be a fertile alternative to the reigning philosophy.

\subsection{Classifying interactions}

As already anticipated in the Introduction, the focus on free cQFTs throughout the bulk of this paper may make the results seem a bit trivial. In a way, this focus was unavoidable, as free theories (in the sense of Subsection \ref{subsec definition}) are natural starting points for defining cQFTs in the first place. But interactions \emph{can} be added to this analysis. This was mentioned in Subsection \ref{subsec EFTs} when discussing effective field theories. Now, \emph{repetita juvant}, the same point will be made in a slightly different way, following the discussion of perturbations of the Ising CFT given in \cite{Radicevic:2019mle}.

There are three classes of interactions that should be distinguished:
\begin{description}
  \item[Smooth (continuum) interactions] are obtained by adding only operators at momenta $k \in \Pbb\_S$ to the free Hamiltonian. None of these perturbations change the expectation values $\avg{n_k}$ for $k \notin \Pbb\_S$. All the resulting theories thus have the same entanglement pattern at short distances.

      Axiomatic approaches to cQFT often use these kinds of interactions, see e.g.\ \cite{Glimm:1975tw}. In these works, smooth interactions may be called ``normal-ordered'' or ``renormalized,'' to indicate that all the large-momentum operators have been removed. Another sector of literature where the smoothness of interactions is explicit is the study of Luttinger liquids \cite{Haldane:1981zza}, which are obtained by deforming the free Dirac theory by normal-ordered products of four-fermion operators. Many other authors keep implicit the fact that they work with continuum interactions. Indeed, if the interactions are weak and slowly varying, chances are that the modes outside $\Pbb\_S$ will not be sensitive to them.

  \item[Precontinuum interactions] do not change the precontinuum basis but do change occupation numbers $\avg{n_k}$ at all momenta $k \in \Pbb$. The most general precontinuum interactions are shown in the Hamiltonian \eqref{def H general}.

      The simplest examples are masses and chemical potentials. A large mass term can change the ground state of a Dirac fermion from a Dirac sea with $\avg{n_k} = \theta(-k)$ to a trivial state with $\avg{n_k} = 0$.

  \item[Lattice interactions] may redefine what one means by a precontinuum basis. A rather generic example was given in eq.\ \eqref{def H standard}. BCS terms, gauging symmetries, and adding a potential that leads to spontaneous symmetry breaking are other examples.
\end{description}

Unfortunately, this paper does not make it much easier to actually solve a given interacting theory. Even the clock model studied here was not explicitly solved. Perturbation theory, symmetry analyses, numerics, and RG remain indispensable ingredients when studying interactions. This paper does, however, provide a general framework in which lattice and continuum interactions can be meaningfully contrasted to each other.

\subsection{Future directions}

There are myriad directions to pursue using the tools developed here. The upcoming papers in this series will focus on generalizing the lattice-continuum correspondence to higher dimensions. Many questions in $d = 1$ remain to be addressed, however. Here is an incomplete list:

\begin{enumerate}
  \item Conventional lore on continuum limits in lattice QFT often distinguishes between finite and infinite lattices. In this context, both types of lattices are actually infinite --- the lattice spacing $a$ is assumed to be infinitesimal in both cases. What distinguishes the two situations is whether the system size $L$ is also taken to infinity or not.

      This paper has emphasized that $L$ is not a microscopically well defined quantity. (See Subsections \ref{subsec dimensionful} and \ref{subsec eng scale inv} for a refresher, if needed.) It is thus not fundamentally meaningful to ask whether $L$ is infinite or not. The meaningful question is whether the underlying lattice is periodic or not. If $\Mbb$ has boundaries, a host of new effects becomes possible, such as anomaly inflow \cite{Callan:1984sa} and the related existence of symmetry-protected edge modes \cite{Haldane:1982rj, Affleck:1987vf, Kitaev:2001kla, Chen:2011pg}. Even if there are no nontrivial excitations on the boundary of $\Mbb$, the physics in the ``bulk'' will still be different from the one analyzed in this paper. The lattice-continuum correspondence in this case has not been studied yet.
  \item Perhaps the most difficult and important question that remains is that of universality. Why are temporal smoothing or restricting the spacetime torus $\Ebb$ to a disk allowed operations, up to renormalization?  Or, more provocatively, \emph{when} are they allowed?
  \item Part of the Ansatz in the clock model was that the only taming  backgrounds of interest are $\varphi\^{cl}_x$ of form \eqref{taming backgrounds}. Why was this so?  When is it necessary to include nontrivial \emph{smoothing backgrounds} $p\^{cl}_x$?  Recall that these $p\^{cl}$'s appeared in the study of the harmonic oscillator cQM in \cite{Radicevic:1D}, where they were understood as nontrivial spin structures on the target space. When do nontrivial target space spin structures (or their paraspin generalizations \cite{Radicevic:2018okd}) become important in cQFTs?
  \item What is the lattice version of modular invariance?
  \item How does the analysis of Section \ref{sec symmetries} latticize other notions of conformal theory? Which operators satisfy the Virasoro algebra? How does this series synergize with the recent work \cite{Milsted:2017csn, Zou:2019dnc, Zou:2019iwr}?
  \item Abelian bosonization of smooth fields in $d = 1$ was proven in \cite{Radicevic:2019jfe} using the present techniques. In particular, $k\_S$-dependent effects were computed and shown to play a crucial part in making sense of this duality. How does the analogous story work for nonabelian bosonization \cite{Witten:1983ar}?
\end{enumerate}

\section*{Acknowledgments}

It is a pleasure to thank Nathan Benjamin, Luca Delacr\'etaz, Matt Headrick, and Mithat \"Unsal for useful conversations. This work was completed with the support from the Simons Foundation through \emph{It from Qubit: Simons Collaboration on Quantum Fields, Gravity, and Information}, and from the Department of Energy Office of High-Energy Physics grant DE-SC0009987 and QuantISED grant DE-SC0020194.

\bibliographystyle{ssg}
\bibliography{Refs}

\end{document}